# Blockchain for the Cybersecurity of Smart City Applications

Omar Cheikhrouhou [a], *Member, IEEE*, Ichrak Amdouni [b], Khaleel Mershad [c], *Member, IEEE*, Maryem Ammi [d], Tuan Nguyen Gia [e], *Member, IEEE*

*Abstract*—Cybersecurity is an inherent characteristic that should be addressed before the large deployment of smart city applications. Recently, Blockchain appears as a promising technology to provide several cybersecurity aspects of smart city applications. This paper provides a comprehensive review of the existing blockchain-based solutions for the cybersecurity of the main smart city applications, namely smart healthcare, smart transportation, smart agriculture, supply chain management, smart grid, and smart homes. We describe the existing solutions and we discuss their merits and limits. Moreover, we define the security requirements of each smart city application and we give a mapping of the studied solutions to these defined requirements. Additionally, future directions are given. We believe that the present survey is a good starting point for every researcher in the fields of cybersecurity, blockchain, and smart cities.

## CONTENTS

Corresponding author: Omar cheikhrouhou
[a] CES Laboratory, National School of Engineers of Sfax, University of Sfax, Sfax 3038, Tunisia.
[b] National Computer Science School (ENSI), Cristal Laboratory, University of Manouba, Tunisia.
[c] Computer Science and Mathematics Department, School of Arts and Sciences, Lebanese American University (LAU), Beirut, Lebanon.
[d] Criminal Justice College, Naif Arab University for Security Sciences, Riyadh, Saudi Arabia.
[e] Department of Computing, University of Turku, Finland.

## I. INTRODUCTION

Blockchain is a distributed ledger that permits the tracking and exchange of resources and assets in a secure and transparent manner [282]. Assets are anything having a value and can be tangible (such as houses, cars, moneys) or intangible (patient vital signs, copyright, digital documents) [225].

Thanks to its inherent characteristics, Blockchain technology has gained a lot of attention world wide. The number of scientific papers and industrial projects dealing with blockchain is significantly increasing [85]. According to

Gartner report, Blockchain will support the global movement and tracking of 2 trillion of goods and services annually by 2023 [85]. Moreover, Blockchain technology will continue its evolution and integration in industrial and business projects and it will reach its maturity by 2030 [85].

Although there are several works dealing with blockchain, there is no single survey that gives an overview about the use of blockchain in smart cities applications and the mapping of these solutions to the requirements of these applications. The present survey tries to fill this gap by presenting an exhaustive survey about the use of blockchain for the cybersecurity of smart cities applications and present future directions and open issues that need to be addressed by the industrial and scientific communities. We believe that this survey is a good starting point for researcher working around the cybersecurity, smart cities, and blockchain fields.

### A. Review of Related Surveys

As an emerging technology, blockchain has gained a lot of attention from researchers and, therefore, several surveys have been proposed. Some surveys are related to the use of blockchain in IoT such as [19, 223, 80, 79, 114]. For instance, authors in [19] presented a comprehensive survey about applications of blockchain in IoT. They first presented blockchain fundamentals and the consensus algorithms. Then, they presented the motivation for integrating blockchain with IoT. Moreover, they reviewed the proposed solutions to achieve security in IoT, including privacy, identity management, trust, and main security goals. The paper [223] analysed unique features of blockchain technology and outlined various ways of integrating IoT and blockchain. For this purpose, the authors discussed the challenges, benefits and open issues of the IoT-blockchain integration. Moreover, they reviewed existing IoT-blockchain platforms and applications. Similarly, [80] surveyed existing blockchain protocols for IoT networks and provided a classification of threat models. The authors in [79] described the use of blockchain in IoT-based applications including sensing, smart living applications, intelligent transportation systems, wearable devices, supply chain management, mobile crowd sensing, etc.

The survey [225] studied the security services provided by blockchain technology independently from smart cities applications. They focused on the use of blockchain technology to ensure secure network services and outlined associated challenges with the proposed blockchain-based approaches.

Some surveys on blockchain focused on a single application of smart cities such as smart home [186], smart healthcare [53, 231, 76], and smart transportation [107]. Note that, these surveys will be cited in their corresponding section in this paper.

The widespread adoption of blockchain technology in smart cities applications has led to the appearance of several surveys papers. The most related ones to our work are presented in Table I.

The authors in [182] reviewed the technical implementation of Blockchain technology in different academic and industrial fields including healthcare, financial, IoT, government, power grid, transport system, commercial world, cloud computing, reputation, E-business, and supply chain. Although the authors claim to present a wide area of applications, their analysis is not deep enough. Moreover, the defined areas present an overlap between them, such as the IoT area with other smart cities applications, and the financial area with the commercial and e-business ones.

TABLE I: **Blockchain for Smart Cities** Existing Surveys

| Ref | Scope | Description | Limits |
|---|---|---|---|
| [182] | Smart cities applications including healthcare, financial, IoT, government, power grid, transport system, commercial world, cloud computing, reputation, E-business, supply chain. | Utilisation of blockchain technology in several smart cities applications and their associated security and privacy issues have been discussed | Authors loosely presented these areas. |
| [48] | A systematic review of blockchain adoption in several applications including education, supply chain, business, healthcare, IoT, privacy, and data management. | Present a systematic literature review of blockchain-based applications including education, supply chain, business, healthcare, IoT, privacy, and data management. | ● The studied applications are loosely reviewed ● The cited solutions are neither presented nor discussed. ● The relation between reviewed solutions and smart cities security is not discussed. |
| [162] | General overview of the blockchain technology and its integration in some applications | The paper reviews the main applications based on the blockchain. | Only a general overview was presented. Lack of solutions' description. |
| [241] | Review of blockchain-based solutions in business, healthcare, and IoT. | Presented fundamental concepts of core blockchain architecture and its applications in three major areas: business, healthcare, and IoT. | Limited to only three categories of applications of smart cities. |
| [265] | Blockchain technology applied to smart cities including smart citizen, smart healthcare, smart grid, smart transportation, and supply chain management. | First, a brief overview of smart cities and blockchain was presented. Then the applications of blockchain to some smart cities applications was reviewed. | The security aspects were not addressed, and how blockchain can fill up the security requirements of smart cities applications is not answered. |
| [42] | Blockchain for smart cities communities including healthcare, transportation, smart grid, supply chain management, financial systems, and data center networks. | ● Benefits of blockchain for smart cities ● Security requirements and challenges of smart cities ● Review of existing blockchain-based solutions for smart cities | ● Only general benefits of blockchain were addressed. ● Relation between blockchain-based solutions and security requirements is not discussed. ● Only few references are presented. |
| [8] | Smart cities applications including healthcare, smart grid, intelligent transportation, IoT, data center networking, financial, and voting system | ● Blockchain Technology overview ● Applications of blockchain for different smart cities applications. ● Process models ● Communication infrastructure supporting Blockchain | Did not present smart cities security requirements |
| [65] | Blockchain 3.0 applications including healthcare, identity management systems, access control systems, decentralised notary, supply chain management, and electronic voting system | ● Distributed Ledgers and Blockchain overview ● Applications of blockchain for the selected smart cities applications. | Did not focus on the security aspects |

In another work, Xie et al. [265] surveyed the state-of-the-art of blockchain technology as a solution that improves the security, efficiency, smartness and performance of smart cities. The authors first gave a brief overview of smart cities and blockchain. Then, they reviewed the applications of blockchain in some smart cities applications including smart citizen, smart healthcare, smart grid, smart transportation, and supply chain management. However, this work did not address the security aspects of smart cities, and did not discuss how blockchain can fill up the security requirements of smart cities applications.

Authors in [42] reviewed the use of blockchain in the context of smart cities. They first presented the security requirements and challenges of smart cities applications. Then, they reviewed existing blockchain-based solutions for smart cities applications including healthcare, transportation, smart grid, supply chain management, financial systems and data center networks. However, the authors presented only the general benefits of blockchain and did not focus on the security aspects. Moreover, the relation between blockchain-based solutions and smart cities security requirements is not discussed. Another issue with this survey is that it only presents a few references for each category of applications. For example, the authors presented only seven references in the healthcare field, however in our survey we presented more than twenty references.

TABLE II: Comparison of Present Work with Existing **Blockchain for Smart Cities** Surveys

| Survey | Features | | | | | | Smart Cities applications | | | | | |
|---|---|---|---|---|---|---|---|---|---|---|---|---|
| | Bgrd | Bfts | Reqs | ES | FD | MS | HC | IT | SG | Ag | Sp | SH |
| [182] | ✓ | ✗ | ✗ | ✗ | ✗ | ✗ | 8 | 3 | 3 | ✗ | 3 | ✗ |
| [162] | ✓ | ✗ | ✗ | ✗ | ✓ | ✗ | 11 | ✗ | ✗ | ✗ | ✗ | ✗ |
| [241] | ✓ | ✓ | ✗ | ✗ | ✓ | ✗ | 12 | ✗ | 3 | ✗ | 4 | ✗ |
| [265] | ✓ | ✗ | ✗ | ✗ | ✓ | ✗ | 9 | 6 | 12 | ✗ | 10 | 4 |
| [42] | ✓ | ✓ | ✗ | ✗ | ✓ | ✗ | 7 | 12 | 8 | ✗ | 6 | ✗ |
| [8] | ✓ | ✗ | ✗ | ✗ | ✗ | ✓ | 17 | 17 | 16 | ✗ | 3 | ✗ |
| [65] | ✓ | ✗ | ✗ | ✗ | ✗ | ✗ | 7 | ✗ | ✗ | ✗ | 8 | ✗ |
| Present Study | ✓ | ✓ | ✓ | ✓ | ✓ | ✓ | 25 | 72 | 30 | 30 | 45 | 13 |

**Features notations:**
Bgrd: Blockchain Background, Bfts: Blockchain Benefits, Reqs: Requirements, ES: Existing surveys, FD: Futur directions,
MS: Mapping with smart cities security requirements
**Applications notations:**
HC: HealthCare, IT: Intelligent Transportation, SG: Smart Grid, Ag: Agriculture, Sp: Supply chain, SH: Smart Homes,
✓: considered, ✗: Not considered

The paper [8] surveys the use of blockchain technology in smart communities and smart cities applications, including healthcare, smart grid, intelligent transportation, IoT, data center networks, financial, and voting systems. However, the authors missed to present the requirements of smart cities applications and the survey lack the smart agriculture and smart homes applications.

The reference [48] is a systematic literature review of blockchain adoption for several applications including education, supply chain, business, healthcare, IoT, privacy, and data management. Although the authors presented an exhaustive list of blockchain applications, they have loosely reviewed them. Moreover, they addressed these applications independently from the context of smart cities applications and therefore they did not consider the security concern of smart cities applications.

Authors in [241] presented fundamental concepts of core blockchain architecture and its applications in three major areas: business and vehicular industry, healthcare, and IoT. For each area, they presented the challenges and solutions that have been proposed by the research community and industry.

Lu et al [162] presented a general overview of the blockchain technology and its integration in some applications such as healthcare and IoT. However, the authors did not explain the proposed solutions and their contributions.

The authors in [65] review the use of blockchain 3.0 in some selected smart cities applications including healthcare records management, identity management systems, access control systems, decentralised notary, supply chain management, and electronic voting system. For each application, the authors describe the problem formulation that blockchain tries to solve and then give some blockchain-based solutions description. However, the authors only give a general idea of the selected applications and present only some selected references. Moreover, the authors did not focus on the security aspects.

### B. Contributions

Differently from previous surveys that have either addressed the security of smart cities independently from blockchain technology or those that have focused on a single smart cities application, this survey enriches this state of the art by discussing the blockchain based solutions for the cybersecurity of multiple smart cities applications, namely: smart healthcare, smart transportation, smart agriculture, supply chain management, smart grid, and smart homes (Figure 1). Table II highlights the key differences between the present survey and the most recent related surveys. It is clear from this table that some features such as Blockchain benefits, requirements, and the citation of existing surveys for each field of smart cities are missing in previous surveys. Moreover, the mapping of smart cities requirements with the reviewed solutions is slightly studied by only paper [245]. In Table II, we also indicated the number of papers reviewed in each survey for each smart cities applications. It is worth noting that the number of papers discussed in the present survey is significantly bigger than the one of the previous surveys. Moreover, some applications such as agriculture and smart homes are not addressed in previous surveys.

More precisely, the present survey provides the following contributions:

- We present a detailed overview of the blockchain technology including fundamentals, architecture, characteristics, consensus algorithms and types.
- We present a comprehensive review of existing blockchain-based solutions for the main smart city applications, namely: smart healthcare, intelligent

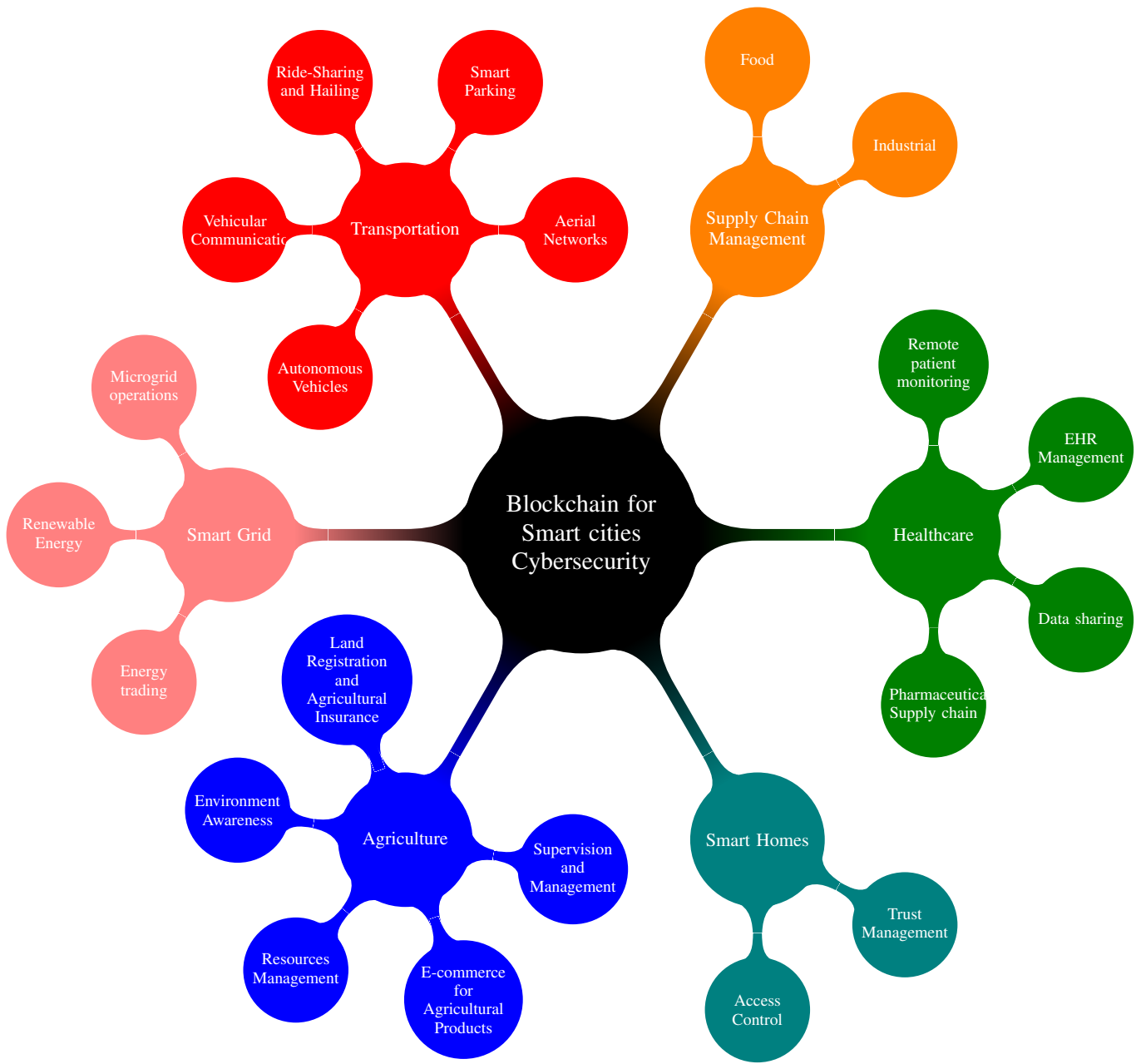

Fig. 1: The Studied Smart Cities Applications

transportation, smart agriculture, supply chain management, smart grid, and smart homes.

- We provide a mapping between existing solutions and the security requirements of smart cities, and we discuss how the proposed solutions can answer these security requirements.
- For each smart city application we present future directions and open issues.

### C. Structure of the Paper

The roadmap of this paper is as follows. First we start by presenting an overview of the blockchain technology (Section 2). Then we review the blockchain solutions for each smart city application in a separate section. The studied smart city applications are smart healthcare (Section 3), smart transportation (Section 4), smart agriculture (Section 5), supplychain management (Section 6), smart grid (Section 7), and smart homes (Section 8). For each smart city applications, we start by presenting the existing surveys, then we discuss the benefits of using the blockchain technology, then the application requirements, then we give a mapping of the reviewed solutions with these defined requirements, and finally we discuss future directions. Figure 1 presents the studied smart city applications with their use cases.

## II. BLOCKCHAIN OVERVIEW

Blockchain can be considered as a distributed ledger that keeps track of all transactions. The key characteristic of blockchain is that it is an append only database, in other words we can add a new transaction but we cannot update or delete an existing one. These transactions are bundled into a block, and blocks are linked together through a cryptographic hash to form the blockchain. Moreover, the distributed ledger is generally replicated at every node in the blockchain network. These characteristics make blockchain a trustworthy and immutable record keeping service.

### A. Blockchain Structure

As shown in Figure 3, Blockchain is a chain of blocks. A block consists of a block header and a block body, as illustrated in Figure 2. The block header includes metadata such as the block version, nonce, timestamp (the block time creation), height (the block position in the chain), Merkle tree root hash (computed over all transactions), parent block hash, and current block hash. The block body contains a set of transactions and the transaction counter.

The hash is computed for each block and inserted in the subsequent block, which permits to cryptographically link blocks together and make the blockchain immutable. As shown in Figure 3, if an attacker tries to modify the value of a transaction in block $i$, the hash of this block will be invalid and therefore all subsequent blocks will be also invalid as they contain (or use) this hash. The Merkle tree root permits the fast searching of a transaction. The nonce is a random number resolved by the block miner (the node that creates the new block) and which must produce a hash below a specific target.

### B. Blockchain Characteristics

Blockchain technology has received a lot of attention from both industry and academia due to its distributed, decentralized, immutability, and transparency properties.

- **Distributed**: In blockchain, the storage of data is done in a distributed manner. A distributed ledger is stored in different nodes in the blockchain network. Moreover, the decision to add new data (new block) is taken through consensus protocol and not by a central entity.
- **Decentralization**: The blockchain system does not require a centralized third party to operate in a P2P manner. Transactions are endorsed in a decentralized manner by the peer-to-peer network and without the intervention of a central entity.
- **Immutability**: The distributed ledger in Blockchain is composed of a set of blocks linked together through a cryptographic hash as shown in Figure 3. Alteration of block data is impossible as it will make the modified block invalid and all subsequent blocks. This makes blockchain immutable and secure. Moreover, transactions are signed with the help of digital signatures (Figure 3).
- **Transparent**: Before being added to the ledger, transactions are validated by participant nodes. Therefore, participants can track and view the changes on the blockchain.

### C. Blockchain Types

Blockchain systems can be broadly classified into three types as illustrated in Figure 4.

*1) Public:* A public or permissionless blockchain is a decentralized open source platform that facilitates every individual to join and perform mining. There is no access restriction and every participating node is able to write, read, validate and mine blocks. Generally, a set of transactions are bundled into a new block and participants compete to found the right nonce (the mining process) in order to earn the reward. The most well-known public blockchain is bitcoin and Ethereum.

*2) Private:* A private or permissioned blockchain is a decentralized network that allows private data sharing amongst a specified group of people or within an organization. The access to a private blockchain is controlled and generally restricted to a specific group of people with the same affiliation or field.

Private blockchain also controls who can participate in the consensus protocol. Although a private blockchain is restricted and controlled by an authority it still benefits from other blockchain properties such as transparency, distributed ledger, and consensus. The most known private blockchains are Hyperledger, MultiChain [88], and Ripple.

*3) Consortium:* A consortium blockchain is a hybrid version of private and public blockchain in which a group of organisations control the access to the blockchain, and the consensus and block validation decisions. As examples of consortium blockchain platforms, we can cite Hyperledger Fabric [27], Quorum [170] and Corda [46], that can be used to deploy either consortium or private blockchains [156].

### D. Consensus Mechanisms

The consensus mechanism in the blockchain is an important component for the operation of the blockchain system as it decides whether a block will be added or not. The first consensus mechanism which was proposed with the bitcoin system is PoW. Due to the inefficiency of the PoW consensus as it requires a lot of computation and energy power, several new consensus methods were proposed.

*1) Proof-of-Work (PoW):* In this consensus algorithm the miner nodes compete to find the right nonce that gives a hash value lower than a specified threshold. Therefore, the node with more computational power will have more chance to find the appropriate nonce first. Upon finding the valid nonce, miners broadcast the block to all other network nodes to verify it. If all the miners approve the block, it will be appended to the existing chain. The first miner who finds the valid nonce will be rewarded and its block will be added to the chain. The problem with this consensus algorithm is that high computational power is wasted in solving the mathematical puzzle.

*2) Proof of stake (PoS):* In this consensus mechanism the nodes with more resources (more wealthy: having good stake) have more chances to add a block to the network. Chances of doing a block validation depends on the wealth of the participating node i.e, its stake in the system. However, this approach favours the wealthier nodes as they will receive more block validation opportunities, and as a consequence they become more dominant in the network which results in centralization or unfair distribution.

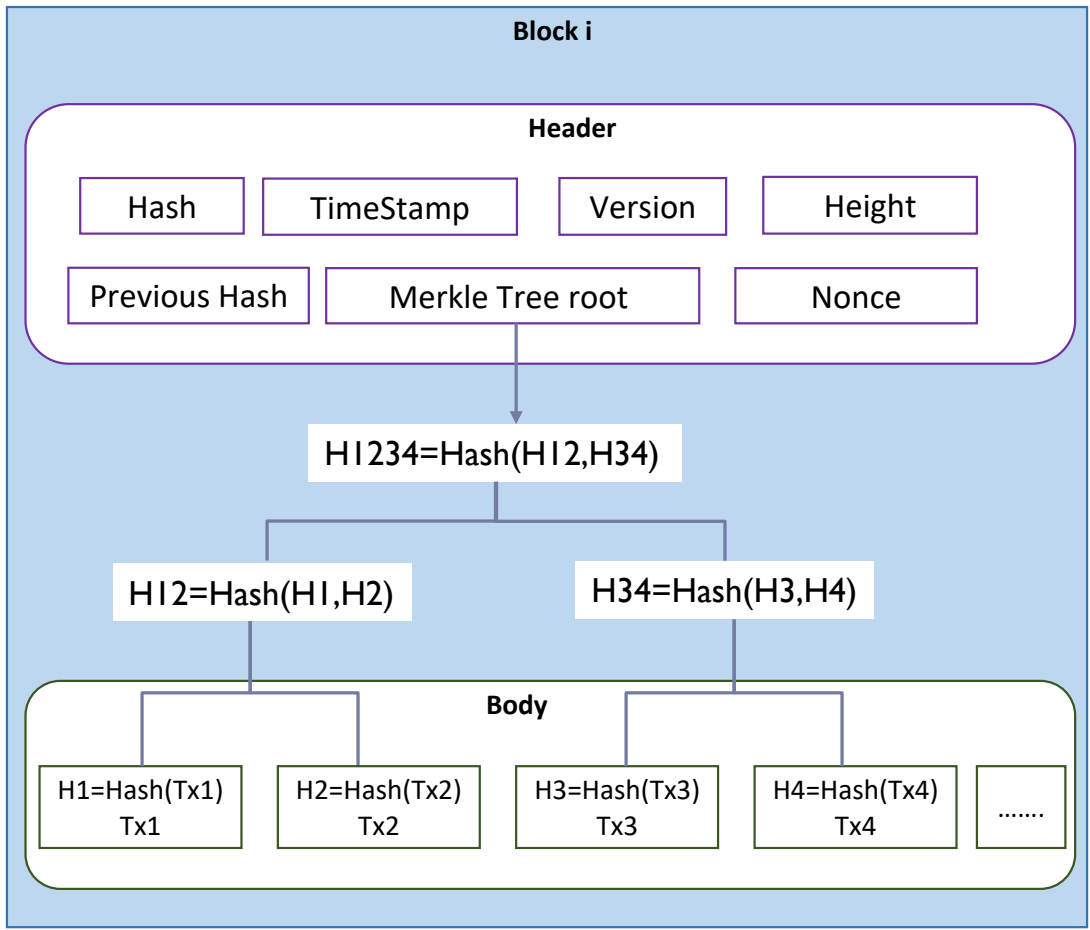

Fig. 2: The block structure in the blockchain

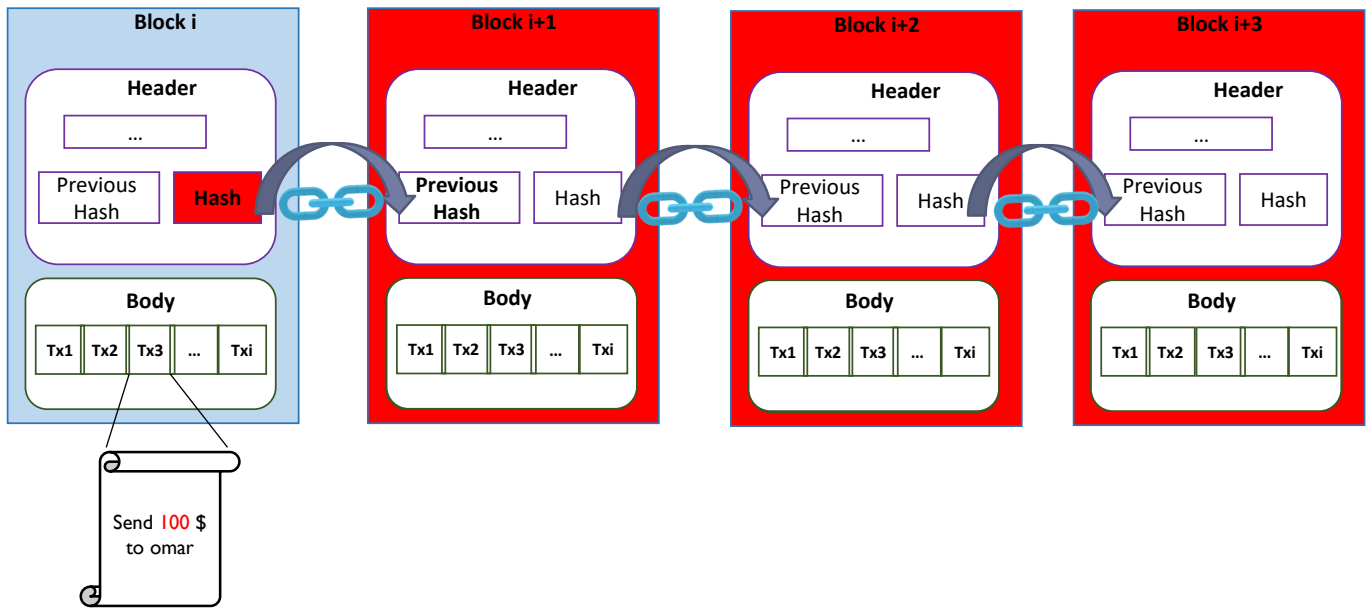

Fig. 3: Blocks are cryptographically linked, which makes them immutable

*3) Delegated proof of stake (DPoS):* To overcome the previously mentioned drawbacks of the PoS consensus

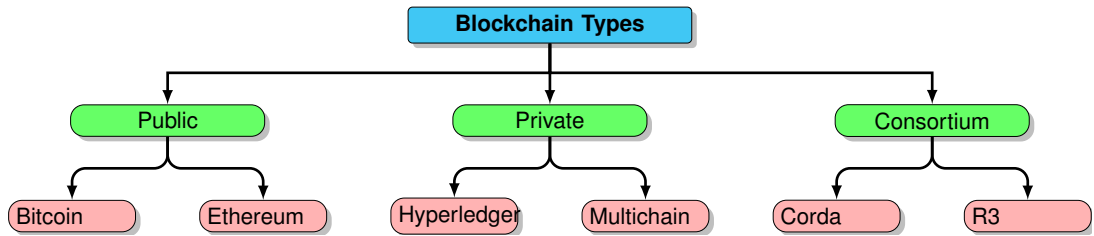

Fig. 4: Blockchain Types

TABLE III: Comparison of blockchain types.

| Properties | Public | Private | Consortium |
|---|---|---|---|
| **Nature** | Decentralized and open | Restricted and controlled | Restricted and controlled |
| **Consensus protocols** | PoW, PoS, DPoS | PBFT, RAFT | PBFT |
| **Transaction throughput** | Low | Medium | High |
| **Participant type** | Anonymous | Trusted and Identified | Trusted and Identified |
| **Permissions** | Permissionless | Permissioned | Permissioned |
| **Energy consumption** | High | Low | Low |
| **Scalability** | High | High | Low |
| **Efficiency** | Low | High | High |
| **Example** | •Bitcoin<br><br>•Litecoin<br><br>•Dash<br><br>•Ethereum<br><br>•Factom<br><br>•Blockstream | •Ripple<br><br>•Multichain<br><br>•Hyperledger | •Quorum<br><br>•R3<br><br>•Corda |

algorithm, the delegated PoS was proposed, which can be considered the democratic version of PoS. DPoS is an elective consensus scheme in which every node with a stake in the network can vote for validator nodes. The validator nodes called also witnesses will in turn valid new blocks.

*4) Proof of Space:* PoSpace is similar to PoW except that the puzzle requires a storage space instead of computation power. A miner proves its ability to create a new block by allocating the required storage space to perform the mining process. In other words, instead of having a high computational capability, the mining node needs to show a high storage capability.

*5) Practical byzantine fault tolerance (PBFT):* PBFT is a consensus algorithm used where members are partially trusted. PBFT tries to reach an agreement between participants even in presence of malicious nodes. However, the number of malicious nodes must not exceed one-third of the total number of participants. PBFT consists of three stages, which are pre-prepare, prepare and commit. In all these three stages, a node would move to the next stage only if it receives the same reply from two-thirds of all the nodes in the network. This enables the PBFT consensus mechanism to run effectively even under the presence of some malicious byzantine replicas. PBFT suffers from scalability problems due to the exponentially increasing messages with new nodes addition.

*6) Proof of Elapsed Time (PoET):* In the PoET consensus algorithm, each node generates a random number and waits for that random number. The first node that completes its waiting time has the opportunity to generate the new block.

*7) Proof of authority (PoA):* The proof of authority is based on the identity of validators instead of their stakes. In other words, the node with the well-reputed identity will be selected to validate the new blocks.

## III. BLOCKCHAIN FOR SMART HEALTHCARE

The integration of emerging IT technologies into healthcare 4.0 has raised the security concern. In this section, we identified the main security requirements for next generation healthcare applications and then we reviewed existing blockchain based solutions and highlight how they can achieve these security requirements. Finally, we give some open issues that need to be addressed as future directions. First of all, we start by presenting the state of the art of existing surveys dealing with blockchain for healthcare.

### A. Existing surveys Blockchain for healthcare

Although there are several surveys related to the use of blockchain in healthcare applications [212, 171, 76, 231, 53], [98, 6], [67], [295], [69] , [131], [96], [102], all of them do not address the security aspects of the reviewed solutions, except the paper [231] who has partly addressed the security aspects, however, this paper was limited to only one use case of blockchain in healthcare, which is the EHR management. We summarised in Table V existing surveys focusing on the application of blockchain for healthcare.

### B. Benefits of Blockchain to healthcare applications

The blockchain is a promising technology that will play a vital role in empowering and developing next generation healthcare applications. The various benefits of the blockchain to the smart healthcare applications can be summarised as follows:

- **Decentralization**: healthcare applications are generally distributed over several stakeholders, which requires a distributed management system. Blockchain can provide this decentralized management, where all participants and stakeholders can control access to patients' data, without the need for a central authority.
- **Improved data security and privacy**: Blockchain technology is immutable and therefore helps protect the patients' data from alteration or corruption. Moreover, the real identity of patients are hid through the use of cryptographic keys, which help to protect the privacy of patients.
- **Health data ownership**: Through smart contracts, blockchain can deploy user centric healthcare applications, where the patient can control the access to his health data. Thanks to well-defined smart contracts, the user can decide to which medical staff he/she will give access and the access validity.

- **Availability and robustness**: The data is stored on the blockchain in a distributed manner and is replicated on multiple nodes. This permits to guarantee the availability of the data and increases the system robustness.
- **Transparency and trust**: By nature, the blockchain is an open and transparent system, which increases trust between the different participants and stakeholders.

### C. Healthcare Applications Requirements

Based on the previously cited surveys (mainly [231]) and a new security-based survey for healthcare 4.0 [97] , we have identified the following requirements for a next generation healthcare application (**RH** for **R**equirements for **H**ealthcare application):

- RH1: **Authentication**. Authentication is the first line of defense for any healthcare application. Indeed, to be secure a healthcare application needs to carefully authenticate every participant in the system including patients, doctors, care givers, etc.
- RH2: **Access Control**. Access control methods permit to specify who can access the healthcare data and the privilege level (read only, read and write, etc.). In classical healthcare applications, patient data was managed centrally by the hospital. Thanks to blockchain, the patient can manage his data with a fine grained access control method.
- RH3: **Privacy**. The privacy of patients' data needs to be preserved. This can be done through cryptographic techniques such as homomorphic encryption. Privacy defines in which situation patient data might be accessed, utilized, and disclosed to a third party.
- RH4: **Integrity**. Patient stored data needs to be protected against any unauthorized modification. Moreover, any modifications or alteration of data need to be detected.
- RH5: **Traceability**. Also known as auditability or accountability, tracks and audits who accesses the patient data, with what aim, and the time-stamping of any operation in the entire life cycle.
- RH6: **Availability**: we mean here ubiquitous availability of data. More precisely, patient data can be accessed from anywhere and distant access is possible.
- RH7: **Interoperability**. This requirement guarantee that patient data issued by different organisations and with different formats can be understood by each other. This facilitates data sharing for research and educational purpose.
- RH8: **Patient-Centric access control**. This requirement indicates that the patient has the right to own his healthcare data and to control it. More precisely, the patient will control which data is accessible to whom.
- RH9: **Scalability**. Healthcare applications generally involve big data such as X-ray images, clinical data, etc. Therefore, healthcare solutions designers need to take into consideration the large volume of data generated.

### D. Blockchain for Smart Healthcare Use Cases

Additionally, based on the previous surveys we have identified four use cases of the applications of blockchain in healthcare including secure EHR management, secure health data sharing, secure remote patient monitoring, and Pharmaceutical supply chain management.

TABLE IV: Comparison of various consensus protocols.

| Consensus Protocol | Background | Language | Resource Consumption | Processing Speed | Energy Efficiency | Limitations |
|---|---|---|---|---|---|---|
| PoW | Nodes solve a computationally difficult puzzle in order to receive mining opportunity. | Solidity, C++, Golang | High | Slow | Low | High power consumption and Less secure |
| PoS | The opportunity for block validation is proportional to the node stake in the system. | Native | Low | Fast | High | The wealthier nodes become more dominant in the network |
| DPoS | Every node with a stake in the network employs 'voting'. | Native | Low | Fast | High | Constraints on the number of token holders |
| PoSpace | Utilize the hard drive space of the nodes. | – | High | Slow | High | Nodes with more disk space receive more stake |
| PBFT | The decisions are made considering the majority votes where nodes communicate in order to prove the integrity and origin of the message. | Java, Golang | High | High | High | High communication overhead |
| PoET | Each node generates a random number in order to estimate its waiting time | Python | High | Medium | High | Same nodes are elected as a leader every time. |
| PoA | Requires the validators to have a monetary stake on the blockchain. | Java, Solidity | High | Medium | Low | Scalability issues |

TABLE V: **Blockchain for Healthcare** existing Surveys

| Ref | Scope | Description | (+)Pros/(-)Cons |
|---|---|---|---|
| [212] | Healthcare IoT | blockchain as an emerging technology for future healthcare IoT | |
| [171] | Limited to few solutions | ●Healthcare Industry Requirements <br>●Blockchain features, applications and limitations <br>●Review of 9 solutions <br>●Potential research directions are discussed | -Only few solutions are presented. <br>+Presented solution are deeply explained. |
| [76] | Blockchain platform for healthcare | ●Blockchain overview <br>●Start-up companies involved in blockchain healthcare solutions. <br>●Review of 11 solutions <br>●Potential research directions are discussed | +Industrial companies are presented. |
| [231] | Blockchain for EHR management | ●Healthcare Systems Requirements <br>●Blockchain features, applications and limitations <br>●Review of xx solutions <br>●Potential research directions are discussed | -Limited to only EHR storage solutions <br>+Mapping between reviewed solutions and healthcare requirements |
| [53] | Systematic Review of Blockchain in Healthcare. | ●Healthcare Systems Requirements <br>●Blockchain overview <br>●Review of xx solutions <br>●Potential research directions are discussed | -Solutions are not discussed. |
| [131] | A comprehensive review | The paper reviews blockchain applications in healthcare and classifies them to three broad categories; data management, supply chain management, internet of medical things. | -Some redundancies exist. |
| [102] | Systematic Review | A comprehensive systematic review about the use of blockchain in healthcare. Several classifications are presented including the addressed problems, the proposed solutions, and the security objectives. Moreover, recommendations to different stakeholders are presented. | -Several redundancies and several sections could be merged. |
| [6] | Systematic Review | The authors have classified reviewed papers to six use cases including EMR management, remote patient monitoring, pharmaceutical supply chain, biomedical research, health insurance claims, and health data analytics. | -Future directions are not presented. |

*1) Secure EHR management :* Patient health data are generally saved into an Electronic Health Record (EHR). One of the popular applications of blockchain in healthcare is the security of this EHR. In literature, the term EHR was used interchangeably with Electronic Medical Record (EMR), and Personal Health Record (PHR). The security of the EHR needs to be achieved during its creation, storage, management, and sharing. The main blockchain based EHR secure systems are presented in Table XXIX and are discussed as follows.

An early conception of the use of blockchain to provide confidentiality, authentication, and accountability of EHRs was described in [31], where the authors suggested using the blockchain for medical data access control and permission management.

The authors in [178] proposed an identity and access management system to provide EHR authorisation and authentication. The proposed system was implemented using Hyperledger Fabric Framework. [253] designed BHEEM, a blockchain-based framework to securely store and maintain EHRs.

[238] proposed a decentralized attribute-based signature scheme for privacy-preserving of user identity during signature verification. The proposed scheme was deployed for the security of EHR in healthcare. The decentralised EHR storage system based on blockchain guarantees integrity, auditability, and availability.

The authors in [256], similarly [272], combined attribute-based encryption, identity-based encryption, and identity-based signature in one crypto-system, to provide authentication, confidentiality integrity and traceability of medical data records in EHR. However, no implementation was done for the evaluation of the proposed system.

To overcome the scalability problem, due to the large volume of healthcare data, some solutions proposed to store the healthcare data on the cloud and to keep on the blockchain only the pointers to that data, along with their hashes [73] [13][216].

HealthChain [216] is a secure and scalable EHR management system. HealthChain is based on two blockchain networks; Private Blockchain: for intra-regional communication and, Consortium Blockchain: for inter-regional communication.

The authors in [56] proposed Ancile, which is built on the Ethereum blockchain platform and utilizes smart contracts to achieve access control, data privacy and interoperability of electronic medical records. The QuorumChain Consensus algorithm [214] was adopted to determine the next block to be added to the chain. Additionally, the authors used the concept of proxy-encryption to store keys and small encrypted records directly on the blockchain. Moreover, Ancile defined different smart contracts for each function of the system.

The work in [130] discussed and implemented different medical workflows using Ethereum smart contract system for secure EHR management. These healthcare workflows involve complex medical procedures.

The authors in [243] used Hyperledger Fabric to evaluate the performance of blockchain-based EHR systems. The authors have also developed the algorithm for access control and patient, clinician and lab interaction with the blockchain network.

Recently, the authors of [282] proposed a new blockchain-based healthcare data management system. They have used blockchain to control access to the stored patient's medical data in a decentralized database which is OrbitDb with IPFS. They have adopted two channels: the devices channel and the consultation channel. In fact, the concept of channels enhances data privacy. Moreover, they have adopted a patient-centric access control through the execution of smart contracts to allow or deny access to the patient's health data. Therefore, data confidentiality is preserved. Note that only the hash of data stored in OrbitDB with IPFS is saved in the blockchain to ensure data integrity and auditability. Besides, the adopted approach is fully decentralized providing no single point of failure which ensure data availability.

*2) Secure health data sharing:* The authors in [263] [264] proposed MedShare to securely share medical data stored on the cloud. The proposed system implements smart contracts and an access control mechanism to provide traceability and permissions on data.

The authors in [75] proposed MedBlock to share healthcare data for clinical and research purposes. MedBlock combines access control protocols with symmetric encryption to provide high-level security. One limitation of MedBlock is that it focuses only on hospital medical records of patients collected from medical examinations and does not store the vital signs of patients.

To transcend this drawback, Shen et al. [230], proposed MedChain: a blockchain based healthcare data sharing system. In addition to the EHRs of patients, MedChain shares also their vital signs collected from the IoMT devices.

FHIRChain [288] was proposed to meet the Office of the National Coordinator for Health Information Technology [195] requirements for secure and scalable clinical data sharing. These requirements include user identifiability, user authentication, controlled data access, secure data exchange, consistent data formats, and system modularity. In FHIRChain each participant possesses a public/private key pair. The public key will serve for user identity and the private key for user authentication. To address the scalability requirements, FHIRChain suggests keeping protected data off-chain and only pointers of it are stored on the blockchain.

In [106], the authors proposed a blockchain based framework method called BiiMED. It permits to manage and share Electronic Health Records (EHR) stored on the cloud between different medical organisations. Additionally, a Trusted Third Party Auditor (TTPA) based on blockchain technologies was introduced. TTPA permits to validate the exchanged data and ensures data interoperability and integrity.

*3) Secure remote patient monitoring:* To secure the remote patient monitoring systems Griggs et al.[90], proposed to use a consortium blockchain based on Hyperledger. The proposed RPM system Gateway implements a smart contract to analyse collected data (from sensors) and sends notification to the medical staff. The action of data read or doctors commands are considered as transactions, which are stored in the blockchain. One challenge of this solution is how to perfectly choose the transmission time of the aggregated data to the blockchain network.

The authors in [250] proposed a two-tier architecture; the first one ensures vital sign streaming and storage, whereas the second one is responsible for key management. The lower tier includes a patient centric agent (PCA), which is connected to a blockchain network and a cloud system. The PCA is a software that needs to be executed on a computer or server and ensures three functionalities: medical Data Management Module(DMM) responsible for storage management and compression, Security Service Module(SSM) responsible for key management, and Miner Management Module(MMM) is responsible for miner selection and blockchain interaction. The proposed system used a modified version of the PoW consensus algorithm, where only one miner is selected based on its characteristics to add a new block to the network. Moreover, the authentication of the different components of the system was proposed and which is mainly based on XOR operation and hash functions.

TABLE VI: Summary of **Blockchain for Secure EHR management** solutions

| Ref | Contribution/ Purpose | Blockchain Type | Framework | Consensus | Storage | Validation Tools | Merits/Limits |
|---|---|---|---|---|---|---|---|
| MedRec: [31] | Decentralized blockchain based EHR access control and permission management | Public | Ethereum | PoW | Local database | Smart contracts | -Using local database represent a single point of failure. |
| [178] | Proposed an identity and access management mechanism to support EHR authorisation and authentication. | Consortium | Hyperledger Fabric | - | CouchDB | Javascript + Hyperledger Fabric +PostgreSQL. | +Implementation of the system was done. |
| [253] | Blockchain-based framework for efficient storage and management of EHRs was proposed. | Consortium | Ethereum | - | - | - | -No validation of the proposition was done. |
| [238] | a decentralized attribute-based signature to provide privacy for blockchain-based EHR security. | Consortium | - | PBFT | on-chain and off-chain storage | Security analysis + Prototype imp. in C | -They do not use standard framework for evaluation. |
| HealthChain: [216] | a secure and scalable EHR management system. | Two Blockchains; one Private and one Consortium | Ethereum | PoA | IPFS + Cloud | Mathematical analysis | -No implementation is done. |
| Ancile: [56] | a privacy-preserving framework for the security and interoperability of EHR management. | Private | Ethereum | QuorumChain | Local database | - | -No implementation or validation is done. |
| [130] | Design and implementation of different medical workflows | Private | Ethereum | PoW (Ethash) | Back-end distributed file system (DFS) | solidity language+ Remix and Kovan test network | -Only partial system is developed |
| [243] | Performance evaluation of blockchain-based access control mechanism for EHR | Consortium | Hyperledger Fabric | BFT | CouchDB | Hyperledger Fabric +Hyperledger composer+ Docker. | +Description of the algorithms including access control. |
| [14] | Proposed MediBchain: a patient centric blockchain based EHR management system | Consortium | Hyperledger Fabric | BFT | on-chain on cloud | Solidity+ Java | +Using ECC. |
| [282] | Proposed HealthBlock: A secure blockchain-based healthcare data management system | Private | Hyperledger Fabric | PBFT | OrbitDB with IPFS | Hyperledger Fabric +Hyperledger composer | +Using two channels to enhance privacy. |

TABLE VII: Summary of **Blockchain for Secure health data sharing** solutions

| Ref | Contribution/ Purpose | Blockchain Type | Framework | Consensus | Storage | Validation Tools | Limits |
|---|---|---|---|---|---|---|---|
| MedShare: [264] | Sharing healthcare data in a trust-less environment such as cloud | Consortium | NA | NA | Cloud Database | JMeter. | - No implementation is done |
| MedBlock: [75] | Sharing healthcare data for clinical and research purposes. | Public | NA | An endorser is elected by more than half of the nodes. Then, the endorser validates the transactions. | Local Database | Security analysis. Latency of service provider requests | - The endorser is a central entity which minimizes the advantages of the distributed nature of blockchain. - No implementation is done |
| MedChain: [230] | Healthcare Data Sharing | Consortium blockchain | Java | BFT-SMaRt [237] | Local Database | Implementing using Java | -The frequent intervention of healthcare providers is needed. |
| FHIRChain: [288] | Secure and scalable healthcare data sharing for collaborative clinical decision making. | Private | Ethereum | - | Hybrid on-chain/ off-chain | Javascript + solidity | -Limited to only healthcare system supporting FHIR [77]. |
| BiiMED: [106] | Healthcare data sharing with a decentralized Trusted Third Party Auditor (TTPA) | Private | Ethereum platform | Two Ethereum nodes deployed in Amazon servers responsible for mining | Cloud database | Testnet of Ethereum+ Solidity language | -The use of cloud servers might lead to security issues |

To provide anonymity and authenticity,[68] proposed to use a lightweight privacy-preserving ring signature scheme [167]. In ring signature, the signature is mixed with other groups (named ring), to keep the identity of the signer private. Moreover, the authors introduced the concept of clustering the blockchain network to provide scalability. More precisely, nodes are organized into clusters and in each cluster the cluster head is responsible for the addition of new blocks. However, the authors did not precise how clusters are formed and did not evaluate their work.

The authors in [110] proposed an IoT-based blockchain platform for the secure remote monitoring of patients physiological parameters. One limit of this solution is that it stores the high volume of data generated by the medical sensors into the blockchain nodes. This design choice requires nodes with big storage space and leads to scalability problems.

Attia et al. [30], proposed to use two separate blockchains to secure the remote monitoring of patients. One blockchain manages the medical wearable devices and stores their collected data, and the other blockchain manages the consultations and contains all the history of patients records. moreover, the NDN paradigm [22] was used to retrieve data from the patient wearable devices. A prototype of the proposed architecture was implemented using Hyperledger Fabric Framework.

Recently, the authors of [283] proposed a new architecture for a remote patient monitoring (RPM) system based on blockchain technology. The overall architecture of the proposed RPM system is composed of a perception layer, a network layer and an application layer deployed in the cloud. In the application layer, the authors used Hyperledger fabric integrated with Hyperledger Composer to implement the business model of the RPM system in the blockchain network. Note that the ledgers and transactions are deployed in the cloud to ensure the scalability of the proposed system. Moreover, the wearable health devices and the IoT gateway are considered assets of the blockchain network which ensure data integrity and auditability. Besides, the participants which are patients and doctors must be registered and enrolled in the blockchain network to benefit from RPM services. This fact ensures data confidentiality. Finally, the proposed system is fully decentralized with no single point of failure which ensures data availability.

TABLE VIII: Summary of **Blockchain for Secure RPM** solutions

| Ref | Contribution/ Purpose | Blockchain Type | Framework | Consensus | Storage | Validation Tools | Limits |
|-----|----------------------|-----------------|-----------|-----------|---------|-----------------|--------|
| [90] | Sharing healthcare data for clinical and research purposes. | Private/ Consortium | Ethereum | PBFT | Designed EHR storage database | Solidity + Smart contracts | Sensors data are sent regularly to the patient phone, which will consume its energy and make solution not practical and dependent on the user phone |
| [250] | Continuous patient monitoring using body sensors network. Security and blockchain based functions are integrated into a PCA. | Private | - | PoW with miner selection based on capacity | on-chain+ Cloud | Java + Ethash | -Partially evaluated on non standard platform. -The PCA needs a computer to run which limits patient mobility. |
| [68] | Use ring signature for anonymity and organize nodes into cluster for scalability purpose. | - | - | Cluster Heads are the miners | Cloud database | - | -How cluster are formed is not described, -No validation is done |
| [30] | Remote Patient Monitoring | Private | Hyperledger Fabric | PBFT | Blockchain database | Go language + Application SDK | -The capacity limitation of direct storage in blockchain |
| [110] | Vital signs real time remote monitoring | Private | Hyperledger Fabric | PBFT | Distributed ledger technology (DLT) | Hyperledger composer + Caliper | - The storage of patients data in blockchain network is heavy. |
| [283] | New architecture of a blockchain based RPM system | Private | Hyperledger Fabric | PBFT | Cloud | Hyperledger composer + Caliper | - The storage of patients data in cloud might reveal the privacy of patient data. |

*4) Pharmaceutical supplychain:* Several contributions have been proposed for the supplychain management in the context of healthcare.

The authors in [123] present an overview of Pharmaceutical Supply Chain Management (SCM) systems. They discussed their benefits, issues (including counterfeiting, improper labeling, improper temperature controls and handling, transportation and storing issues), and challenges.

The authors in [109] proposed a blockchain based SCM to check the drug integrity. The proposed system enables the medical staff, patients, and pharmacists to manage, control access, and share personal medical records and the complete patient drug life cycle in a secure and transparent way. The system was implemented using Hyperledger fabric and tested using Hyperledger composer [55].

The authors in [1] proposed a drug supply chain management and recommendation system. The authors used blockchain to track the drug delivery and detect the counterfeit drugs. Moreover, a machine learning module was proposed to recommend the best medicines. The authors used Hyperledger Fabric as a blockchain framework and the N-gram, LightGBM models as the machine learning method.

The authors in [235] proposed an SCM for drugs that takes into consideration the temperature of the drug during transit and storage. The solutions necessitate the presence of a sensor in the drug box. Moreover, a QR code was used to store drug information. To address the scalability issue the bloXroute [135] servers are introduced. bloXroute are a scalable blockchain distributed network. Moreover, the authors suggested to use the Raft consensus protocol. The authors made a security analysis. However, the solution requires the presence of several pieces of hardware which can be difficult to provide such as smart transportation box, sensor nodes, and dynamic QR code in each packet.

TABLE IX: Summary of **Blockchain for Secure Pharmaceutical SupplyChain Management** solutions

| Ref | Contribution/ Purpose | Blockchain Type | Framework | Consensus | Storage | Validation Tools | Limits |
|---|---|---|---|---|---|---|---|
| [109] | Blockchain based drug management to check drug integrity | Private | Hyperledger Fabric | - | couchDB + Offchain | Hyperledger Composer/Caliper | - The consensus algorithm and how consensus manager is selected are not specified. |
| [1] | Blockchain based drug management to detect counterfeit drug + A machine learning based drug recommendation module | Private | Hyperledger Fabric | - | couchDB + Offchain | implementation in Hyperledger | - The consensus algorithm is not specified. |
| [235] | Convergence of IoT and Blockchain to monitor drug temperature and prevent counterfeit pharmaceutical | Private | Hyperledger Fabric | Raft | Cloud | Security Analysis | - No implementation done - Requires different components. |

### E. Mapping of existing solutions to Security requirements and Future Directions

Table X shows the mapping of the previously reviewed solutions to the specified healthcare requirements described in subsection III-C. It is worth noting from this table that some requirements such as authentication, access control, integrity, and auditability are well fulfilled by the proposed solutions. This might be explained that these security services are inherent to the blockchain technology. However, the interoperability requirement is rarely fulfilled by the existing solutions. This is due mainly to the different formats of data used in healthcare. Moreover, proposed healthcare solutions use generally private blockchain, which fails to scale to very high number of participants. Therefore, scalability is another issue that needs to be resolved in healthcare solutions. Additionally, the majority of healthcare solutions store the patient data either on the blockchain or in the cloud. However, blockchain cannot fit the large volume of patient data. This explains why some solutions opted for the cloud storage solutions. However, storing data in the cloud might reveal the privacy of the patient and expose the system to several attacks. Finding a storage solution that overcomes these limits is another future research direction.

The following open issues still need research focus for the large deployment of blockchain based healthcare systems.

- Storage: Healthcare systems generate a large volume of data that current blockchain systems fail to store directly in the distributed ledger due to block size constraint. Some solutions opted to store data in the cloud. However, cloud based solutions might breach the privacy of patients as some cloud systems might eavesdrop on patient data for some business organisations. Another alternative was to store data in the distributed InterPlanetary File System (IPFS) and to store the hash produced by the IPFS system in the blockchain. This is a hybrid off-chain/on-chain solution, where patient data is stored off-chain and its metadata (pointers and references to data) is stored on-chain. Although, these solutions can partially solve the storage problem of blockchain based ERHs management systems, they remain inefficient, especially for RPM systems that are composed of IoMT devices that generate huge volumes of data at high speed. Some pioneer proposals for storage efficiency need to be explored for next generation healthcare systems [206, 57, 270].

- Scalability: In addition to the storage limit, current blockchain based healthcare systems did not scale to the large community due to the inefficiency of existing consensus algorithms. For example, the PoW consensus algorithm consumes a lot of energy and has a low throughput. The PBFT presents an $O(n^2)$ messages and therefore cannot scale to a large number of nodes. Moreover, existing blockchain based healthcare systems are tested for few number of nodes and therefore, more effort must be put to fit these solutions to a large scale. The following surveys can be a good starting point to address the scalability problem [52, 292].

- Type of blockchain: although there are solutions for different types of blockchain (public, private, and

consortium). The consortium blockchain seems to be the right choice for healthcare system. Indeed, public blockchain breaches the privacy of patients as it allows every participant to read data. Private blockchain is generally limited to one organisation and does not support medical data sharing. However, consortium blockchain can be established between different medical organizations and hospitals and can impose access control rules on the shared data.

- Interoperability: one problem of the healthcare systems is that they use different formats for the medical data recommended by different standards such as the Health Level Seven International (HL7), Digital Imaging and Communications in Medicine (DICOM) and European Committee for Standardisation (CEN). This characteristic obstacle the data sharing between different stakeholders. Indeed, data stored in blockchain need to have the same format to be used by different stakeholders. More on blockchain interoperability can be found here [40, 116].

- Standards and regulations: To improve compliance and interoperability between developed blockchain based healthcare systems standards and regulations must be set up. Moreover, regulations must encourage blockchain adoption between stakeholders. One objective of these standards is to unify the EHR structure of the patient to facilitate data storage and search when manipulated by different entities.

- Privacy: Privacy is a major challenge in healthcare industries. Patients cannot share their medical data if they feel there is a risk of data leakage. Therefore, healthcare systems need to preserve the privacy of their users. Due to the transparency characteristic of blockchain, exchanged transactions can be read and analysed by participants in the blockchain network. If these transactions carry sensitive healthcare data, this will lead to patients' privacy breaching. Some mechanisms have been put in place to defeat the privacy problem such as ring signature, blind signature, and other techniques that are discussed in the following survey[78].

## IV. BLOCKCHAIN FOR SMART TRANSPORTATION

Connecting vehicles securely and dependably is essential to the employment of modern Intelligent Transportation System (ITS) applications in smart cities. With the incessantly emerging security risks, vehicles could be subjected to a range of malicious attacks that could jeopardize the safety of passengers, services, and information. In this section, we first identify the benefits of the blockchain model to the ITS applications. Next, we describe the major security necessities for ITS services. The main part of this section will comprise an analysis of the systems that were proposed so far to integrate the blockchain into the ITS. These systems will be divided into six use cases: security and privacy of communications, autonomous vehicles, smart parking, ridesharing and ride-hailing, security and management of unmanned aerial vehicles (UAV), and miscellaneous applications. Lastly, we will discuss the challenges and future research directions of the blockchain in smart city ITS (BITS). However, we will start by briefly outlining the existing surveys on BITS and the significance of our study as compared to them.

### A. Existing BITS surveys

Table XI provides a summary of the most important papers in the literature that provided comprehensive overviews of the blockchain implementation and utilization in smart city ITS. Wang et al. [254] studied the implementation of the blockchain in the Internet of Vehicles (IoV) and analyzed it from seven aspects, which are architecture, privacy, security, trust management, data management, certificate management, and data monetization. Xie et al. [265] focused on analyzing three sides of the implementation of the blockchain in smart city ITS, which are decentralized architecture, management of vehicles' communications, and charging of electric vehicles. Sharma and Kaushik [229] identified 27 different kinds of attacks on the ITS and discussed various security solutions, including blockchain mechanisms, that were implemented to defend and thwart these attacks.

Peng et al. [205] categorized the blockchain systems in the literature based on their positions in the IoV networking layers, which are the Perception Layer; Networking Layer; and Application Layer. For each layer, the authors classified the blockchain mechanisms based on their purposes, such as decentralization, incentive, security and privacy, or audit. The authors of [177] discuss the various blockchain types, consensus mechanisms, smart contracts applicability, computation, communication, and storage overhead in blockchain-based vehicular communications. Mollah et al. [185] identified eight main applications of the blockchain in the IoV, which are Data Protection, Data Trading, Resource Sharing, Vehicle Management, Ride Sharing, Content Broadcasting, Traffic Control, and Forensics. However, the authors missed some very important blockchain applications, such as those related to autonomous vehicles, Internet of Drones (IoD), and smart parking. In this paper, we identified the five main categories of blockchain applications in the ITS. Each of these categories includes a large number (more than 10) of systems that have been proposed so far for utilizing the blockchain as part of the application. Various other applications (on which a limited number of systems have been proposed) will be discussed in the miscellaneous category.

### B. Blockchain Benefits to ITS

The blockchain is expected to play a vital role in empowering and developing the ITS applications. Several previous works identified various benefits of the blockchain to the smart city ITS. We summarize these benefits as follows:

- **Security and Safety**: one of the most important factors in the success of ITS applications is securing the data, execution, and communications of the applications; and the safety of the application users (drivers and passengers). With its decentralized validation and immutable data, the blockchain can safeguard the ITS

TABLE X: Mapping of **Blockchain for Healthcare** solutions to security requirements

| Ref | Use case | RH1 (Auth) | RH2 (AC) | RH3 (Priv) | RH4 (Intg) | RH5 (Aud) | RH6 (Avai) | RH7 (Itop) | RH8 (PCA) | RH9 (Scal) |
|---|---|---|---|---|---|---|---|---|---|---|
| MedRec [31] | SEM | ✓ | ✓ | ✓ | ✓ | ✓ | ✓ | ✓ | ✓ | ✓ |
| [178] | SEM | ✓ | ✓ | ✗ | ✓ | ✓ | ✗ | ✗ | ✓ | ✗ |
| [253] | SEM | ✓ | ✓ | ✗ | ✓ | ✓ | ✗ | ✗ | ✓ | ✗ |
| [238] | SEM | ✓ | ✓ | ✗ | ✓ | ✓ | ✓ | ✓ | ✓ | ✗ |
| HealthChain [216] | SEM | ✓ | ✓ | ✓ | ✓ | ✓ | ✓ | ✓ | ✓ | ✓ |
| [256] | SEM | ✓ | ✓ | ✗ | ✓ | ✓ | ✓ | ✓ | ✓ | ✗ |
| Ancile [56] | SEM | ✓ | ✓ | ✓ | ✓ | ✓ | ✗ | ✓ | ✓ | ✗ |
| [130] | SEM | ✓ | ✓ | ✗ | ✓ | ✓ | ✓ | ✓ | ✓ | ✗ |
| [243] | SEM | ✓ | ✓ | ✗ | ✓ | ✓ | ✗ | ✓ | ✓ | ✗ |
| [282] | SEM | ✓ | ✓ | ✓ | ✓ | ✓ | ✓ | ✓ | ✓ | ✓ |
| MedBlock [75] | SDS | ✓ | ✓ | ✓ | ✓ | ✓ | ✓ | ✗ | ✓ | |
| MedShare [264] | SDS | ✓ | ✓ | ✗ | ✓ | ✓ | ✗ | ✗ | ✗ | |
| FHIRChain [288] | SDS | ✓ | ✓ | ✗ | ✓ | ✓ | ✗ | ✗ | ✓ | |
| BiiMED [106] | SDS | ✓ | ✓ | ✗ | ✓ | ✓ | ✓ | ✗ | ✗ | |
| [90] | SRPM | | | | | | | | | |
| [250] | SRPM | ✓ | ✓ | ✓ | ✓ | ✓ | ✓ | ✓ | | ✗ |
| [68] | SRPM | ✓ | ✓ | ✓ | ✓ | ✓ | ✓ | ✓ | ✗ | ✓ |
| [30] | SRPM | ✓ | ✗ | ✗ | ✓ | ✓ | ✓ | ✓ | ✗ | ✓ |
| [110] | SRPM | ✓ | ✓ | ✓ | ✓ | ✗ | ✓ | ✓ | ✗ | ✓ |
| [283] | SRPM | ✓ | ✓ | ✓ | ✓ | ✓ | ✓ | ✓ | | ✓ |
| [109] | PSCM | ✓ | ✓ | ✗ | ✗ | ✓ | ✓ | ✗ | ✗ | ✗ |
| [1] | PSCM | ✓ | ✓ | ✗ | ✗ | ✓ | ✓ | ✗ | ✗ | ✗ |
| [235] | PSCM | ✓ | ✓ | ✗ | ✗ | ✓ | ✓ | ✗ | ✗ | ✓ |

SEM:Secure EHR Management, SDS:Secure Data Sharing, SRPM:Secure Remote Patient Monitoring, PSCM:Pharmaceutical SupplyChain Management.

applications from malicious attacks and protect the safety of the participants.

- **Drivers' and passengers' privacy**: when handling private data, the trust aspect is the most important. Several blockchain mechanisms have been proposed that focus on establishing a trust-based environment between the vehicular network members [175, 274, 160]. Implementing an efficient trust-based distributed blockchain mechanism that applies cryptography and hashing operations will assist in keeping private data out of the reach of attackers.

- **Decentralized mechanism removes single point of failure (SPoF) problem**: most centralized management and storage systems suffer from the SPoF problem. In the blockchain, each node stores a copy of the ledger, and all the nodes cooperate in taking the blockchain-related decisions, which eliminates the SPoF issue.

- **Automatization**: the blockchain can be used for life-time management and surveillance of ITS devices, which assists in establishing a self-organized, self-adaptive, and decentralized autonomous ITS ecosystem [280].

- **Providing strong trust for ITS users**: since the blockchain is transparent and all the network exchanges are clearly expressed as immutable transactions within the blockchain blocks, ITS users trust the blockchain data that are utilized by the ITS applications.

- **Providing incentives to various transportation nodes (RSUs, vehicles, drones, etc.) to participate in the ITS applications**: many ITS applications require the network nodes to perform operations that serve the whole network, such as routing other nodes' packets, sending road traffic-related data, performing part of the application computations, etc[174]. Transportation nodes may be reluctant to execute these operations without an incentive. The blockchain solves this problem by providing the required incentive via the blockchain's built-in cryptocurrency system.

TABLE XI: **Blockchain for ITS** existing Surveys

| Ref | Scope | Description | (+)Pros/(-)Cons |
|---|---|---|---|
| Wang et al.[254] | Analyzing the BC in the IoV from seven aspects: security, trust, privacy, architecture, certificate management, data management, and data monetization | ● literature work classified into seven categories<br>● discusses the future enhancements for each category | + accurate classification based on the proposed categories<br>- missing a large number of important literature papers<br>- several BITS applications were not mentioned |
| Xie et al.[265] | Focusing on three aspects of BITS: decentralized architecture, vehicles' communications, and electric vehicle charging | ● papers on some BC applications, such as privacy and trust, are embedded within the three main applications<br>● several important lessons are deduced | + excellent summary of the three targeted applications<br>- missing a large number of important literature papers<br>- several BITS applications were not mentioned |
| Sharma and Kaushik[229] | Focusing on the lightweight security aspect when applying the BC to the VANET | ● IoV applications are discussed and summarized<br>● Studying general security aspects of IoV<br>● classifying the existing possible attacks in the IoV<br>● extensive survey of existing security solutions for each attack, and highlighting the flaws of each security solution | + comprehensive classification of security attacks in VANETs and IoV<br>+ extensive analysis of existing solutions for each attack<br>- does not consider BC applications other than security<br>- limited discussion of BC literature |
| Peng et al.[205] | Focusing on dividing the BC applications in IoV among the three BC layers: Application, Perception, and Networking | ● categorizing BC applications based on their purpose into: decentralization, security, incentive, audit, positioning, and trust<br>● highlighting important issues that must be investigated in order to make blockchain applicable in vehicular environments<br>● discussing how to support blockchain applications by designing efficient vehicular IoT protocols | + identifying the requirements of BC layers<br>+ discussing important BC design requirements<br>- missing a large number of important literature papers<br>- does not mention several BC applications in vehicular networks |
| Mikavica and Kostić-Ljubisavljević[177] | Focusing on classifying BC applications in the ITS based on the security, privacy, and trust requirements | ● discussing various BC applications based on the BC type, consensus protocol, mining nodes, and smart contracts<br>● classifying BC applications based on the vehicular network type (delay-tolerant, social, electric, etc.) | + excellent classification based on the vehicular network type<br>- several BITS applications were not mentioned<br>- missing several important literature papers |
| Mollah et al.[185] | Focusing on vehicular data security, vehicle management, and on-demand transportation services | ● discussing the techniques for integrating the BC into the IoV<br>● discussing several blockchain-empowered IoV architectures<br>● investigating BC integration challenges such as performance, optimized consensus, and incentive mechanisms | + extensive analysis of studied papers<br>+ excellent classification of various BC architectures in the IoV<br>- missing important BC applications in the IoV<br>- focusing too much on less important BC applications |

● **Faster cloud services**: several literature works proposed utilizing edge nodes, such as the RSU network, to provide cloud services (such as money exchange) to ITS users. Integrating the blockchain into such system enables the ITS users to consume these services securely with reduced delay, as compared to similar online services.

● **Scalability**: ITS networks can grow to include a huge number of nodes. Several research works proposed scalability solutions for the blockchain based on efficient approaches, such as partitioning, sharding, and directed acyclic graphs (DAGs) [125, 92]. Implementing similar solutions enables the blockchain-based ITS application to scale to a high degree in order to satisfy the transportation network requirements.

● **Connecting heterogeneous entities**: the blockchain enables various types of nodes to join the blockchain network and take part in the blockchain operations, such as transaction management, consensus, and storage. As long as a node commits to the blockchain protocols, it can act as a blockchain node regardless of its type, structure, and capabilities.

● **Democracy**: the correct implementation of blockchain protocols ensures that each blockchain node has the same rights as other nodes that have the same role. In general, democracy is a very important feature that is desirable by all smart city applications, not only the ITS.

## C. ITS requirements

The extensive review that we made on a large number of ITS-related papers helped us to identify the main features that are required by most ITS applications. These Requirements for Transportation applications (RTs) can be summarized as follows:

- **RT1:** **Authentication**: It guarantees that vehicles that engage in network communications are trustworthy. Without authentication, a legitimate node can be impersonated by an attacker who sends messages on its behalf.

- **RT2:** **Confidentiality**: It certifies that secrecy is granted to sensitive information such that only rightful users can access it. This requirement is vital for almost all ITS applications.

- **RT3:** **Availability**: Information must be obtainable by legitimate users in a timely manner. In many applications, a certain delay would make the information lose its significant value.

- **RT4:** **Data integrity**: This feature means that the content of data must not be changed, whether purposefully or unintentionally, since the time it is generated. Several ITS applications, especially those that are safety-critical, demand that the content of the exchanged messages remains intact.

- **RT5:** **Privacy**: ITS applications generate and process a large amount of information that is related to the privacy of the ITS users. For example, drivers may not want to disclose where they are going or with whom. Hence, ITS applications should keep such information secret to their owners.

- **RT6:** **Access control**: This security feature states that each node should perform its various actions and functions based on the privileges and limits that it is assigned. In order to ensure access control, each node should be given the correct roles and rights to access and use the ITS application.

- **RT7:** **Non-repudiation**: It is the inability to refute responsibility. This feature relates to the process that the receiver of a message obtains a proof of the identity of the sender, and that the sender is provided with a proof that the information has been delivered to the receiver. In ITS applications, compromised nodes can be detected by non-repudiation.

- **RT8:** **Latency guarantees**: most ITS applications are time-sensitive and some of them are real-time. For example, accident avoidance applications fail if the data analysis results are not produced very quickly in order to be able to alert the driver and avoid the accident. These applications require that time constraints that are specified by the application are strictly met.

- **RT9:** **Scalability**: Vehicular networks are characterized by their large number of users, high mobility of vehicles, and dynamic topology of the network. With such attributes, the vehicular network should be scalable in order to enable the ITS applications to execute without disruptions.

## D. BITS Use Cases

A large number of blockchain systems and models have been proposed for the smart city transportation system. These systems span a wide range of applications that belong to many fields/areas. In general, the ITS intersects with different fields that we discuss in other sections of this paper, such as healthcare, smart grid, and supply chain. In this section, we focus on blockchain systems that target ITS-dedicated applications related to driving, vehicular communications, parking, passenger-support, etc. We assign a separate subsection for each use case in which a large number of blockchain systems and frameworks have been proposed. We combine the use cases for which few systems have been previously proposed per use case together in the Miscellaneous sub-section.

### 1) Blockchain for General Vehicular Communications:
A blockchain-assisted ITS (Ba-ITS) that grants privacy protection to users combined with an incentive model that prompts vehicles to provide data to the ITS was proposed by Li et al. [150]. The system utilizes a hierarchical blockchain structure that guarantees the interoperability among different levels of blockchains. In addition, the authors propose several methods that exploit smart contracts to offer ITS services to users. Liu et al. [160] propose a privacy-preserving announcement scheme (BTCPS) in which roadside units (RSUs) calculate the reliability of each packet based on the vehicles' reputation values that are stored in the blockchain. BTCPS utilizes the identity-based group signature method in order to attain conditional privacy. The system exploits logistic regression to boost the sensitivity of reputation values of malicious vehicles. Furthermore, a blended consensus protocol is used to reduce the blockchain overhead. Although the authors claim that their scheme provides privacy, this is not true as the identity of vehicles can be revealed from the public address used in the blockchain. Wang et al. [258] propose TrafficChain that utilizes the blockchain for traffic information collection. TrafficChain is characterized by a two-layer blockchain architecture which comprises a local chain for each road segment and a global chain for the whole network. The system makes use of edge/fog routers that act as local miners of the local chains. Each block on the global chain comprises the comprehensive information of the traffic status of each of the road segments. The authors utilize the Long Short-Term Memory model to defend against Byzantine and Sybil attacks.

Instead of the traditional Proof of Work (PoW), Haouari et al. [94] propose the Proof-of-Useful-Work (PoUW) as a consensus mechanism for transportation systems. PoUW requires the miner to solve an NP-hard optimization problem whose results benefit the supply chain by optimizing emission levels and operational costs. Oham et al. [198] utilize a permissioned blockchain to manage the access to restricted elements of the vehicular ecosystem. The proposed B-FERL protocol uses a challenge–response mechanism between the vehicles and RSUs to observe the internal status of each vehicle and detect cases of vehicle compromise. Lei et al. [142] propose the blockchain as a means to simplify the distributed key management in heterogeneous vehicular networks. The blockchain enhances the key transfer handshake procedure and helps to reduce the key transfer time during a vehicle's handover.

The authors in [266] discuss a blockchain framework to support the vehicular IoT services in SDN-enabled 5G-VANET. Here, a vehicle broadcasts traffic information and the near vehicles rate the packet. The RSU computes the trust value of the packet based on the distance between the scoring vehicle and the sending vehicle, and stores the trust value into blocks. Javed et al. [112] propose a blockchain model for reliable data sharing between resource constrained vehicles and edge service providers. The proposed blockchain uses the Proof of Authority (PoA) protocol to validate transactions. In addition, the system provides incentives to edge nodes via the blockchain. The authors discuss a caching model at the edge servers that can improve the blockchain performance. Another system that aims at providing incentives to vehicles to forward announcements was explained in [146]. The authors propose an announcement network called CreditCoin and a protocol called Echo-Announcement that uses authentication and privacy thresholds to guarantee that anonymous announcements are reliable. Users earn coins by replying to the announcement request of others, and spend coins to make an announcement.

Mershad et al. [175] proposed the Proof of Accumulated Trust (PoAT) consensus protocol that aims at selecting the blockchain miners among RSUs that accumulate enough trust points to become trusted nodes (TNs). The blockchain blocks are mined by one of the TNs who is called the current miner (CM). The CM receives the transactions from the TNs, mines the block, and broadcasts it to the TNs to achieve consensus. The authors in [143] leverage the characteristics of edge computing and blockchain to propose a blockchain-based GPS positioning error evolution sharing scheme for the IoV. The proposed scheme ensures the security of cooperated vehicles and improves the localization accuracy. The scheme uses smart contracts to automate the recording of positioning errors and improve the positioning accuracy. Moreover, mobile edge computing (MEC) nodes are used as miners to maintain the blockchain and achieve the consensus. The authors in [145] combine the blockchain and ciphertext-based attribute encryption (CP-ABE) to create an access control scheme for VANETs (FADB). The system uses Ethereum and the InterPlanetary File System (IPFS) for distributed storage and access control of data.

A joint blockchain and named data networking (NDN)-based vehicle-to-everything (V2X) framework was proposed in [219]. The framework maintains users' privacy by using non-private information such as the plate number of the vehicle in order to ensure the integrity and accountability of the V2X communications. Bao et al. [36] propose a blockchain system for pseudonym management in the IoV. Within the blockchain, a pseudonym certificate shuffling mechanism is applied in order to reuse existing pseudonyms by assigning them to new vehicles. The system utilizes asymmetric cryptography when sending blockchain transactions in order to guard the pseudonym shuffle path.

Wang and Zhang [255] utilize the blockchain for secure data sharing in vehicular networks. In the proposed scheme, the vehicle sends the encrypted message to the RSUs, and the latter invoke the verification nodes to verify the encrypted message via the Ripple consensus. In addition, service providers apply smart contracts to offer multi-dimensional services to vehicles. The latter utilize the attribute-based proxy re-encryption algorithm to search within the blockchain.

Tables XII and XIII summarize the papers that were discussed in this section, while Table XIV maps each paper to the BITS security requirements that were stated in Section IV-C.

*2) Blockchain for Autonomous Vehicles:* An important element of the next-generation ITS is the network of connected autonomous vehicles (CAVs). CAVs are artificial intelligence (AI)-empowered and computer- operated driverless vehicles that can collaborate to monitor the nearby surroundings via sophisticated sensor modules and take instantaneous and appropriate decisions without human involvement. CAVs are susceptible to a large number of safety and security problems. Khoshavi et al. [132] highlight the advantages of utilizing the blockchain in CAVs and demonstrate how the blockchain can be effectively applied to the CAV technology. The authors in [91] propose a blockchain-based architecture to mitigate security and privacy issues in CAVs. The proposed system is divided into four layers which are the data layer, Ethereum layer, edge layer, and cloud layer. The CAVs send their raw and encrypted data to the RSUs. The latter save the transactions into the blockchain and execute the smart contracts to select certain information and send it to the edge servers, while the latter cache the blockchain data that passes between the RSUs and the cloud users. Rathee et al. [218] discuss a blockchain model for autonomous cabs. The blockchain is used to store the reading of the cab sensors in order to protect the autonomous cab from attackers. The IoT devices register on the Blockchain Network by sending a subscription request to the peer manager. During the travel of a user from one location to another, the vehicle's number, location, and current and previous vehicle ratings are captured by the IoT devices and stored in the blockchain.

The fruit fly optimization algorithm (FOA) was utilized by Sharma et al. [228] to distribute the blockchain mining fairly between various miner pools in a CAV. In the proposed algorithm, miners are classified based on their previous contributions to the blockchain. The algorithm selects the best miner nodes that should handle each transaction request when creating a new block. A sensing and tracking architecture for CAVs was presented in [221]. The proposed system ensures the secure sensing and tracking of any object of interest via the blockchain. Wang et al. [261] present a blockchain model to safeguard content delivery in autonomous vehicular social networks (AVSNs). The proposed blockchain utilizes a reputation assessment model to ensure the legitimate behavior of CAVs and improve the content reliability. In addition, the system proposes a proof of reputation (PoR) consensus protocol that implements the Boneh-Lynn-Shacham (BLS) multi-signature method.

The authors in [39] present a blockchain firmware update mechanism for autonomous vehicles. In the proposed system, a consortium blockchain is created by the manufacturers of the AVs. Each manufacturer writes a smart contract that manages the operations that update its AV firmware. In addition, the blockchain is used to incentivize AVs to contribute to the dissemination of new firmware updates from one AV to another. The proposed model applies an attribute-based encryption (ABE) technique to enable manufacturers to apply a policy about who has the right to download and use an update. Pedrosa and Pau [203] utilize the blockchain to securely record and handle the charging of electrical AVs. The charging stations use the blockchain to publish their charging rules and prices. The AV opens a channel to negotiate with the station and commits to the negotiation parameters by publishing them in the blockchain, which locks a number of tokens for the AV during a timestamped period.

TABLE XII: Summary of **Blockchain for General Vehicular Communications** solutions - Part 1

| Ref | Contribution/ Purpose | Blockchain Type | Framework | Consensus | Storage | Validation Tools | Limits |
|---|---|---|---|---|---|---|---|
| [150] | Promoting users and vehicles to provide data to ITS | Hierarchical private and public BC structure | Ethereum | PoW | RSUs, Edge, and Cloud nodes | Ethereum with Truffle and JSON RPC | - Scalability not tested<br>- Use of PoW |
| [160] | Trust management Model combined with conditional privacy-preserving announcement scheme | Public | Not specified | Mixed PoW and BFT | RSUs | Python and Go simulations | - Depends on TA (central point of failure)<br>- No Scalability<br>- Weak consensus model |
| [258] | Utilizing fog/edge infrastructure to collect and store traffic information | Public | Ethereum | PoW | Vehicles, RSUs, and Edge nodes | Simulations using CEbTS dataset | - Using single dataset for testing<br>- Does not compute LSTM overhead<br>- Use of PoW |
| [94] | Proof-of-Useful-Work consensus that puts the vehicles' computing resources to beneficial use | Public | Mathematical optimization | PoUW | Not specified | Mathematical analysis | - Limited to saving nodes' resources<br>- No real testing<br>- BC details are missing |
| [198] | Using a challenge–response between vehicles and RSUs to control access | Permissioned | Custom | Appendable block concept (ABC) | Some transactions in vehicle, remaining transactions in cloud | Simulations via the CORE tool | - Limited number of vehicles in simulations<br>- Consensus replaced with ABC |
| [142] | Distributed key management in vehicular group communications | Public | Custom transactions and blocks | PoW | Not specified | Simulations via OMNeT++ | - Limited to group communications<br>- Use of PoW<br>- Scalability not emphasized |
| [266] | Incorporating software-defined networking into 5G-VANET for secure information gathering and network control | Public | Custom | Combined PoW and PoS | RSUs | Simulations via OMNeT++ and crypto++ | - Limited to video transactions<br>- Assumes all RSUs are trustworthy |

Table XV summarizes the papers that were discussed in this section, while Table XVI maps each paper to the BITS security requirements that were stated in Section IV-C.

*3) Blockchain for Ride-Sharing and Ride-Hailing:* Ride-sharing is an ITS-based service that allows drivers to share the road with other people, which provides many benefits such as joint travel cost and lessening traffic congestion. On the other hand, ride-hailing is a similar service in which the rider specifies to the driver the exact time and place of the ride. Several recent papers proposed a variety of blockchain models for these two services. Baza et al. [38] proposed B-Ride, a decentralized ride-sharing system based on public Blockchain.

TABLE XIII: Summary of **Blockchain for General Vehicular Communications** solutions - Part 2

| Ref | Contribution/ Purpose | Blockchain Type | Framework | Consensus | Storage | Validation Tools | Limits |
|---|---|---|---|---|---|---|---|
| [112] | Edge nodes offer ITS services and use Proof of Authority consensus to validate transactions | Public | Ethereum and Interplanetary File System (IPFS) | Proof of Authority (PoA) | Edge nodes | Simulations via Ganache software and Remix IDE | - Uses RSUs to store vehicles' data - BC is limited to storing vehicles' requests |
| [146] | BC for providing incentives to vehicles to forward announcement anonymously | Public | Custom | Custom based on voting | RSUs | Simulations via PolarSSL and GMP libraries | - Assumes all RSUs are trustworthy - Limited simulations setup |
| [175] | Selecting RSUs based on trust value to play the role of BC miners | Private | Bitcoin-based | Proof of Accumulated Trust (PoAT) | Trusted nodes, which are selected RSUs | Simulations via NS-3 software | - Depends on cloud server - Extra overhead due to redundancy mechanism |
| [143] | A framework for GPS positioning error sharing to improve vehicle positioning accuracy | Consortium | vDLT | BFT-DPoS | Mobile edge computing nodes (MECNs) | Custom-made simulations | - Limited to locations-based applications - Scalability not considered |
| [145] | An access control scheme that combines ciphertext-based attribute encryption, Ethereum blockchain, and IPFS | Public | Ethereum | PoW | RSUs | Simulations using Ethereum, CP-ABE, and the PBC library | - Assumes all RSUs are trustworthy - Use of PoW |
| [219] | Combines BC with named data networking (NDN) to securely hide vehicles' identities | Public | Custom | Not specified | Cluster heads, denoted as cluster block managers (CBMs) | Simulations via NS-3 software | - Implemented BC is not fully described - Consensus not discussed - Focuses on identity hiding |
| [36] | Blockchain for pseudonym management allowing the reuse of existing pseudonyms to different vehicles | Public | Custom | PoW | All network participants | Simulations via OMNeT++ and crypto++ | - BC for pseudonym management only - Use of PoW |
| [255] | Attribute-based proxy re-encryption is used to control data access permissions | Consortium | Custom | Ripple | RSUs | Mathematical analysis | - Large block generation delay - No real testing |

B-Ride enables the rider to post a cloaked pick-up and drop-off location and time to the blockchain. The driver uses a matching technique to examine if the demand falls on his course and then sends the exact trip information encrypted with the rider's public key. To defend against malicious riders who send fake requests and do not commit to them, the system implements a time-locked deposit mechanism by utilizing smart contracts and zero-knowledge set membership proof. Both the driver and rider must demonstrate their good intentions by transferring a deposit to the blockchain. After the ride, the smart contract settles the balances based on the ride results. In addition, the proposed blockchain applies a reputation model to rate drivers and riders based on their previous behavior which incentivizes them to abide by the system rules.

TABLE XIV: Mapping of **Blockchain for General Vehicular Communications** solutions to security requirements

| Ref | RT1 (Auth) | RT2 (Conf) | RT3 (Avail) | RT4 (Intg) | RT5 (Priv) | RT6 (AC) | RT7 (N-rep) | RT8 (Lat) | RT9 (Scal) |
|-----|-----|-----|-----|-----|-----|-----|-----|-----|-----|
| [150] | ✓ | ✓ | ✓ | ✗ | ✓ | ✗ | ✗ | ✗ | ✗ |
| [160] | ✓ | ✓ | ✗ | ✗ | ✓ | ✗ | ✓ | ✗ | ✗ |
| [258] | ✓ | ✓ | ✗ | ✗ | ✓ | ✓ | ✓ | ✗ | ✗ |
| [94] | ✗ | ✓ | ✓ | ✗ | ✗ | ✗ | ✗ | ✓ | ✗ |
| [198] | ✓ | ✗ | ✓ | ✓ | ✓ | ✗ | ✗ | ✗ | ✓ |
| [142] | ✓ | ✓ | ✗ | ✓ | ✓ | ✗ | ✓ | ✗ | ✗ |
| [266] | ✓ | ✗ | ✗ | ✓ | ✓ | ✓ | ✓ | ✗ | ✗ |
| [112] | ✓ | ✓ | ✓ | ✓ | ✓ | ✗ | ✗ | ✓ | ✗ |
| [146] | ✓ | ✗ | ✓ | ✗ | ✓ | ✗ | ✗ | ✗ | ✗ |
| [175] | ✓ | ✓ | ✗ | ✓ | ✗ | ✗ | ✓ | ✗ | ✓ |
| [143] | ✗ | ✗ | ✓ | ✗ | ✗ | ✓ | ✗ | ✓ | ✗ |
| [145] | ✓ | ✓ | ✗ | ✗ | ✓ | ✓ | ✗ | ✗ | ✗ |
| [219] | ✓ | ✗ | ✗ | ✓ | ✗ | ✗ | ✗ | ✗ | ✗ |
| [36] | ✓ | ✗ | ✓ | ✗ | ✗ | ✗ | ✓ | ✗ | ✗ |
| [255] | ✓ | ✗ | ✓ | ✓ | ✗ | ✗ | ✓ | ✗ | ✗ |

A multilayered blockchain architecture for ride-sharing services was proposed in [226]. In the proposed system, the client sends a Request message to a service node. The message includes a nonce that uses the HMAC algorithm to avoid leaking the identity and details of the client. The service node broadcasts the request in the peer-to-peer network. When a driver accepts to offer the ride, it sends an encrypted acknowledgement to the service node. The latter saves the transaction in the blockchain and connects the driver to the rider. After the ride, the transaction data are validated by the service node and a new transaction that contains the ride result is added. The authors in [49] use the Summary Contract (SC), which is a dedicated blockchain smart contract, to store the main information that identifies each user based on his/her sharing criteria. The blockchain uses the details in the user's request to filter the candidate drivers based on the matching attributes within the passenger smart contract (PSC) and driver smart contract (DSC). Li et al. [147] make use of the zero-knowledge proof (ZKP) method and a permissioned blockchain for preserving the identity of passengers and drivers. The system includes a permission issuer node that issues prover keys to drivers and verifier keys to peer nodes. The latter store the blocks and smart contracts of the permissioned blockchain and acts as the data verifier that validate drivers' and riders' identities.

GreenRide, a blockchain-based ridesharing application was presented by Khanji and Assaf in [129]. The proposed application emphasizes environment preservation by promoting carbon emission reduction and enhancing air quality. In order to incentivizing drivers to share rides with passengers, the application rewards the driver with blockchain tokens. A token is given to the driver for every kg of $CO_2$ reduced via sharing rides. A similar application called BlockWheels was proposed by Joseph et al. in [118]. The proposed application uses Drizzle to connect the application to the blockchain using smart contracts in Truffle. In addition, the greedy mechanism is used to match riders with drivers, which enables drivers to acquire passengers who are close to them. A Practical Ethereum Blockchain based Efficient Ride Hailing Service (PEBERS) was proposed in [138]. PEBERS utilizes RSUs as fog nodes that mine the blocks of a consortium blockchain and match passengers with drivers. The blockchain model uses the Delegated Proof of Stake (DPoS) consensus algorithm in which validator nodes cooperate in order to ensure that the blockchain does not fork at any time. A similar approach was proposed by Zhang et al. [290]. Here, the Hash-oriented Practical Byzantine Fault Tolerance (PBFT) based consensus algorithm was used by the authors to avoid the double-spending and Sybil attacks.

Table XVII summarizes the papers that were discussed in this section, while Table XVIII maps each paper to the BITS security requirements that were stated in Section IV-C.

*4) Blockchain for Smart Parking:* Smart cities are rapidly growing and turning into a crowded environment. Within such environment, parking vehicle management is an indispensable requirement for citizens and service-providers. Several recent research works proposed various models for smart parking that utilize modern technologies such as IoT,

TABLE XV: Summary of **Blockchain for Autonomous Vehicles** solutions

| Ref | Contribution/ Purpose | Blockchain Type | Framework | Consensus | Storage | Validation Tools | Limits |
|---|---|---|---|---|---|---|---|
| [132] | Discusses the benefits of BC implementation in CAVs and the obstacles that hinder it | Discusses various BC Types for CAVs | Discusses several BC frameworks for CAVs | Discusses several consensus algorithms for CAVs | N/A | N/A | - missing discussion on BC for CAVs' sensor systems <br> - missing discussion on machine learning solutions for CAVs |
| [91] | Proposes a BC-CAVs architecture consisting of four layers: Cloud, Edge, Ethereum, and Data | Public | Ethereum and IPFS | Proof of Authority (PoA) | Vehicles and RSUs | Case study | - Important details of the cloud and edge layers are missing <br> - No real testing |
| [218] | BC to securely record and analyze the reading of CAVs' sensors | Permissioned | Custom | Not specified | Vehicles | Simulations using NS2 software | - Vehicles as BC miners <br> - Consensus not discussed <br> - Limited simulation setup |
| [228] | A miner node selection model for CAVs based on the fruit fly optimization algorithm (FOA) | Private | Ethereum | Customized PoW (FOA-based) | Edge, fog, and cloud nodes | Simulations using Ethereum and Truffle | - Limited simulations setup <br> - Does not compare with modern consensus protocols |
| [221] | Deploying AI algorithms at the edge servers to secure the sensing and tracking of CAV objects | Public | Ethereum and IPFS | Proof-of-Work (PoW), Proof-of-Burn (PoB), Proof-of-Attorney (PoA) | Vehicles | Simulations using Ethereum | - Simulations setup not described <br> - The usage of AI for intelligent sensing and tracking is not elaborated |
| [261] | Task-based and credit-based reputation models to evaluate the trustworthiness of CAVs and RSUs | Permissioned | Custom | Proof of Reputation | CAVs store block headers, RSUs store full ledger | Custom simulations | - Simulations setup not described <br> - Limited simulations <br> - Overhead of proposed system is not studied |
| [39] | BC for storing the updates to CAVs firmware and providing incentives to CAVs to forward updates | Consortium | Ethereum | Proof of Authority | Cloud nodes | Emulation using Geth Ethereum and Raspberry Pi | - Limited to firmware updates <br> - Scalability not discussed |
| [203] | BC for securely refueling autonomous electric vehicles | Public | Ethereum | PoW | Third party (not specified) | Simulations using Ethereum | - Consensus not discussed <br> - Limited simulations |

5G, ITS, and others to manage and execute the parking operations in an efficient manner. However, smart parking has many challenges, such as communication bandwidth, energy efficiency, integrity, security, privacy, and centralization [236]. A blockchain-enabled framework for Energy-Efficient Smart Parking in smart cities was proposed in [236]. The proposed framework uses Elliptic-curve cryptography (ECC) for data authentication and verification of all smart parking zones and drivers. In addition, a deep Learning LSTM network is utilized at the analysis layer to process cloud data and propose the best parking locations to the driver based on his/her preferences, such as distance, price, and delay. A collaborative blockchain solution with gamification for parking was proposed in [81]. In this approach, drivers who want to park collaborate in order to find which parking spots are empty and satisfy each driver's requirements. Each driver is rewarded with blockchain points for each successful parking in which the driver assists. The points are used by the driver to achieve reduced parking fees or free parking time. The system utilizes Artificial Intelligence (AI) algorithms to reduce the uncertainty of the route optimization algorithm, allowing the system to find optimal paths to the desired parking place.

TABLE XVI: Mapping of **Blockchain for Autonomous Vehicles** solutions to security requirements

| Ref | RT1 (Auth) | RT2 (Conf) | RT3 (Avail) | RT4 (Intg) | RT5 (Priv) | RT6 (AC) | RT7 (N-rep) | RT8 (Lat) | RT9 (Scal) |
|---|---|---|---|---|---|---|---|---|---|
| [132] | N/A | N/A | N/A | N/A | N/A | N/A | N/A | N/A | N/A |
| [91] | ✓ | ✗ | ✗ | ✓ | ✓ | ✓ | ✗ | ✓ | ✗ |
| [218] | ✓ | ✓ | ✗ | ✓ | ✗ | ✓ | ✗ | ✗ | ✗ |
| [228] | ✗ | ✗ | ✗ | ✓ | ✗ | ✓ | ✗ | ✓ | ✗ |
| [221] | ✓ | ✗ | ✓ | ✗ | ✓ | ✗ | ✗ | ✗ | ✗ |
| [261] | ✓ | ✗ | ✗ | ✓ | ✓ | ✗ | ✗ | ✗ | ✗ |
| [39] | ✓ | ✗ | ✓ | ✓ | ✗ | ✓ | ✗ | ✗ | ✗ |
| [203] | ✗ | ✓ | ✗ | ✗ | ✓ | ✗ | ✗ | ✓ | ✗ |

BSFP, a blockchain-based smart parking system with Fairness and Privacy was proposed in [285]. The system utilizes group signatures to achieve anonymous authentication for drivers and parking owners, while granting the traceability of malicious users. In addition, bloom filters are used to preserve the location-privacy of range queries. On the other hand, vector-based encryption is applied to restrict access to blockchain transactions to authorized drivers and owners. A crowdsensing framework that is used to assist the smart parking system was proposed in [134]. The authors describe a multi-blockchain structure that includes a public chain and a private chain. The public chain utilizes the sensing data from several contributors to allow any user to discover the available parking locations and options. On the other hand, the service providers and drivers process the data from the public blockchain to generate parking transactions that are stored in the private chain. A bridge node joins both blockchains and creates channels to link and exchange information between the two blockchains.

Jennath et al. [113] describe Parkchain, a system in which parking pools are created to allow drivers to rent out their unused land for a specified amount of time. The system employs blockchain-based parking tokens to create digital assets that are equivalent to unused land. Smart contracts are used to enforce the terms and conditions between the parking spot owner and the driver, along with the tenure of the hired property and the pricing options to be applied during the lease period. In order to ensure fair parking rates, the system in [32] uses a commitment technique that is applied during the submission of the offers to the driver. In addition, a short randomizable signature is utilized to authenticate drivers during the reservation process while keeping the driver's ID anonymous. A consortium blockchain is used by the drivers to store rating transactions that anonymously rate the parking service and contribute to the reputation of the parking service provider. Ahmed et al. [12] describe the details of Parking-as-a-Service, which provides distinct application interfaces for each type of participants, such as parking owner, driver, and blockchain node. Multiple service providers can utilize Parking-as-a-Service to offer their parking services to drivers. Each service provider has a smart contract to generate the parking offer transaction. When a driver consumes an offer, the resulting transaction is validated by the blockchain nodes to ensure that the transaction matches the details in the service provider offer transaction.

Table XIX summarizes the papers that were discussed in this section, while Table XX maps each paper to the BITS security requirements that were stated in Section IV-C.

*5) Blockchain for Aerial Networks:* A large number of research papers proposed different mechanisms for integrating the BC into the Unmanned Aerial Vehicle (UAV) network, which has been recently labeled the Internet of Drones (IoD). Some of these systems employ the BC as a means of safeguarding the data produced by the drones. Other frameworks utilize the BC as a tool for overseeing the IoD processes and keeping the history of the missions executed by the UAVs for future searching. A third type of systems makes use of the BC to provide UAV services to cloud users by implementing smart contracts and ensuring the security of the users-drones communications.

Barka et al. [37] discuss a model in which each UAV that forwards a packet adds a trust value to it. The BC miners should choose whether to add a new transaction to the next block or not according to the trust values added by the UAVs. Another system by Xu et al. [269] utilizes the UAVs to support the MEC servers. The authors propose a method for permitting resources trading between the Edge servers and UAVs. In this system, the UAVs consume the server resources and the latter is compensated with a certain incentive such as digital currency. The Edge servers act as the BC miners that execute Proof of Work (PoW). A similar system by Luo et al. [163] utilizes the drones to cache data from IoT devices and forward the data to the MEC servers that are not reachable by the IoT network. The MEC servers create a private BC network that stores the transactions of the exchanges between the IoD and the cloud user. The authors of [7] propose a BC framework in which an element labeled Policy Header is added to the BC block header to store the type of actions that can be executed by each user or group on the block transactions. The system classifies the drones' users into groups, where each group has its own BC. The nodes of each group select a forger node to produce the BC blocks. The selection is managed by a set of master controllers based on game theory. The BC is stored within the user layer only.

TABLE XVII: Summary of **Blockchain for Ride-Sharing and Ride-Hailing** solutions

| Ref | Contribution/ Purpose | Blockchain Type | Framework | Consensus | Storage | Validation Tools | Limits |
|---|---|---|---|---|---|---|---|
| [38] | Uses smart contracts and zero-knowledge set membership proof to prevent users from submitting malicious ride requests | Public | Ethereum | Not specified | Public network | Simulations using the Kovan Ethereum test network | - Consensus protocol not specified<br>- Scalability of the system not discussed |
| [226] | BC is used to compute the rides' prices and provide trust between clients and providers | Public | Chord | Not specified | All network nodes | Mathematical analysis and implementation via the Spring framework | - Consensus protocol not specified<br>- BC latency and delay were not evaluated |
| [49] | Ride-sharing service divided into three layers: Registration, Transaction, and Cash; with BC as layers' connector | Public | Ethereum | Not specified | BC distributed among the three layers on all network nodes | Simulations using Geth Ethereum and Solidity | - Consensus protocol not specified<br>- Very limited simulations were conducted<br>- Scalability of the system not discussed |
| [147] | BC is used to store ride logs and zero-knowledge proof (ZKP) records, and perform ZKP verifications | Permissioned | Hyperledger Fabric | Hyperledger Fabric consensus | Service providers' network | Prototype testing using Hyperledger Composer and Caliper | - Consensus details not discussed<br>- Scalability of the system not discussed |
| [129] | BC is used to incentivize drivers to share their rides by rewarding them with BC tokens | Hybrid (part is public and part is permissioned) | Ethereum | Not specified | Cloud nodes | Testing via Google Cloud services | - Consensus protocol not specified<br>- Simulations included cost savings only<br>- BC role is limited |
| [118] | BC is used to store users' and rides' information and to perform matching and fare calculation | Public | Custom (Ethereum-based) | Proof of Stake (intended) | All willing nodes | Theoretical analysis | - System was not evaluated<br>- Consensus protocol not discussed<br>- BC was not fully discussed |
| [138] | Drivers deploy smart contracts to offer ride-hailing services to riders | Consortium | Ethereum | Delegated Proof of Stake (DPoS) | RSUs | Testing via an Ethereum Virtual Network (EVN), Ganache, and Truffle | - Simulations focused on gas consumption only |
| [290] | Proposing a hash-oriented Practical Byzantine Fault Tolerance model for ridesharing delivery | Public | Custom | Hash-oriented Practical Byzantine Fault Tolerance | Mobile edge computing (MEC) servers | Simulations | - Simulation setup and configuration were not explained |

TABLE XVIII: Mapping of **Blockchain for Ride-Sharing and Ride-Hailing** solutions to security requirements

| Ref | RT1 (Auth) | RT2 (Conf) | RT3 (Avail) | RT4 (Intg) | RT5 (Priv) | RT6 (AC) | RT7 (N-rep) | RT8 (Lat) | RT9 (Scal) |
|---|---|---|---|---|---|---|---|---|---|
| [38] | ✓ | ✓ | ✗ | ✓ | ✓ | ✗ | ✗ | ✓ | ✗ |
| [226] | ✓ | ✓ | ✗ | ✓ | ✗ | ✓ | ✗ | ✗ | ✓ |
| [49] | ✓ | ✗ | ✗ | ✓ | ✓ | ✗ | ✗ | ✗ | ✗ |
| [147] | ✓ | ✓ | ✗ | ✓ | ✓ | ✓ | ✗ | ✓ | ✗ |
| [129] | ✗ | ✓ | ✗ | ✗ | ✓ | ✗ | ✗ | ✗ | ✓ |
| [118] | ✓ | ✓ | ✗ | ✗ | ✓ | ✗ | ✗ | ✗ | ✗ |
| [138] | ✓ | ✓ | ✗ | ✗ | ✓ | ✓ | ✗ | ✗ | ✓ |
| [290] | ✓ | ✓ | ✗ | ✓ | ✓ | ✗ | ✗ | ✓ | ✓ |

Several research works present BC-based systems for 5G- or 6G- UAV networks. Aloqaily et al. [21] propose a cloud service, named Drones-as-a-Service, in which UAVs are used on-demand to deliver a high QoS connection to users in areas where cellular base stations cannot offer the required QoS. A user who desires to utilize the services of a certain drone is given access to the drone keys in order for the user and drone to communicate securely. A model for anonymous authorization in IoD was proposed in [24]. The model conceals the identities of UAVs inside the network packets and permits only the receiver to deduce the ID of the sender. In this system, UAVs are lightweight BC nodes that save only the header part of the BC blocks. A zone-based architecture that utilizes a tailored consensus algorithm, named the Drone-based Delegated Proof of Stake (DDPOS), was presented in [275]. In this framework, the smart city is partitioned into zones where each zone can include several drones and a single drone controller (DC). The BC is saved and supervised by the drone controllers. Each DC certifies the transactions within its zone and authorizes the drones' tasks during their missions in the smart city.

Nguyen et al. [192] present a system in which drones are grouped into clusters where each cluster contains a drone leader. The latter is selected as one of the powerful drones in the cluster. Drones in each cluster communicate with the leader only, and the leaders collaborate with each other and with the edge servers. Leaders gather, manage, and send important information to the edge servers, and the latter use the RAFT consensus algorithm to organize and store the BC. Unlike the system in [192], Tan et al. [242] propose a model that makes use of the cluster heads (CH) as BC miners. The cluster drones save only the BC state, which includes the list of cluster members. The system uses a lightweight election mechanism to select the CH that produces each BC block. Singh et al.

[234] propose ODOB, a model in which an Aviation Authority (AA) outlines the access rights of each node and makes security-related resolutions. The authors propose that the BC data should be decoupled from the BC. When producing the hash of the previous block, only the block header is utilized in the process. This technique makes the BC block modifiable and permits adding new transactions to it when required. The authors of [86] present two major changes to the IoD BC: first, they implement the Keccak lightweight cryptography hashing to sign transactions and produce the Merkle hash. Second, they insert into the block two new elements: the Policy List and Reputation Tree, which specify the access rights and the reputation of each drone.

The authors in [215] propose a flight compliance system for Drone Service Providers with two key goals: 1) predefining the flight path of each drone according to the drone's mission, storing the path within the drone's BC, and ensuring that the path does not contain restricted or private areas, and 2) monitoring each drone to ensure that it follows its flight path in order to avert accidents. A similar method is proposed in [202]. Here, drone owners register their drones with a UAV central authority. The latter provides UAV-as-a-Service to customers through virtualization. The central authority outlines the description of each service, generates a smart contract for the consumer, and stores it in the BC. Liao et al. [152] present the Proof-of-Security-Guarantee (PoSG) consensus algorithm, which assigns a higher mining probability to drones with superior security guarantees. In PoSG, a specified amount of resources should be provided by a drone, and each resource must satisfy a minimum threshold for the drone to join the BC network.

Tables XXI and XXII summarize the papers that were discussed in this section, while Table XXIII maps each paper to the BITS security requirements that were stated in Section IV-C.

*6) Blockchain for Miscellaneous ITS Applications:* In addition to the five main use cases that were discussed in the previous subsections, several authors proposed BC-based systems for specific applications. In this section, we discuss the most important miscellaneous applications of the BC in transportation. Li et al. [143] propose a GPS positioning error evolution sharing framework that utilizes the BC to store and distribute the evolution of positioning errors. The system enhances the vehicle positioning accuracy and ensures the security and credibility of positioning data. The system makes use of the edge servers that execute a deep neural network (DNN) prediction algorithm to obtain the positioning error evolution. The authors in [33] present the concept of tradable mobility permit (TMP) to reduce traffic congestion. In this model, traffic authorities utilize the blockchain to issue a limited number of mobility credits that are distributed equally among the users. The latter can exchange and trade the mobility permits based on their travel needs. Users can also use the TMPs for transport-related payments such as toll payments, parking fees, public transport tickets, etc. The authors utilize Ethereum smart contracts to test the trading of TMPs in the conducted simulations.

TABLE XIX: Summary of **Blockchain for Smart Parking** solutions

| Ref | Contribution/ Purpose | Blockchain Type | Framework | Consensus | Storage | Validation Tools | Limits |
|---|---|---|---|---|---|---|---|
| [236] | BC virtualization is combined with Deep LSTM to analyze the parking data and offer the required parking space | Public | Hyperledger Fabric | Proof of Authentication (PoA) | RSUs | Simulations using SUMO, PyPML, TraCI, and Hyperledger Fabric | - Assuming all RSUs are secure and trusted<br>- System scalability is not fully analyzed |
| [81] | Users collaborate to report free parking spaces and receive free parking minutes rewards | Public | Custom | Not specified | Cloud nodes | Android application | - BC details are missing<br>- BC implementation is not discussed |
| [285] | Group signature is used for anonymous authentication, and Bloom filters are used to preserve the location privacy of range queries | Public | Hyperledger Fabric | Hyperledger Fabric consensus | Third-party network | Simulations using Hyperledger Fabric and PBC Library | - BC network is not described<br>- BC overhead and cost are not analyzed |
| [134] | A multi-blockchain structure that utilizes mobile crowdsensing for smart parking | Hybrid (contains a public BC and a private BC) | Ethereum and Hyperledger Fabric | Not specified | Vehicles for public BC and service providers for private BC | Testing using the Ethereum test network and the Hyperledger Fabric environment | - Consensus protocol not specified<br>- Scalability of the system not discussed |
| [113] | BC is used to allow users to search parking pools and rent out the unused parking spots | Consortium | Ethereum | Multi-party consensus | Governmental bodies and agencies | System implementation using Geth Ethereum, Solidity, and Truffle | - BC role is limited<br>- No testing was performed |
| [32] | Uses a time-locked anonymous payment technique to discourage drivers from not committing to their reservations | Consortium | Custom | Raft consensus algorithm | Parking lots | Mathematical analysis and evaluations | - No real testing was made<br>- System scalability is not fully analyzed |
| [12] | Integrates BC into Parking-as-a-Service in order to update parking spaces and offer cost information to users | Public | Custom | Not specified | Car parking service providers | Theoretical analysis | - Details of BC network are missing<br>- No real testing was made |

Utilizing the BC for smart traffic light system was proposed by Zeng et al. in [284]. In the proposed system, traffic light nodes collect traffic data, filter and aggregate them, and send the results to the edge servers. The latter add the results to a new BC block, cooperate to obtain consensus, and analyze each new block to produce traffic light decisions. Zhang et al. [289] propose a similar smart traffic light system that includes a credibility mechanism to prevent vehicles from broadcasting mendacious messages and malicious requests. The deployed BC system utilizes ElGamal encryption and group signature algorithm to ensure the confidentiality, privacy, and non-repudiation of traffic information. In addition, the authors propose a mixture of (artificial intelligence, computational experiments, and parallel execution) to achieve the optimization of signal light management. Yang et al. [273] propose a traffic event validation scheme based on blockchain. The authors introduce Proof-of-Event as a new consensus mechanism to confirm the event occurrences and detect false warning messages. The traffic data are collected by the RSUs, and each passing vehicle validates the correctness of the traffic data at the RSU when it receives the event notification. Moreover, the BC transactions are divided into two consecutive stages, first synchronizing the local blockchain then the global blockchain. This enables the delivery of warning messages in the correct region and time.

TABLE XX: Mapping of **Blockchain for Smart Parking** solutions to security requirements

| Ref | RT1 (Auth) | RT2 (Conf) | RT3 (Avail) | RT4 (Intg) | RT5 (Priv) | RT6 (AC) | RT7 (N-rep) | RT8 (Lat) | RT9 (Scal) |
|---|---|---|---|---|---|---|---|---|---|
| [236] | ✓ | ✓ | ✗ | ✓ | ✓ | ✗ | ✓ | ✓ | ✗ |
| [81] | ✓ | ✗ | ✗ | ✗ | ✓ | ✗ | ✗ | ✗ | ✗ |
| [285] | ✓ | ✓ | ✓ | ✓ | ✓ | ✗ | ✓ | ✗ | ✓ |
| [134] | ✗ | ✓ | ✗ | ✓ | ✓ | ✓ | ✗ | ✓ | ✗ |
| [113] | ✗ | ✓ | ✗ | ✗ | ✓ | ✗ | ✗ | ✗ | ✗ |
| [32] | ✓ | ✓ | ✓ | ✓ | ✓ | ✗ | ✗ | ✓ | ✗ |
| [12] | ✓ | ✓ | ✗ | ✓ | ✓ | ✗ | ✗ | ✗ | ✗ |

TABLE XXI: Summary of **Blockchain for Aerial Networks** solutions - Part 1

| Ref | Contribution/ Purpose | Blockchain Type | Framework | Consensus | Storage | Validation Tools | Limits |
|---|---|---|---|---|---|---|---|
| [37] | Trust management solution for UAVs that uses the Bayesian inference approach | Private | Custom | PoW | Cloud servers | Simulations via the NS3 tool | - Use of PoW<br>- Very high latency<br>- BC Scalability was not studied |
| [269] | A resource pricing and trading scheme to allocate edge computing resources to UAVs | Public | Custom | PoW | MEC servers | Mathematical analysis and simulations | - Use of PoW<br>- Very low number of UAVs in simulations |
| [163] | UAVs offload IoT readings to MEC servers that create the blockchain | Private | Ethereum | Not specified | MEC servers | Ethereum-based simulations | - Consensus protocol not specified<br>- Assuming that the consensus time is negligible<br>- Limited testing |
| [7] | A forger node is selected based on game theory to create the BC blocks | Public | Ethereum | Proof of Stake | Master controllers (selected among user noes) | Mathematical analysis | - Threat model does not consider BC attacks<br>- No real testing |
| [21] | Public and Private BCs are used to offer Drone-as-a-Service to 5G users | Hybrid (public and private) | Custom | Not specified | Cloud, fog, and edge nodes | Simulations using NS3 software | - Authors claim scalability but fail to prove it (maximum of 100 nodes in simulations) |
| [24] | Non Interactive Zero Knowledge Proof (NIZKP) is used to provide UAV anonymity | Public | Custom | Lightweight orderer consensus | Drones store light BC and GCSs store full BC | Mathematical analysis | - Focus on drone anonymity only<br>- No real testing |
| [275] | Network divided into zones based on drone groups. System allows automatic handoff | Public | Custom (Bitcoin-based) | Drone-based Delegated Proof of Stake (DDPOS) | Drone controllers (ground stations) | Simulations using Bitcoin and NS3 | - Limited to zone-based framework<br>- Fails to consider high-mobility issues |

TABLE XXII: Summary of **Blockchain for Aerial Networks** solutions - Part 2

| Ref | Contribution/ Purpose | Blockchain Type | Framework | Consensus | Storage | Validation Tools | Limits |
|---|---|---|---|---|---|---|---|
| [192] | BC is used in maritime search and rescue scenarios to secure drones' communications | Permissioned | Hyperledger Fabric | Hyperledger Fabric consensus | Edge servers | Real testing using two types of drones | - Limited to maritime scenarios<br>- Limited testing using five drones only |
| [242] | Blockchain-based distributed key management for heterogeneous FANETs | Public | Custom | Custom | Cluster head drones | Custom Python-based simulations | - BC is used for key management only<br>- Assumes that cluster head drones are trustworthy |
| [234] | Decoupling block body from block header to enable drones to store a lightweight BC copy | Permissioned | Custom | Custom (trust-based voting) | GCSs store full copy while drones store lightweight copy | Mathematical analysis | - Proposed BC is not immutable<br>- Drones still access GCSs for old transactions<br>- No real testing |
| [86] | A reputation-based consensus model and data access control policies based on transactions' types | Private | Custom | Combined DPoS and reputation evaluation scheme | Drones | Simulations using NS3 and the UB-ANC Emulator | - Drones use DPoS (energy consumption)<br>- Limiting the size of the BC transaction to 1KB |
| [215] | Establishes pre-allocated flight paths for drones and restrict their access to unauthorized areas | Private | Ethereum | Not specified | Not specified | Ethereum-based simulations on Amazon AWS EC2 | - Consensus is not discussed<br>- A maximum of 10 BC nodes were considered<br>- Huge latency |
| [202] | Providing UAV-as-a-Service for industrial Applications | Permissioned | Custom | Combination of PoAh and PBFT | Drones | Mathematical analysis | - Drones as BC nodes<br>-BC role is limited<br>- No real testing |
| [152] | Ensuring trusted collaboration between controllers of software defined IoD (SD-IoD) | Consortium | Custom (utilizes a new Cooperation Coin) | Proof of Service Guarantee (PoSG) | SDN controllers | Custom simulations | - Limited to SDN-based network<br>- Limited testing in simulations |

Deng and Gao [64] propose two electronic payment schemes based on the BC. The members involved in a transaction could be a vehicle and a single RSU (V-R transaction); such as in the park toll management system, or a vehicle and multiple RSUs (V-Rs transaction); such as in electronic toll collection. The authors in [50] adopt the blockchain for storing the energy trading transactions that are exchanged between electric vehicles (EVs). The BC miner nodes that validate the EVs requests are selected based on their energy requirements, time of stay, dynamic pricing, and connectivity record. The system implements a software-defined network in order to efficiently route the EVs requests to the BC network. The proposed system allows EVs to trade energy among themselves or with the service provider/utility. Li et al. [148] propose the idea of a BC gateway that resides between two neighboring blockchain-based traffic management systems to switch traveling vehicles from one blockchain into another. The gateway utilizes the non-interactive zero-knowledge range proof (ZKRP) scheme to verify the information of an incoming vehicle without revealing any sensitive data.

TABLE XXIII: Mapping of **Blockchain for Aerial Networks** solutions to security requirements

| Ref | RT1 (Auth) | RT2 (Conf) | RT3 (Avail) | RT4 (Intg) | RT5 (Priv) | RT6 (AC) | RT7 (N-rep) | RT8 (Lat) | RT9 (Scal) |
|---|---|---|---|---|---|---|---|---|---|
| [37] | ✓ | ✗ | ✗ | ✓ | ✗ | ✗ | ✗ | ✓ | ✗ |
| [269] | ✗ | ✓ | ✓ | ✗ | ✓ | ✗ | ✗ | ✗ | ✗ |
| [163] | ✗ | ✓ | ✗ | ✗ | ✓ | ✓ | ✗ | ✓ | ✗ |
| [7] | ✓ | ✓ | ✗ | ✓ | ✓ | ✗ | ✗ | ✓ | ✗ |
| [21] | ✓ | ✓ | ✓ | ✓ | ✓ | ✗ | ✗ | ✗ | ✗ |
| [24] | ✓ | ✓ | ✗ | ✗ | ✓ | ✗ | ✗ | ✗ | ✗ |
| [275] | ✓ | ✗ | ✗ | ✗ | ✓ | ✗ | ✗ | ✓ | ✓ |
| [192] | ✓ | ✓ | ✗ | ✓ | ✗ | ✗ | ✗ | ✓ | ✗ |
| [242] | ✓ | ✓ | ✗ | ✗ | ✓ | ✗ | ✗ | ✓ | ✗ |
| [234] | ✓ | ✓ | ✗ | ✗ | ✗ | ✓ | ✗ | ✗ | ✗ |
| [86] | ✓ | ✓ | ✗ | ✓ | ✗ | ✗ | ✓ | ✓ | ✗ |
| [215] | ✗ | ✓ | ✗ | ✓ | ✓ | ✗ | ✗ | ✓ | ✓ |
| [202] | ✓ | ✓ | ✗ | ✗ | ✓ | ✓ | ✗ | ✗ | ✗ |
| [152] | ✗ | ✓ | ✗ | ✗ | ✓ | ✗ | ✗ | ✗ | ✓ |

The BC is used to implement Mobility-as-a-Service (MaaS) by the authors in [193]. The blockchain-based MaaS consists of two main types of participants: 1) the transportation providers who act as BC miners and manage and maintain the BC, and 2) travelers or clients who receive routes' information from the providers to create their own route as a smart contract (based on their travel preferences). Once receiving the smart contract of a client, providers collaborate to verify and confirm the smart contract before adding it to the next block. A model for enhancing road safety by integrating the IoT and BC was proposed in [209]. The system utilizes the hashgraph technology to create communication networks between the various types of vehicles in the network and other relevant nodes such as the RSUs and cellular base stations. The proposed system has the flexibility of selecting the order and speed of the transactions, postponing them, or stopping them from reaching the block. A timestamp consensus is used in hashgraph to prevent nodes from changing the order of transactions, which is a major issue faced by the IoT and blockchain. The authors describe various consensus models that can be used with hashgraph, such as the Gossip protocol, virtual voting, and round creation.

Table XXIV summarizes the papers that were discussed in this section, while Table XXV maps each paper to the BITS security requirements that were stated in Section IV-C.

### E. BITS Future Directions

From our study of a large number of BC-based systems for the smart city intelligent transportation, we observe the important role that the BC plays in securing the data, communications, and operations of the ITS. The BC has been utilized in a large number of ITS applications, including vehicles' and drones' network security, ride-sharing and ride-hailing, smart parking, traffic light system, electric vehicles' charging, etc. However, several challenges still exist in the BC integration into ITS. We describe the important ones below, highlighting the future research possibilities stemming from these challenges:

- **Data availability:** among the large number of studied papers, very few took data availability into consideration, and fewer papers proposed reliable solutions to ensure that ITS data in the BC will be sufficiently available to all ITS participants. With the highly demanding computations and cryptographic operations of the BC, it becomes quite a challenge to satisfy the time demands of ITS participants related to storing, retrieving, and analysing the BC data.

TABLE XXIV: Summary of **Blockchain for Miscellaneous ITS Applications** solutions

| Ref | Contribution/ Purpose | Blockchain Type | Framework | Consensus | Storage | Validation Tools | Limits |
|-----|----------------------|-----------------|-----------|-----------|---------|------------------|--------|
| [143] | A GPS positioning error evolution framework to improve vehicle positioning accuracy | Consortium | vDLT | BFT-DPoS | MECNs | Custom simulations | - BC need not fully justified<br>- BC role is limited |
| [33] | Tradable mobility permit (TMP) via mobility credits that are used to consume ITS services | Public | Ethereum | Not specified | All network nodes | Numerical analysis and testing scenario | - Application limited to the proposed TMP model<br>- Scalability was not studied |
| [284] | Traffic lights system that uses Edge Intelligence and BC to reduce the communication and time costs | Public | Ethereum | Combination of Byzantine and Rayleigh wave | Traffic lights nodes and edge nodes | Simulations using the VISSIM tool | - Limited simulation setup<br>- Does not compare with recent consensus algorithms |
| [289] | A credibility mechanism to prevent vehicles from broadcasting malicious messages | Consortium | Custom | AlgoRand consensus algorithm | Vehicles, RSUs, and Traffic Department | Mathematical analysis | - Vehicles participate in consortium blockchain<br>- Limited simulations |
| [273] | Proof-of-event consensus in which passing vehicles verify the traffic information at RSUs | Public | Custom (BTEV) | Proof of Event | RSUs | Simulations using NS3 software | - Assumes all RSUs are trustworthy<br>- BC scalability not studied |
| [64] | Two BC-based payment schemes for park toll management and electronic toll collection | Public | Ethereum | Custom | RSUs and payment platform | Mathematical analysis and Ethereum-based simulations | - Consensus details not discussed<br>- Scalability not discussed<br>- Limited testing and simulations |
| [50] | An energy trading scheme for electric vehicles in which the BC validates EVs' charging requests | Consortium | Custom (SDN-based) | PoW | Vehicles and SDN nodes | Mathematical analysis | - Use of PoW<br>- Vehicles act as BC miners<br>- Limited testing scenarios |
| [148] | A gateway that switches traveling vehicles from one blockchain into another | Permissioned | Hyperledger Fabric | Hyperledger Fabric consensus | Edge nodes | Simulations via the Hyperledger Fabric, Ursa, and Caliper tools | - Assumes all RSUs are trustworthy<br>- BC network overhead was not studied<br>- BC scalability not studied |
| [193] | Offers Mobility-as-a-Service to passengers who consume MaaS via smart contracts | Consortium | Custom | Custom (voting-based) | Transportation providers | Theoretical discussion | - Consensus details not discussed<br>- No testing was made |
| [209] | Hashgraph-based BC to schedule vehicles' requests according to their priorities | Private | Custom | Hashgraph virtual voting | Vehicles | OMNeT++ simulations | - BC is limited to IoT-based transactions<br>- Limited testing |

TABLE XXV: Mapping of **Blockchain for Miscellaneous ITS Applications** solutions to security requirements

| Ref | RT1 (Auth) | RT2 (Conf) | RT3 (Avail) | RT4 (Intg) | RT5 (Priv) | RT6 (AC) | RT7 (N-rep) | RT8 (Lat) | RT9 (Scal) |
|-----|-----|-----|-----|-----|-----|-----|-----|-----|-----|
| [143] | ✗ | ✓ | ✗ | ✓ | ✗ | ✓ | ✗ | ✗ | ✗ |
| [33] | ✓ | ✗ | ✗ | ✗ | ✓ | ✗ | ✗ | ✓ | ✗ |
| [284] | ✓ | ✗ | ✗ | ✓ | ✓ | ✗ | ✗ | ✓ | ✓ |
| [289] | ✓ | ✓ | ✗ | ✓ | ✓ | ✗ | ✓ | ✓ | ✗ |
| [273] | ✓ | ✓ | ✓ | ✓ | ✓ | ✗ | ✓ | ✗ | ✓ |
| [64] | ✓ | ✓ | ✗ | ✗ | ✓ | ✗ | ✓ | ✓ | ✗ |
| [50] | ✓ | ✓ | ✗ | ✗ | ✓ | ✓ | ✗ | ✓ | ✗ |
| [148] | ✓ | ✗ | ✗ | ✓ | ✓ | ✓ | ✗ | ✓ | ✗ |
| [193] | ✗ | ✓ | ✗ | ✗ | ✓ | ✗ | ✗ | ✗ | ✗ |
| [209] | ✓ | ✓ | ✗ | ✗ | ✗ | ✗ | ✗ | ✗ | ✓ |

- **Computational overhead:** The BC operations demand heavy processing and communications to validate transactions, generate the BC blocks, and reach consensus. In a highly dynamic environment such as that of the ITS, it becomes a challenge to ensure that mobile nodes with limited resources can perform the BC operations reliably and efficiently. This is true taking into consideration the ITS characteristics, such as high mobility, intermittent connectivity, and limited resources. A wide-scale real testing of the BC within the vehicular and UAV networks is still missing in order to study these issues.

- **Artificial Intelligence:** Several research works attempted integrating the BC with A.I. models to improve the ITS applications' performance and reliability. An example of such integration is federated learning that focuses on mobile nodes' collaboration while implementing the machine learning algorithm. However, much of this domain hasn't been explored yet, and a lot of promising progress in the ITS can be expected from future systems that exploit the benefits of the BC and A.I. together.

- **Scalability:** Similar to data availability, very few systems attempted testing the BC in ITS while scaling the network to a realistic level that resembles real life vehicular networks in urban environments. It is still questionable how the BC will perform in a real city scenario where tens of thousands of vehicles communicate with hundreds of RSUs and cloud nodes to execute the ITS applications.

## V. Blockchain for Smart Agriculture

The report called "Agriculture 4.0 – The Future Of Farming Technology" [169], launched by the World Government Summit, claims the urgency to modernize the agriculture activity. The report states that, by 2050 we will need to produce 70% more food, an ambitious number which is very far from the current state. Indeed, 8% of the world population will still suffer from undernourishment by 2030. These indicators urge us to develop a new vision for the agriculture summarized as the agriculture 4.0. Agriculture 4.0 focuses on the precision agriculture while using new technologies like the internet of things (IoT), the big data as well as blockchain to bring greater business efficiencies to face the rising populations and climate change [199].

### A. Existing Surveys about Blockchain for Smart Agriculture

Table XXVI reviews some recent surveys dealing with the use of blockchain technology in the agriculture field.

### B. Benefits of Blockchain Technology in Smart Agriculture

Blockchain technology has many benefits with respect to its application in the agricultural sector:

1) **Trusted data availability**: as a trusted means of storing data, it facilitates the exploitation of data-driven technologies to make farming activities smarter. It even promotes the agricultural research which relies on big part on the data sharing between researchers.

2) **Fraud prevention**: as it allows the tracking of the provenance of food and crops, blockchain technology builds trust between producers and consumers and avoids counterfeiting.

3) **Financial transactions efficiency and security**: it facilitates financial transactions between stakeholders by means of smart contracts that can be triggered upon the update of the data appearing in the ledger.

4) **Cost and time reduction**: it reduces costs and saves time. The most representative example here is the insurance industry where checking claims (e.g damage) is time and effort consuming. However, smart contracts provide an automatic way that avoids the heavy administrative procedures.

5) **Data security**: blockchain technology is able to increase data security. For example, the decentralized public ledger stores time-stamped transactions relative to lands

TABLE XXVI: **Blockchain for Smart Agriculture** Existing Surveys

| Ref | Scope | Description | (+)Pros/(-)Cons |
|---|---|---|---|
| [247] | Integrating IoT with Blockchain | •Focuses on the integration of IoT with blockchain in precision agriculture<br>•Discusses opportunities in IoT-based precision agriculture like food safety, supply chain and farm overseeing. The paper proposes an architecture for each use case.<br>•Reviews existing blockchain agriculture platforms<br>•Discusses opportunities and challenges in this integration (managing identities of IoT devices, securing communications, optimizing parameters like energy consumption, etc)<br>•Reviews three different patterns on how to integrate IoT with blockchain | + Comprehensive analysis<br>- Platform description lacks comparison between these platforms. |
| [267] | Applications of blockchain in agriculture | •Describes four applications of blockchain: agricultural insurance, smart agriculture, food supply chains, and agricultural e-commerce.<br>•Discusses limitations regarding mainly the practical development aspects. | + Provides a classification of agriculture blockchain applications<br>- Reviewed solutions are not detailed |
| [156] | Review of applications, blockchain platforms and current implementation efforts in the agricultural field | •Explains blockchain technical building blocks (including cryptographic methods, data structure and consensus methods)<br>•Provides a classification of applications<br>•Reviews blockchain existing platforms and provides comparison between them<br>•Reviews agricultural blockchain systems<br>•Discusses challenges (scalability, data privacy and security, legacy system integration) and proposes solutions to them.<br>•Demonstrates a food supply chain management system in post-COVID-19 conditions. | + Comprehensive analysis, both conceptual and practical aspects are addressed, with a use case demonstration<br>- Solutions to described challenges are only conceptual (architectures), need to know how development could face these challenges. |
| [84] | Review of agri-food applications which use blockchain | •Reports existing works on the subject and classifies them into clusters.<br>•Reviews the scientific and commercial works | + Detailed review of cited papers<br>- The review lacks comparison between cited papers |
| [181] | Review of security threats and solutions in IoT-based green agriculture | •Classifies the security threats into five categories<br>•Reviews access control and privacy oriented solutions as well as consensus mechanisms | + Different security aspects are addressed<br>+ Good technological assessment<br>- The practical integration of blockchain in the proposed architecture is not discussed |

trade. Thus, it becomes possible to protect such transactions and track the property transfer.

### C. Agriculture Applications Requirements

We identify the following requirements for smart agriculture.

1) **Auditability/Traceability** of products is essential for tracking crops provenance for instance.
2) **Availability**: sharing data in a permanent way is a requirement for some agriculture applications like land ownership validation.
3) **Securing data**: stored data are more or less sensitive. For sensitive data, it is necessary to protect them from malicious access.
4) **Integrity**: must be guaranteed to prevent and detect malicious modifications of data. To achieve this integrity, users must be authenticated, their access must be controlled, and their privacy must be protected.
5) **Trust** is necessary when we speak about e-commerce for agriculture products for instance.

From these different requirements, we deduce that old management methods and technologies are not always relevant to make smart agricultural services. In the next section, we argue about the role of the blockchain technology in this evolution.

### D. Blockchain for Smart Agriculture Use Cases

From our study, we can synthesise the state of the art solutions and classify them into the following four categories as illustrated in Figure 5:

1) Supervision and management including the food supply chain, data sharing (for traceability or research) and supervising the workers contracts.
2) E-commerce for agricultural products.
3) Resource management and environment awareness like managing water resources and collecting wastes.
4) Land registration and agricultural insurance.

In the following we detail these categories.

*1) Supervision and Management:* By providing secure data, blockchain technology is able to strengthen the supervision

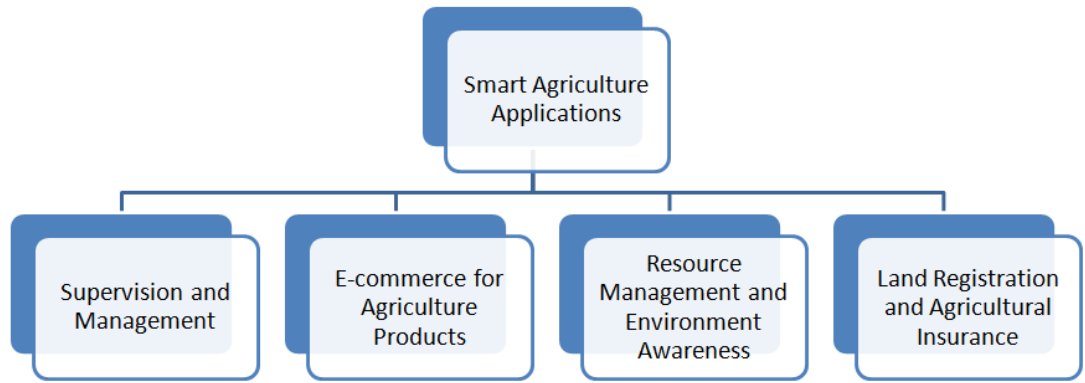

Fig. 5: Classification of Blockchain Agriculture Solutions

of agricultural products quality and facilitate the cooperative agricultural work. First of all, as previously highlighted in Section VI, food supply chain can be efficiently supervised through blockchain [108, 281, 176]. Blockchain enables both consumers and farmers to track agricultural crops, livestock and machinery. This will increase consumers' as well as stakeholders' trust in the agricultural products. This ensures food safety and reduces food fraud. There are many projects along with platforms that are based on the use of the blockchain technology like Provenance [211], Agridigital [10, 268], IBM Blockchain [105] and Carrefour [4].

Agricultural collaborative research benefits from the blockchain transparency and security to increase confidence in research results and makes information sharing easier as proposed by [164].

Blockchain is a means of protecting agriculture workers by supervising their contracts or temporary agreements. It is the aim of [26] which proposed a model for assisting employers in the contract processing, and for ensuring the respect of the contracts (fair and legal remuneration). Many companies have already started to apply this strategy like CoCaCola and Unilever [43].

*2) E-commerce for Agricultural Products:* In this category, blockchain technology is a means of securing financial transactions by means of solid key-based authentication mechanisms. It also constitutes a digital payment solution without rates. Furthermore, cryptocurrency based transactions will reduce costs more substantially, increase transparency and gain easier access to global markets.

As examples, we can cite the Spanish platform OlivaCoin which is a platform for the trade of olive oil [20]. Similarly, the platform [10] ensures the transfer of financial transactions and virtual currency between consumers and farmers. FTSCON [66] is a food trade system with a consortium Ethereum blockchain. It includes 300 agri-food enterprises in Shandong (China). Foodcoin [83] is a new blockchain ecosystem for creating a food and agricultural products global marketplace. It is based on the use of Ethereum smart contracts. The food deals are triggered by smart contracts, while transactions employ a proper cryptocurrency named FoodCoin.

*3) Resources Management and Environment Awareness:* The proper management and use of agricultural fields, water resources and soils is very crucial. Blockchain is useful in this regard as it allows tracing this information for the public. For instance, the automatic rationale is that watering is necessary for areas that are known for their severe hot weather. Additionally, common people do not have much knowledge on how to grow plants in an appropriate way [187]. Following this direction, [157, 71] focus on the use of the blockchain in smart irrigation systems. In particular, [157], a blockchain model is proposed for monitoring water distribution for a set of smart farms. Mainly, the blockchain technology is used to store and share, in efficient ways, data like water quality, energy used, weather conditions, etc. [71] focuses on smart irrigation system using photovoltaic energy generation systems. Blockchain technology and smart contracts are used to manage the energy trading using the SolarCoin cryptocurrency. The paper presents a prototype to showcase the proposed architecture. [44] addresses water scarcity and water distribution transparency issues in rural areas. For this, the authors propose blockchain technologies to provide trust among community members and to coordinate the use of water. Smart contracts are exploited to enable irrigation execution. This paper provides a first implementation of the proposed architecture based on Ethereum and some sensing devices (soil and water monitoring, etc). [187] proposes a Smart Watering System (SWS) for smart water consumption in small and medium-scale gardens and fields. SWS is based on: i) sensors to provide real-time measures about plants and environmental conditions like soil moisture and temperature, ii) fuzzy logic to take decisions about watering schedule based on sensors information, iii) blockchain to provide security by allowing only the trusted IoT devices to access and manage the proposed SWS. Also, a mobile application is developed to enable humans to monitor the watering process.

Blockchain has also been used as a backbone infrastructure for waste management. In [124], Plastic Bank employs a blockchain based token to enable the collection and recycling of plastic waste to clean up agricultural lands. [286] proposes a blockchain based-model for collecting agricultural waste. This waste is then converted to clean energy and agricultural by-products like fertilizer, and animal feedstock. The waste trade is based on a digital coupon or cryptocurrency. As examples of commercial solutions for waste management, we can cite: Recereum [220] and Swachhcoin [240].

*4) Land Registration and Agricultural Insurance:*
Blockchain could also help for securing farmers against weather conditions or natural disasters that impact their crops. Blockchain is then used in the insurance programs especially via employing the smart contracts which are able to automate insurance policies and thus reduce administrative management [16]. Indeed, in insurance companies, processing and verifying claims is known to be a long process. This does not encourage farmers to establish insurance contracts. Meanwhile, the removal of risk through insurance can increase smallholder investment and income. The project ARBOL [29] is an example that allows farmers to receive payments in case of droughts, floods, etc.

Blockchain technology has also been used as a means of land registration and ownership management. Current classical systems do not track the whole chain. Information is fragmented in multiple offices of the government. These offices are not sufficiently synchronized and corrupted people can modify legal documents. About (70–80%) of worldwide land transactions are not formally registered in any national platform [247]. Also, in many countries, hundreds of thousands of civil cases take place every year concerning land ownership. That is why blockchain technology has been adopted to create a trust atmosphere where property information is shared in a transparent and secure way. Many countries start using blockchain in land registration like Sweden [51], Bangladesh [15] (a prototype system using Ethereum and smart contracts) and Ghana [210].

### E. Future Directions

From our study, we can conclude that there are many efforts on applying the blockchain in the agriculture field (e.g various applications, commercial solutions). However, we believe that there are still many things to do.

- Infrastructure issues: The use of the blockchain technology in smart agriculture requires that rural areas be equipped with networking infrastructure like for instance 4G/5G stations. This is also required for applications that integrate IoT devices or for agricultural e-commerce applications in under development areas.
- Technical challenges: Scalability is a real issue in blockchain networks. The size of the blockchain grows in proportion to the number of nodes and thus the number of transactions. For an agriculture application, transactions could be generated continuously, the ledger size tends to grow rapidly. This limits the performance of the platform in terms of transaction download time and mining memory space as confirmed by [247]. Consensus algorithms complexity is also a limiting feature for real implementations as it requires many energy resources and processing time. Applying them for low capacity peers like IoT devices is very challenging. For instance, the Proof-of-Work (PoW) consensus approach does not fit the IoT environment as it requires both computing power and electric energy to a great extent.
- Incentives and regulations: We believe that further research is required on the incentives that could

be provided to the involved entities to integrate a blockchain ecosystem. Working on policies, governance and regulation rules are fundamental to attract more stakeholders and even small farmers [120]. Also, a concrete involvement of stakeholders is required to promote commercial solutions.

## VI. Blockchain for Supply Chain Management

### A. What Does Supply Chain Management (SCM) Mean?

There are various definitions of the Supply Chain Management (SCM) in the literature [239, 213]. Some common keywords found in the literature are: *planning, activities, logistics, manufacturing operations, network, collaboration, coordination, integration, suppliers, intermediaries, raw material, costumer, etc.* The reader can refer, for instance, to the definition of the Council of Supply Chain Management Professionals (CSCMP) [239].

For us, based on the above cited references, we can synthesize the following definition:

*A supply chain is a network of individuals, organizations, businesses, resources as well as technologies that combine together and perform activities related to the manufacturing of a product or service. The management of the flow of goods, services, and information involving the storage and movement of raw materials, building products as well as finished goods from one point to another is referred to as supply chain management. It implies processes coordination and activities that involve services such as marketing, sales, design, finance and information technology.*

### B. Existing Surveys about Blockchain for Supply Chain Management

This section sheds some light on existing surveys dealing with SCM. Readers can refer to these surveys, in Table XXVIII, for more detailed information on the topic.

### C. Benefits of Blockchain in SCM

Let us recall the most distinctive features of blockchain that make it highly applicable for the SCM security:

1) **Traceability**: it is a major requirement to secure supply chains, especially for sensitive products like medicines and drugs. Traditional tracking methods range from wireless sensors, Radio Frequency Identification Devices (RFID) or Near Field Communication (NFC), Barcode readers, secure tags, firm specific traceability mechanisms, etc (see for instance [246, 60]). However, barcodes can be damaged or unreadable because of environmental factors like moisture and temperature variations. Also, though RFID is small and compatible with many products, it may be very expensive for stakeholders [222]. Data provided by such means are either tracked manually, or digitally stored in databases prone to failures and security threats. Besides, tracing some information when needed may take a long time. Surely, tracking devices should continue to be used in the SCs to identify products, machines, etc., but the data must be secured. Thanks to its immutability and the authentication procedures required for ledger access, the blockchain technology provides a means to store and share data securely.

TABLE XXVII: Summary of **Blockchain for Smart Agriculture**

| Ref | Scope | Conceptual paper / Implementation or Proof of Concept (PoC) | Features |
|-----|-------|------------------------------------------------------------|----------|
| [157] | Monitoring water distribution | Conceptual | • Describes a model for Blockchain use to provide farmers with useful information to help them manage the water distribution. |
| [71] | Smart irrigation | Ethereum smart contracts used to trade photovoltaic energy in farms | • Use of SolarCoin, a digital coin based on Proof-of-Stake for the simulation of solar electricity production |
| [44] | Automatic management of irrigation | A prototype with Ethereum smart contracts and sensing devices | • Main motivation: provide trust among rural communities<br>• Measures the response time |
| [187] | Smart watering system for small farms and gardens | A prototype with sensor devices, development of the fuzzy model, and mobile application for remote monitoring. | • When a user wants to launch irrigation, a transaction is generated and validated by the blockchain. |
| [15] | Land title management in Bangladesh | Prototype with Ethereum and smart contracts | • The blockchain proposed includes many actors like government, court, financial organizations, etc |
| [164] | Blockchain for collaborative research | Conceptual | • Proposes a generic architecture for sharing research data<br>• Use of smart contracts for research protocols definition (data collection, automatic processing and analysis, etc) |
| [26] | Employers contracts management | Conceptual | • The aim is to process the contract efficiently, and to protect the right of the parties involved in the contract, especially temporary workers. |
| [66] | Food trace system | Consortium Ethereum based blockchain | • Proposes an improved consensus method based on PBFT (Practical Byzantine Fault Tolerance)<br>• Studies a use case including 300 agri-food enterprises in Shandong (China) |
| [286] | Agricultural waste collecting and trading | Case study in China | • Proposes different forms of incentives for farmers and entrepreneurs to make the model attractive.<br>• The model is integrated with IoT (E.g. QR codes, smart sensors). |
| [16] | Insurance farms | Conceptual | Discusses also the use of blockchain in smart energy grids |

2) **SCM time and cost reduction**: monitoring supply chains is complex and costly as it involves many entities and hugely depends on the type of industry itself. When a distributed system is used, costly transactions are avoided by eliminating the need for intermediaries. Also, a blockchain can make faster and simpler tracking items and transactions in the supply chain by an estimated 85% when used in conjunction with IoT technology, cutting administrative and logistic timelines in shipping [23].

3) **Trust enhancement**: knowing that the provided information is certified, immutable and not owned by a single entity but totally distributed, cooperating stakeholders will enjoy an atmosphere of trust. From customers, information like product origin, production, and modifications, brings needed assurance about the products they are consuming.

4) **Frauds prevention**: having detailed information about products prevents forged products from the entry in the market and provides good quality indicators (e.g in case of food, for instance, sharing important data about the origin, whether the food is organic or non-GMO (genetically modified organism), crop yields, etc).

5) **More efficient decision making**: blockchain technology helps stakeholders to make business decisions or predict market trends.

6) **IoT-based applications compatibility**: the blockchain has proved that it fits well with the IoT technology: many blockchain-based solutions are integrated with IoT devices (devices to identify products, smart devices used by consumers to investigate some products, etc).

7) **Payment securing**: the blockchain technology adoption can also guarantee payment security to sellers.

*D. SCM Requirements*

From the definition presented in Section VI-A, we identify the following requirements for the SCM in the context of industry 4.0.

1) **Auditability/Traceability**: is at the core of supply chains as it enables the tracking of the whole lifecycle of products. Furthermore, collecting **real time** information is crucial as it enables fast decisions when needed

TABLE XXVIII: **Blockchain for SCM** Existing Surveys

| Ref | Scope | Description | (+)Pros/(-)Cons |
|---|---|---|---|
| [63] | Agriculture SC traceability | • Definitions of relative keywords<br>• An overview of some solutions classified according to the type of the blockchain<br>• Overview of traditional traceability methods<br>• Detailed literature review of solutions classified according to agriculture sector and products<br>• Overview of commercial solutions<br>• Discussion on future blockchain challenges | + Focus on both academic and industrial efforts in the field. |
| [23] | Reviews the existing literature on blockchain | • An overview of blockchain and smart contracts<br>• States references on the potential of blockchain in the SCM<br>• Explains the paper methodology | - only one reference database is considered.<br>+ Highlights the growing number of publications relative to the subject and classifies them based on the SC monitoring level |
| [207] | Review of the works from both business and technical perspective | • Classification of references depending on the type of the supply chain and the approach (conceptual/implementation)<br>• Discusses adoption barriers, challenges and benefits for manufacturers.<br>• Reviews the implementations and use cases | + Dealing with real development is a strong point. |
| [233] | Systematic review | • Classification of references according to their scope (distributed ledger, security, business model, etc)<br>• Classification of papers according to industrial application sector (agriculture, food, drugs, etc)<br>• Discussion on the current state of the adoption of blockchain in industry, use of smart contracts, etc | - As the analysis is systematic, there is no detailed description of the references<br>+ Good overview of the perspective of papers when dealing with blockchain |
| [41] | Bibliometric review | • Explains the review methodology (database collection, citation analysis, etc)<br>• A bibliometric literature review from 2016 to January 2020<br>• It adopts a citation network analysis and a cocitation analysis using dedicated tools (Gephi, PageRank).<br>• Classifies the existing literature into five research clusters (theoretical sensemaking, conceptualizing and testing blockchain applications, digital SCM, the design of BCT applications targeting real-world scenarios, framing BCT in supply chains | + Good entry point for academics and practitioners into the topic<br>+ Classification of research interests into clusters |
| [101] | Overview of the use of blockchain in trade SC. | • Highlights the complexity of the international trading process while taking importation to Dubai as an example<br>• Explains the research methodology (planning, execution and reporting)<br>• Classification and description of the papers: electronic trading, validation, SCM | + Papers are relatively well discussed |
| [260] | Explore blockchain capability to transform SC, a study based on interviews. | • Highlights the benefits of Blockchain in SC<br>• Describes the potential in terms of applications<br>• Identifies the future challenges<br>• Describes the research methodology (individual interviews) | -Individual interviews could be enhanced with diverse range of participants<br>+Good explanation of blockchain in SCM |

(e.g removing fish destined to market whenever its temperature drops during the transportation).

2) **Availability**: means the permanent availability of traceability product data that can be accessed anywhere and at any time. Imagine a case of food contamination, the history of its supply chain is crucial for post-checks.

3) **Integrity**: supply chains data are more or less sensitive depending on the nature of the products, type of information, etc. But in any case, data security is essential. For instance, data integrity must be guaranteed to prevent and detect malicious modifications of data.

4) **Authentication**: every participant has to be authenticated in order to access the supply chain.

5) **Access control**: rules aim at managing access depending on users profiles and roles.

6) **Privacy**: there is a need to preserve the privacy of the different entities (e.g consumers don't like to disclose their food habits, companies tend to hide their information because of competitions, etc).

7) **Trust**: SCM involves many entities that need to trust each other. Essentially, the consumer needs to trust tracked information to be encouraged to buy products. Some hints to establish the trust are to guarantee transparency and to remove any single central authority.

8) **Guarantee of the quality**: of the products delivered to final consumers.

9) **Minimal cost and delays**: through the supply chain.

### E. Blockchain for SCM Use Cases

*1) SCM in Food Industry:* One of the most targeted applications of blockchain regarding the SCM is the food industry. This is due to the following features:

1) Food products have more sensitive and vulnerable supply chains compared to other industries [115]. Indeed, food has changing quality that could be highly impacted by outer conditions such as temperature and transport.

2) Food security is fundamental, and requires a high degree of accuracy and efficiency. Indeed, in case of supply chain problems, consequences are very dangerous [190]: products failure or counterfeit, food poisoning and disease, low quality food, or mislabelling and undeclared ingredients after production.

3) Consumers are more and more requiring awareness of what they eat (origin, ingredients, and so on). It is necessary for stakeholders to present food quality guarantees for their consumers to drive out competitors.

Blockchain is able to secure the food supply chains by transparently tacking food. People will have equal chances to access food data. Indeed, authorities and food companies tend to share only selective data with the public (e.g Halal food [168]). Information credibility might be questionable due to marketing purposes, or when companies want to escape from taking the responsibility in case of problems [176]. Besides, solutions integrating blockchain with IoT devices allow real time information updates which enables the involved entities to monitor the supply chain efficiently (act immediately in case shipping conditions are not respected). Finally, blockchain can reduce food waste. Indeed, by having accurate and immutable information about food consumption, stakeholders can forecast the market needs and manage the production efficiently.

As previously cited, blockchain and IoT fit well for food SCM applications. It is the case for instance in [108, 281, 176] which integrates IoT with blockchain in order to track the food chain. In particular, [108] focuses on monitoring the temperature of fish using IoT devices interacting with an Ethereum-based blockchain. In the proposed architecture, each IoT device has its own gateway device and can communicate with a close RPC server which allows it to interact with the blockchain via its Web3 Ethereum API. Like that, IoT devices can for instance access the information stored in the chain or launch smart contracts. The gateway has a network identification and signs the transaction, using its private-key, before sending it to the RPC server, in order to prevent transaction tampering. For the use case presented, an IoT device is installed into a refrigerator, connected to the Internet through a mobile network connection, and monitors the fridge temperature and truck position using a temperature sensor and a GPS module, respectively. Moreover, the simulation has been carried out using the Quadrans blockchain [61].

[281] proposes a blockchain-IoT-based food traceability system (BIFTS) which is based on three modules architecture: IoT monitoring module (devices, connectivity, application),

fuzzy food quality evaluation module (to evaluate the food quality based on collected data), and the blockchain management module. For consensus, the authors propose the PoSCS algorithm that is developed to select validators who are the stakeholders (instead of miners) in a probabilistic way. This algorithm considers transit time, stakeholder analysis, and shipment volume in the supply chains, instead of computational power and wealth. A case study was conducted in a retail e-commerce company. However, the architecture relies on a cloud server, which makes the system a bit centralized and prevents data from being authenticated inside the cloud.

Authors in [176] present AgriBlockIoT, a decentralized traceability system for the Agri-Food SCM. AgriBlockIoT is organized in layers and provides a REST API interface for authorized users. Unlike [108], this paper relies on full blockchain nodes which results in scalability issues. The data storage relies on the cloud. The authors present a practical farm-to-fork use case where the information provided ranges from raw material to consuming details. The development supports both Ethereum and Hyperledger Sawtooth. However, the use of IoT devices was not practically implemented; it is only simulated.

The paper [127] focuses on soybean SC and proposes a solution based on Ethereum smart contracts. The authors present a description of the smart contracts to perform business transactions and to restrict access to some users when needed. All transactions are stored in the blockchain ledger with links to a decentralized file system (IPFS).

Many papers discuss the potential of blockchain in the food industry using different methodologies. For instance, the references [291, 276] base their evaluation on an analytical model and by systematic review consolidated by companies' interviews respectively. Among its finding, the paper [291] confirms that when the traceability awareness of consumers is high, the supply chain should adopt the blockchain technology. Similarly, [276] witnesses the suitability of blockchain technology in the era of food supply chain in industry 4.0.

Both [222, 281] consider that the challenge for the food industry is to provide consumers with high quality food to have a superior business advantage. For this, they propose a blockchain based prototype for food quality indicator (FQI) based on fuzzy logic and a developed model respectively. In particular, [222] targets the pork restaurants and bases the FQI computation on three major quality parameters: Refrigeration/storage time, product shelf life and nutritional value. This computation takes into account standard storage and handling regulations specified by food safety authorities. The authors find that the storage conditions of pork meat are the major element for its quality indicator. In this paper, a particular package is traced by using unique identifiers like barcode, RFID transmitter or QR code. The paper assumes that such data is available and base their model validation on it.

[62] proposes a solution for the agricultural products' traceability in India. The solution is developed based on Ethereum smart contracts. Like in [127], all of the transactions

are recorded and stored in a centralized interplanetary file system database (IPFS). The proposed solution shows a throughput of 161 transactions per second with a convergence time of 4.82 seconds.

Although the blockchain technology is relatively young, there are serious industrial efforts in the food market to produce real applications [276]. We can cite for instance the project in [180] which focused on tuna fish tracking using QR codes or RFID tags. IBM is investigating in this sector with others partners. [111] is based on a big consortium (IBM, Walmart, Nestle and Unilever) where food security is the major challenge. With Maersk, IBM has also a project to use Blockchain in order to make digital trade workflows and end-to-end shipments [103, 194]. Carrefour also has a famous application launched in 2018 [4, 47] to monitor some products through the whole supply chain. Initially the solution was tested with chickens, then deployed with other products like milk, tomatoes and eggs. And, for this big company, the objective is to apply this technology to all food products by 2022.

*2) SCM in Other Industries:* Blockchain technology has also been applied to monitor the SC in other domains [101, 233] like drugs [1], textile [155], energy trading [151], wood [82], etc. Table XXX highlights some of recent solutions in these different domains.

Li et al. [151] proposed a consortium blockchain-based secure energy trading system for Industrial IoT (IIoT). In the proposed architecture, the electricity traded is computed by a smart meter. Energy nodes form the network (e.g smart building) while energy aggregators work as brokers to facilitate transactions they receive from the server. Furthermore, each node has a smart wallet to store its coins.

Because medicines are very sensitive products and are prone to counterfeiting, blockchain technology has been used to secure the drug or vaccine supply chain like in [1]. Section III-D4 from the Healthcare part of this paper focuses on the medical supply chain as it is a form of applying blockchain in the medical sector.

The textile domain has been a focus of many works. Indeed, the supply chain in the textile and clothing industries is complex and fragmented [155], and it involves numerous actors from very different sources and operations. Also, there is not a standard path for garments: e.g: clothes went through different processes each can be in a different country. Usually papers divide the textile related activities/information that need to be tracked into primary and secondary [139, 155]. The primary one is related to logistics (operations, sales, receiving, collecting, storing, disseminating products), while the secondary one includes human resource management, technological development, accounting, planning, etc. References [245, 155, 166] discuss the advantages of blockchain technology focusing mainly on the textile traceability. The chapter [245] consolidates its discussion with a case study dealing with cotton fiber. The paper [155] proposes a framework for ready-to-wear clothing SCM. Considering a women's shirt company use case, the authors describe different components of the blockchain solution: structure of blocks, how blocks are hashed together, how data is retrieved by participants (RFID, NFC, QR code, etc), etc. However, the paper does not provide a real development of the proposal, the authors only focus on the distributed ledger and how it could answer the challenges.

In [166], the authors propose a blockchain-based framework to detect defective batches. The objective is to mainly enhance transparency. Actors include farms, transporters, factories, retailers and consumers. The transactions identify the processes and machinery used and manufacturers record the purchase of a certain batch of material. Also, manufacturing machines are equipped with sensors that communicate the measurements to the blockchain and ensure remote control. Like that, the framework allows the machine to use the information shared in order to automatically stop processing the material and require human intervention. Authors focus on products quality, and show how shared information can enable real time detection of defective products. Also, using simulations, they measure the quality impact (as a function of blockchain information and all available information) and show that missing information leads to quality degradation. However, the paper only provides a very vague description, no practical details about blocks structure, nor cryptographic procedures are given.

Blockchain technology has also been proposed as a solution to secure the aircraft supply chain in [54, 248, 165]. In the IATA (The International Air Transport Association) study "Future of the Airline Industry 2035" [104], blockchain technology has been identified as a technology that may have a great impact on the future of aviation. Many recent projects have already adopted this technology in their aviation services. For instance, the Hainan Airlines (HNA) group implemented a blockchain-enabled E-commerce platform[278]. Other projects are also described in [217]. [54] uses the blockchain to build a "digital twin" for the metal additive manufacturing of aircrafts (manufacturing processes starting from a 3D model). A digital twin of an aircraft can be defined as a collection of digital data representing this physical aircraft. The advantages are: the traceability of the process, compliance with standards regulations, control of the authenticity and quality, as well as safety conditions of the components installed on every airplane. The paper explains conceptually, how blockchain can be applied during the different steps of the manufacturing process. This conceptual explanation was strengthened via a case study showing design information, block creation and hash verification.

[248] approaches the aircraft industry from another angle. The authors address the risk analysis of the global supply chain using the mean-variance method. They consider that the blockchain is able to facilitate the use of information for enhanced air logistics operations (e.g forecasting for demand and supply, enhanced revenue management) and better assessment of risk sensitivity by examining previous decisions and transactions.

In [165], the authors focus on the aviation industry and highlight the benefits of blockchain. They illustrate their thoughts by sketching the process of acquiring and assembling parts for air-crafts using a blockchain.

More recently, [217] highlights and discusses the opportunities of blockchain technology in different aviation applications, like digitizing crew certificates, securing customer loyalty programs, and maintenance, repair, and overhaul (MRO) operations. The authors proposed a generic Blockchain-based framework for the aviation industry based on smart contracts. Also, the paper presents five case studies using the blockchain in the aviation industry.

TABLE XXIX: Summary of **Blockchain for Supply Chain Management** Solutions

| Ref | Contribution/ Purpose | Framework | Consensus | Storage | Validation Tools | Merits (+)/Limits (-) |
|---|---|---|---|---|---|---|
| [108] | Sea-food temperature monitoring | Quadrans | - | Remote | Quadrans, PoC using Quadrans Testnet | +Integration with IoT device<br>- No details on how consensus could be done. |
| [281] | IoT-based blockchain for online commerce with integrated consensus | - | PoSCS | IBM cloud | Online e-commerce real use cases | - Centralized cloud storage |
| [176] | Traceability system for farm-to-fork use case | Ethereum or the Hyperledger Sawtooth | Default ones in Ethereum or the Hyperledger Sawtooth | Cloud | Farm-to-fork traceability use case | - IoT devices are only simulated<br>- Centralized cloud storage |
| [222] | Pork quality estimation based on Blockchain data | - | - | - | Food quality estimation based on developed model | - The model should take into account availability time of data and evaluate its impact on the estimation |
| [127] | Soybean business transactions tracking and traceability | Ethereum | - | In the blockchain and IPFS (decentralized file system) | Soybean traceability use case | - Crop images are stored in the IPFS, thus scalability issues may arise |
| [62] | Monitoring of agricultural SC in India | Ethereum | - | In the blockchain and IPFS (decentralized file system) | PoC (thirty clients, ten controllers, and thirty validators) | + A PoC applied to the Indian context |

Although there are few works focusing on the application of blockchain in engineering management, there are some works that discuss the potential of this technology to revolutionize the supply chain of this sector [119, 58, 249, 252]. This discussion is either conceptual like in [252], or via case more scientific approaches. For instance, [58] applies a case study and reveals a potential cost reduction from blockchain deployment at 8.3%. Also, [252] gives some keys to using a maturity model approach to evaluate the blockchain role.

### F. Future Directions

The state of the art in food supply chain management is abundant. It includes conceptual/theoretical papers as well as papers with development, Proof of Concepts (PoC) or prototypes. A large number of these papers end by discussing the future challenges of the blockchain technology. We believe that some challenges need to be addressed like:

- Performance enhancement: The current research effort on blockchain for the supply chain is not yet on the performance evaluation of the proposals. It is rather put on the potential business use cases. However, blockchain adoption in supply chain management implies technical challenges like scalability, consensus complexity, and the ledger size [247]. We need to observe and evaluate real performance under real conditions in big supply chains.

- Interoperability and integration with other systems: The supply chains, mainly the food supply chains, are complex and involve many stakeholders. Each one of these involved entities uses its own tools for traceability. Thus, blockchain platforms need to integrate into these existing systems such as warehousing management and manufacturing execution systems, and enterprise resource planning [267]. For instance, the food supply chain can be distributed through large geographical areas, even in different continents. This fact implies significant interoperability and deployment obstacles [63].

- Privacy: The reference [247] mentions many privacy attacks that the blockchain technology can suffer from, especially in the supply chain management field. The privacy issue is more challenging for sensitive information like big companies commercial data. Thus, to take advantage of the technology, special attention should be given to security and in particular privacy issues.

- Standardization and regulation: There is a need to deploy regulations, incentives and rules to attract more parties to

TABLE XXX: Summary of **Blockchain for Supply Chain Management** Solutions in Different Industries

| Ref | Field | Conceptual / Implementation (PoCs) | Blockchain Opportunities | Features |
|---|---|---|---|---|
| [155] | Textile | yes/framework (mainly ledger description) | •Traceability and transparency<br>•Clothing authenticity<br>•Reliability and integrity<br>•Validity of the retail final products | •Describe the ready-to-wear SC<br>•Presents a framework and a case study discussion |
| [166] | Textile | yes/framework and simulation | •Transparency<br>•Quality, and defective batches detection<br>•Traceability | •Describe a theoretical framework with the associated workflow<br>•Simulate the impact of using BC on textile quality<br>• |
| [245] | Textile | yes/ case study (cotton fibre) | •Traceability | •Highlight the limitations to use BC in the traceability system like cost and technology immaturity |
| [151] | Energy | yes/detailed framework | •Blockchain provides secure energy trading | •Describe a detailed framework based on BC for energy trading in IIoT (Industrial IoT) applications<br>•Propose a consensus mechanism<br>•Propose a credit-based payment method<br>•Propose a pricing strategy using Stackelberg game<br>•Validate the proposal with simulation (to measure the transaction time for instance) |
| [54] | Aircraft | yes/yes | •Traceability and security of data<br>•Compliance with aeronautical regulation<br>•Easier functions such as monitoring<br>•Maintenance and repair overhaul | •Digital twin building using blockchain<br>•Focus on the additive manufacturing process<br>•Conceptual design on how blockchain can provide security<br>•Simplified implementation to showcase block creation and securing |
| [248] | Aircraft | yes/no | •Transparency of shared information<br>•Blockchain technologies enable faithful information sharing and updating.<br>•Better management of demands<br>•Fosters trust by providing permanent and trustworthy information<br>•Automate contractual agreements and arrangements, as well as the partnerships with air logistics companies. | •Review the literature about the air logistics operations like demand and supply management<br>•Explain how the blockchain technology can be used to facilitate the realisation of mean-variance risk analysis for supply chain operations. |
| [165] | Aircraft | yes / Small use case description | •Easy tracking of faults in final products<br>•Easy monitoring and planning of the manufacturing progress<br>•Provide a transparent network of supply chain for aircraft parts and reduce the risk of selling aircraft parts on the black market. | •Highlight challenges of smart supply chain management<br>•Sketch the process of acquiring and assembling parts for air-crafts on a blockchain. |
| [217] | Aircraft | yes / Smart contracts-based architecture | •Securing customers loyalty programs<br>•Digitizing crew certification<br>•Real-time baggage and cargos tracking<br>•Securing maintenance, repair, and overhaul (MRO) operations<br>•Securing identity management and air ticketing<br>•Automating airport collaborative decision making | •Discusses the opportunities brought by blockchain technology in various aviation applications<br>•Presents a smart contract based framework for aviation services<br>•Presents real blockchain based projects<br>•Future challenges |
| [58] | Construction | yes / Case study | •Cost reduction | •Explores the implications, risks and applications of blockchain technology<br>•Identifies opportunities for future research on blockchain applications in construction. |
| [252] | Construction | yes / no | •Enhancement of collaboration, trust, transparency and regulation | •Discussion on the feasibility of blockchain applications within the construction industry. |

the supply chain management [120]. For instance, there is a fundamental need to govern existing platforms. Another issue that is limiting a more massive blockchain adoption is the lack of standardization. Standards are required for any technology to have a scalable adoption across the globe [172].

## VII. Blockchain for Smart Grid

Blockchain has been applied in many smart grid applications due to its benefits such as decentralization, trust, and improved security and privacy. Particularly, participants of a smart grid can directly carry on a business trade (e.g., energy trading) without the need of a third party or a centralized organization such as an energy supplier. The blockchain-based approaches for smart grid applications have been investigated and presented in several surveys. Some of the most recent surveys are presented as follows.

### A. Existing Blockchain Surveys for Smart grid

In [185], the authors present a comprehensive survey of blockchain for a smart grid. In the survey, the authors discuss the security challenges of smart grid scenarios which can be overcome by applying blockchain. The authors show and analyze different related projects that apply blockchain for smart grid applications. Finally, the authors discuss research questions, challenges, and future directions of utilizing blockchain for solving security issues in smart grid applications.

In [189], the authors present a survey of blockchain-based smart grids. The authors discuss the advantages, challenges, and approaches of applying blockchain to the smart grid. In addition, the authors unveil how blockchain can be utilized as the cyber-physical layer of the smart grid. Last but not least, the authors provide industrial blockchain-based smart grid applications and present the directions for future works.

In [35], the authors perform a survey of blockchain-based approaches for energy sectors such as decentralized storage and control in a power grid, peer-to-peer trading in smart grid, and electrical vehicles. The authors discuss the requirements of distributed energy systems (e.g., in a smart grid), which are decentralization, anonymity, transparency, democracy, and security. Finally, the potential applications of blockchain for the energy sector and smart grids are also discussed.

In [294], the authors perform a survey of blockchain for cybersecurity in a smart grid. The survey covers architecture and implementation techniques related to blockchain applications that help deal with cybersecurity in the smart grid. Finally, the survey discusses the potential future research directions.

In [257], the authors present a survey of electrical power systems utilizing blockchain for energy trading. The survey classified the existing blockchain-based approaches for energy trading into three categories based on their primary focus including system optimization, consensus mechanism, and energy transaction. The survey also presents the challenges and unsolved problems in the existing approaches, and future directions.

In [89], the authors provide a survey of blockchain applications in microgrids. The authors focus on the projects and different start-up approaches for microgrids. The authors compare these approaches in terms of blockchain type, consensus mechanism, hardware development, and open source. In addition, technical background and concepts related to the blockchain, and future directions are also discussed.

In [25], the authors present a systematic review of blockchain technology in the energy sector. The authors provide an overview of the main principles of blockchain technologies including system architecture and consensus algorithms. The authors selected 140 blockchain-based approaches and projects. These works had been classified into different categories based on the field of activity, implementation platform, and consensus strategy. The authors present the opportunities, potential challenges, and limitations of the existing blockchain-based systems ranging from IoT, peer-to-peer energy trading, electric vehicles to e-mobility. Finally, the authors provide the future directions for blockchain-based approaches in the energy sector.

In [95], the authors present a survey of different blockchain-based approaches and platforms for smart energy domains. The authors discuss the advantages and disadvantages of blockchain technology solutions based on different categories including blockchain network management, data management, consensus management, identity management, and automation management.

In [11], the authors perform a review of the blockchain-based approaches for a microgrid. The authors develop an analytical framework based on both expert interviews and literature review. The framework focuses on aspects related to the technologies, economy, social issues, environment, and institutional dimensions. The authors also discuss directions for future work in terms of academia and industry.

In [2], the authors present a review of peer-to-peer energy trading approaches The authors cover several topics such as demand response optimization models, distributed energy trading architecture, power routing, plug-in electric vehicles, and security. The authors also investigate different enabling technologies including energy Internet, blockchain, digital grids, and software-defined networks that can be exploited to provide an efficient and secure peer-to-peer energy trading system.

In these surveys, technical issues of blockchain and blockchain applications for the smart grid are mainly focused whilst some of the requirements of smart grids such as availability and autonomy are not considered. Although the blockchain has many advantages that help improve the quality of smart grid services, blockchain cannot overcome all the existing challenges and problems. Therefore, this survey analyzes the security requirements and other important requirements of the smart grid. In addition, this section shows the mapping between the blockchain technology/applications and the requirements of the smart grid. Furthermore, the section aims to show the existing challenges even after applying blockchain technologies. This helps to develop a more advanced smart grid that utilizes blockchain, other security techniques, and state-of-the-art technologies to improve the quality of service.

### B. Security requirements of Smartgrid applications using blockchain

Based on the previously cited surveys and requirements of smart grid in the context of industry 4.0, we have identified the requirements for a next generation smart grid application. These requirements are presented as follows:

- R1: **Confidentiality**: Information such as business transactions including money transfer needs to be maintained confidentially. Only partners who are involved in the business transactions are eligible to access the information. To deal with this, symmetric or asymmetric cryptographic algorithms such as Advanced Encryption Standard (AES) or Elliptic-curve cryptography (ECC) can be applied before inserting the data into the blockchain as transactions.

- R2: **Integrity**: All of the records in the smart grid must be original and not manipulated by any partner. Some of the widely used methods for ensuring the integrity of the records in a smart grid are applying time-stamps and hash-function.

- R3: **Authentication**: is required in all the smart grid applications. Authentication is the process of verifying the credentials of a user or a participant. Some of the authentication factors that are often used to verify the user's credentials are knowledge, possession, inherence, location, and time factors [224]. The knowledge factors are "something you know" such as username, password, and PIN. The possession factors are "something you have" such as a one-time-use token, the one-time password generated by a mobile app, or a smartcard. Inherence factors are "something you are" such as face recognition, fingerprints or other biometric identification forms.

- R4: **Auditability**: Public records and information related to a smart grid are in the public domain which can be publicly available and audited by any user or participant.

- R5: **Authorization and Access Control**: Only authorized users can have permission to access specific resources. Energy seller has full control of the energy generated by their renewable systems.

- R6: **Privacy:** Customer or smart grid participant's privacy must be protected. Information of a customer or a smart grid participant cannot be revealed without the concern or acceptance of the person.

- R7: **Trust:** In the conventional systems, trust has been built between energy sellers/buyers and the centralized system. When selling energy to a centralized energy company, energy sellers trust that the correct amount of money will be transferred to their account. In the smart grid, trust is not placed in the centralized system but is distributed among the smart grid's participants.

- R8: **Transparency**: In a smart grid, transactions, records, events, and logs need to be transparent. Participants should be able to track and check public information and all the information related to the works or sections in which they are involved.

- R9: **Availability**: Connections between the smart grid's participants must be ensured. Participants such as individual energy sellers should be able to connect to both energy companies and also other individual users such as specific energy buyers.

- R10: **Automaticity**: Business transactions between an energy seller and an energy buyer should be done automatically. For example, when an individual house generates a particular amount of energy that exceeds the energy usage, the surplus energy will be automatically transferred to an energy company and the selling money will be sent to the seller's account.

### C. Security solutions of Smartgrid applications using blockchain

The blockchain technology has found its way to many smart grid applications. In most of these applications, energy storage, energy distribution, and peer-to-peer energy trading are focused. These factors are discussed as follows:

- *Energy trading:* Currently, centralized energy systems have many disadvantages such as low level of fault tolerance, lack of transparency, and energy inefficiency caused by energy loss in long-distance transmission and distribution [144]. Correspondingly, it can increase the energy cost. In addition, these centralized energy systems may act as a single point of failure [191]. For instance, when the system has a problem in terms of security, all consumer data can be leaked. Some researchers proposed blockchain-based energy trading approaches for overcoming some of the existing problems of the centralized energy systems [287, 204, 140]. In the proposed decentralized approaches, renewable energy can be generated from wind or solar panels and is locally stored in each participant's house. When a participant has more energy than needed, he/she can sell it to other consumers. Each consumer can proceed with peer-to-peer energy transactions without the need for a centralized management system such as an energy controller and supplier [191]. The approaches support distributed transaction ledgers which are tracked and managed by each consumer. Therefore, transparency and privacy are provided. In these approaches, a user's request of dynamic pricing and energy requirements is validated by a miner node [191]. In other words, the miner node will add a transaction with a smart contract into a blockchain. When the predefined conditions are met, the smart contract will be automatically executed.

- *Renewable Energy:* plays an essential role in energy production. By utilizing natural resources such as sunlight; wind; and ocean waves, renewable energy can be generated. Nowadays, renewable energy is generated not only by large centralized factories but also by local and individual houses. Particularly, a solar panel system can be equipped at home for collecting energy from sunlight. Depending on the location, an amount of the collected energy can exceed the need of each house. The surplus energy can be sold to a centralized energy company or other houses. This causes the renewable energy distribution networks to become more complex [277]. Correspondingly, the conventional centralized energy distribution and control may not be suitable for many cases. Therefore, a new type of energy distribution grid that can deal with the complex renewable energy distribution networks is required. The blockchain offers distributed control and management which makes it a suitable candidate for solving the existing problems of conventional systems.

TABLE XXXI: **Blockchain for SmartGrid** existing Surveys

| Ref | Scope | Description | (+)Pros/(-)Cons |
|-----|-------|-------------|-----------------|
| [89] | Microgrid | Overview of current blockchain technology based projects | - Solutions are not presented |
| [257] | Energy trading | Blockchain-based energy trading in the electrical power system | - Solutions are not presented |
| [185] | Systematic review | Comprehensive survey on the application of Blockchain in smart grid | + Significant security challenges of smart grid scenarios are discussed |
| [189] | General | Review of blockchain applications and new frameworks in smart grid | - Solutions are not presented |
| [25] | General | Systematic reviews of blockchain in the energy sector, including 140 blockchain research projects and startups | + Drawback of the existing blockchain-based approaches for the energy sector is presented |
| [95] | Technologies | Requirements of smart energy systems and customized blockchain technologies for various smart energy applications | + Advantages and disadvantages of blockchain technology solutions are presented |
| [11] | Microgrid | Holistic review of factors of blockchain-based P2P microgrids | + Multi-dimensional analytical framework is proposed |
| [2] | Energy trading | Review of peer to peer energy trading systems | + Different enabling technologies are investigated |
| [35] | Energy Sector | Survey of blockchain-based approaches for energy sectors such as decentralized storage and control in a power grid, peer-to-peer trading in smart grid, electrical vehicle | + Requirements of distributed energy systems are discussed |
| [294] | Smart grid | Survey of ideas, architectures, and techniques related to blockchain's application in the smart grid for cybersecurity | + Major issues and recent advances together with future directions are discussed |

- *Microgrid operations:* Microgrid is an essential component of the development of a smart grid. A microgrid can be defined as a small-scale power/energy system with distributed energy resources [227]. A microgrid can be in one of four conditions which are grid-connected, decoupling, island-mode, and recoupling [89]. The microgrid is in the grid-connect mode when it transfers or exchanges energy with the distribution grid. It is in the island mode when the energy is locally transferred between loads, decentralized energy resources, and energy storage systems. The conventional centralized system is not suitable for these modes. Therefore, Blockchain with distributed management and control has been widely applied for the microgrid. Some of the state-of-the-art approaches are shown as follows.

In [262], the authors presented an approach that utilizes the blockchain for dealing with security issues in peer-to-peer energy trading. The authors mentioned that although the blockchain helps in solving some of the existing problems of peer-to-peer energy trading, the blockchain trilemma still has some uncertainty such as scalability, maintaining decentralization with an off-chain transaction, and security aspects (such as the aspects that blockchain handling overwhelms the network and applying suitable protocols on the side-chain). Therefore, the proposed approach utilizes both the blockchain and second-layer solutions to enhance the quality of services of peer-to-peer energy trading. The second layer is another blockchain which is an intermediary between a decentralized app and the main chain. The second layer is lightweight and can produce faster responses with lower costs. Transactions are carried out on the second layer instead of the main chain. When all the transactions are successfully done on the side chain, the side chain and the main chain are synchronized. One of the reasons causing the side chain to have fast responses is that it has fewer nodes. The proposed approach has been modeled by using data captured from a trial case study. The result shows that the proposed approach helps improve the scalability without a negative impact on security.

In [72], the authors present a decentralized platform using blockchain for peer-to-peer energy trading. The authors mentioned that the proposed platform helps provide automation, a high level of security, and real-time settlements via the advantages of smart contract implementation. The platform function was simulated with several operations such as energy trading, market clearing, smart contract operations, and blockchain-based settlements.

TABLE XXXII: Summary of **Blockchain for Smart Grid** solutions

| Ref | Contribution/ Purpose | Blockchain Type | Framework | Consensus | Storage | Validation Tools | Limits |
|-----|----------------------|-----------------|-----------|-----------|---------|------------------|--------|
| [173] | Proof-of-concept of a simple Local Energy Market LEM scenario on a private blockchain | Private | Ethereum | PoW | N/A | Simulation | N/A |
| [201] | Optimization model to maximize the profit in peer-to-peer trading in smart grid. | N/A | N/A | N/A | N/A | Simulation | No details related to blockchain |
| [158] | Blockchain-based energy trading Model for EV Charging | N/A | N/A | N/A | N/A | Simulation | No details related to blockchain |
| [122] | Localized peer-to-peer electricity trading model to buy and sell electricity among plug-in hybrid electric vehicles | Consortium blockchain | N/A | PoW | local aggregators | Security analysis and simulation using real dataset | No implementation |
| [9] | EnergyChain, a decentralized energy trading system using blockchain to store and access the data generated by smart homes | N/A | N/A | PoW | Cloud | Security analysis and simulation | No implementation |
| [197] | P2P trading platform to buy and sell electric power | Private | Multichain | N/A | Local | Implementation | No contribution and only generic implementation |
| [161] | decentralised architecture based on smart contract for energy transaction | N/A | N/A | N/A | Local | Simulation | No implementation |
| [100] | efficient method based on smart contract for energy trading | N/A | N/A | N/A | N/A | Simulation | No implementation |
| [279] | strategy based on smart contract for reliable energy trading | N/A | N/A | N/A | N/A | Simulation | No implementation |
| [208] | an approach based on distributed ledger and smart contract for energy demand and production | N/A | Ethereum | N/A | N/A | Implementation | N/A |
| [196] | optimization method for energy management | private | Ethereum | N/A | N/A | Implementation | N/A |
| [259] | a hierarchical framework based on blockchain and smart contract for energy demand management | N/A | N/A | N/A | N/A | Implementation | N/A |
| [117] | an application of blockchain for peer-to-peer energy trading and sharing in smart grid | N/A | N/A | N/A | N/A | Implementation in IBM platform | N/A |
| [149] | a hybrid system for management of energy demand and renewable operation | N/A | N/A | N/A | N/A | Simulation | No implementation |

In [121], a blockchain-based energy trading platform for smart homes in a microgrid is introduced. The microgrid platform based on a private blockchain network was implemented by using Ethereum and smart contracts. Each node of the network is a smart home and can be categorized into consumer and producer. Each smart home has an energy storage system that can store harvested energy and energy received from other smart homes. The platform is secured, automated, and decentralized. The authors implemented two nodes of consumer and producer by using a client distributed by Ethereum while a smart contract was implemented by using solidity language.

In [201], the authors presented an auction-based mechanism for peer-to-peer energy trading in a smart grid. Particularly, the authors propose an optimized model together with a sealed-bid auction that aims to achieve the maximized profit for peer-to-peer energy trading. Several aspects such as transmission costs, buyers bid offering, and operating costs had been considered in the model. Two scenarios including peak time and non-peak time trading were used to evaluate the proposed model. The results show that the proposed model helps sellers achieve maximized profit while satisfying buyers.

In [161], the authors propose an architecture based on the blockchain and smart contracts for reliable and cost-effective energy transactions. The presented architecture is secure, reliable, and distributed. Transactions of energy are safely stored in the blockchain to avoid replication or tampering by other parties. The proposed architecture includes a feature to mitigate cyberattacks that target vulnerabilities of smart meters. Although the proposed architecture has many benefits, the authors do not provide the necessary information to build and implement it. In [100], the authors present a method for efficient energy trading via the utilization of smart contracts. In the method, encourage-real-quotation rules are applied to determine the trading parties and energy price. The method guarantees that the trading information is secured, transparent and fair. The proposed method is useful for small-scale transactions in the energy trading market.

In [279], the authors present a strategy based on smart contracts for reliable energy trading. The authors aim to reduce the energy cost by shifting the demand of users. Particularly, end users can buy energy from other residential users via a blockchain network. The authors present and deploy a demand response model to prove the validity of the blockchain-based energy trading strategy. The model is simulated and the results show that by applying the proposed strategy, the energy cost is reduced by around 30%.

In [208], the authors investigate the decentralized blockchain mechanisms that provide a high level of transparency, security, reliability, and flexibility. The authors propose an approach based on a distributed ledger that stores information of energy consumption and trading in a secure manner. The approach applies smart contracts to define the energy flexibility, rewards, penalties, and the rules for balancing energy demands. The proposed approach is validated via a prototype that is implemented within the Ethereum platform. The results show that the proposed approach is suitable for energy demand and production and helps reduce the energy flexibility required for convergence.

In [196], the authors present a blockchain-based optimization method for energy management. The Alternating Direction Method of Multipliers (ADMM) is used for building distributed optimization. However, ADMM requires building an aggregator that collects computation results in each agent. By using the blockchain, an aggregator can be avoided as the distributed ledger can offer the function of an aggregator. The method is tested via numerical experiments to clarify the effectiveness and limitations in terms of tamper tolerance and computation time.

In [117], the authors present an application of blockchain for peer-to-peer energy trading and sharing in the smart grid. The presented blockchain applies smart contracts for automatic trading when requirements are fulfilled. A use case of peer-to-peer energy trading is implemented on the IBM platform using Hyperledger Composer. In addition, the authors discuss the challenges in utilizing the blockchain for energy trading.

In [149], the authors present a distributed hybrid energy system based on the blockchain and smart contracts for the management of multi-sectorial energy demand and renewable generation. The system consists of a hierarchical framework for managing energy demand via the information of peer-to-peer energy trading and exchange. In the framework, receding horizon optimization is applied to handle the uncertainty of renewable resources. A use case of energy microgrid in Singapore is used to test the proposed system and framework. The simulated results show that the presented system is secure, transparent, and efficient.

In [259], the authors present a study of the design and management of distributed energy systems. The authors first propose a hierarchical framework for energy demand management that utilizes the information related to peer-to-peer energy exchange. The framework is based on the blockchain and smart contracts to ensure a seamless and efficient energy trading system. The results show that the proposed approach helps reduce peak load, increases the energy market's efficiency and saves the economy.

*D. Mapping of existing smart grid solutions to Security requirements and Future Directions*

Table XXXIII shows the mapping of the previously reviewed solutions to the specified smart grid requirements described in subsection VII-B. It is noted that the complete end-to-end application is analyzed. When a part of the application does not fulfill a specific requirement such as privacy or confidentiality, the whole application is considered as a non-support of the requirement. In addition, when a requirement is not mentioned in the discussed application, the application is also considered as a non-support of the requirement.

Currently, most public blockchain applications lack confidentiality, data privacy, and data availability. To deal with these problems, symmetric and asymmetric encryption should be used whenever necessary. For example, energy data should be encrypted before inserting into the blockchain. A secret key for encryption and decryption can be distributed off-chain during an initial key exchange. The secret key or information needed to generate the key must not be shared on the blockchain. Only participants who have the secret key can access the information retrieved from the transactions. This solution protects confidentiality but it has some drawbacks. For example, when using encryption, a key needs to be shared off-chain. If the key management is not properly maintained, a private or secret key is disclosed and the encryption mechanism completely fails. In addition, smart contracts cannot interpret data without a secret key. This limits the utility of smart contracts. Therefore, it is required that an advanced approach for key sharing and management should be developed to maintain a high level of security. It is noted that the latency and energy consumption of each encryption mechanism (e.g., AES256) are really small when compared with the latency and energy consumption of data transmission. Therefore, a decision of using an encrypted mechanism will not mainly depend on these parameters. The decision relies on the particular application in which some of the aspects can have higher priorities.

The future smart grid needs to support the following aspects (1) allowing user categorization in which only authorized users can access the information, (2) ensuring integrity, (3) having fault-tolerant approaches that ensure data availability and overcome network problems, (3) supporting flexible scalability, (4) supporting automaticity e.g., by using smart contracts and other similar technologies, (5) ensuring transparency and democracy among participants, (6) supporting a hybrid approach that allows using encryption when needed, (7) support advanced authentication approaches to verify the users, (8) promoting trust between participants, (9) supporting data privacy, (10) having more efficient monitoring solutions, and (11) promoting flexibility of feature support. However, it is difficult to support all the mentioned features. Therefore, it is recommended that some requirements should be categorized with a higher priority than others. The solutions focus on the requirements with the highest priorities first. Then, requirements with lower priority can be implemented as long as they do not conflict with the higher priority requirements.

## VIII. BLOCKCHAIN FOR SMART HOMES

*A. Smart Homes*

Various research studies have attempted to outline a definition of what a smart home is. Indeed, a smart home is an ensemble of IoT devices interconnected via a network [200] to manage home appliances while improving security and energy consumption. IoT refers to the various connected devices and physical objects including sensors, microcontrollers and software components that allow their connections [74]. Daily life is becoming basically dependent on these connected devices which can be controlled remotely. The interconnection between devices and real objects makes life easier and helps in saving time and costs greatly. Moreover, those IoT devices might serve different purposes and uses at the home, such as: Smart door lock system, Security home system, Smart air conditioning, Lighting control system, Energy management system, Appliance control system, Smart thermostat, Smart bathroom appliance, Garage door opener, Patient monitoring. The smart home is composed of different modular functionalities that can be removed, or added upon new options. It offers the advantage of creating different scenarios depending on the equipment. All those applications tend to make life at home comfortable and convenient. Thus, many research studies are proposed for improving these applications while preserving the security and privacy aspects. Thus, designing a smart home requires usually the following components[159]:

- **Sensors**: the sensor nodes are used to monitor the state of the parameters subject to change such as: temperature, pressure, ..etc. These environmental or physical parameters are meant to be perceived, collected and processed with the network coverage by the coordinates of each node.
- **Sensors network**: consists of multiple wired and wireless networks. Its configuration depends on the nodes. It involves sensors, embedded computing, and wireless communication.
- **Wireless communication technology**: many technologies can be used in this context. Each wireless communication has some specific features related to the parameters of this technology. Among these parameters are network scalability, battery life, transmission rate, band, terminal equipment costs, integration and reliability, and cost. Examples of those technologies are Zigbee, WIFI, and mobile communication.
- **Power line communication**: the power lines used for data and information transmission.

Furthermore, the modern production of home appliances equips them with devices that enable their monitoring and their control, which can be done remotely through dedicated control systems. As a result, these devices become smart in a sense of having the self-management capability to make decisions on how to deal with different variations of the parameters (for example, based on the temperature degree, setting the air conditioner at a certain temperature and certain mode).

TABLE XXXIII: Mapping of **Blockchain for Smartgrid** solutions to security requirements

| Ref | RH1 (Conf) | RH2 (Integ) | RH3 (Authen) | RH4 (Aud) | RH5 (Author) | RH6 (Priv) | RH7 (Trust) | RH8 (Transpa) | RH9 (Avai) | RH10 (Auto) |
|---|---|---|---|---|---|---|---|---|---|---|
| [173] | ✓ | ✓ | ✓ | ✗ | ✓ | ✗ | ✓ | ✓ | ✗ | ✗ |
| [201] | ✗ | ✗ | ✗ | ✗ | ✗ | ✗ | ✗ | ✗ | ✗ | ✗ |
| [158] | ✗ | ✓ | ✓ | ✗ | ✓ | ✗ | ✓ | ✓ | ✗ | ✗ |
| [122] | ✗ | ✓ | ✓ | ✗ | ✓ | ✗ | ✓ | ✓ | ✗ | ✗ |
| [9] | ✗ | ✓ | ✓ | ✗ | ✓ | ✗ | ✓ | ✓ | ✗ | ✗ |
| [197] | ✓ | ✓ | ✓ | ✗ | ✓ | ✗ | ✓ | ✓ | ✗ | ✗ |
| [161] | ✗ | ✓ | ✓ | ✗ | ✓ | ✗ | ✓ | ✓ | ✗ | ✓ |
| [100] | ✗ | ✓ | ✓ | ✗ | ✓ | ✗ | ✓ | ✓ | ✗ | ✓ |
| [279] | ✗ | ✓ | ✓ | ✗ | ✓ | ✗ | ✓ | ✓ | ✗ | ✓ |
| [208] | ✗ | ✓ | ✓ | ✗ | ✓ | ✗ | ✓ | ✓ | ✗ | ✗ |
| [196] | ✗ | ✓ | ✓ | ✗ | ✓ | ✗ | ✓ | ✓ | ✗ | ✗ |
| [259] | ✗ | ✓ | ✓ | ✗ | ✓ | ✗ | ✓ | ✓ | ✗ | ✗ |
| [117] | ✗ | ✓ | ✓ | ✗ | ✓ | ✗ | ✓ | ✓ | ✓ | ✗ |
| [149] | ✗ | ✓ | ✓ | ✗ | ✓ | ✗ | ✓ | ✓ | ✗ | ✓ |

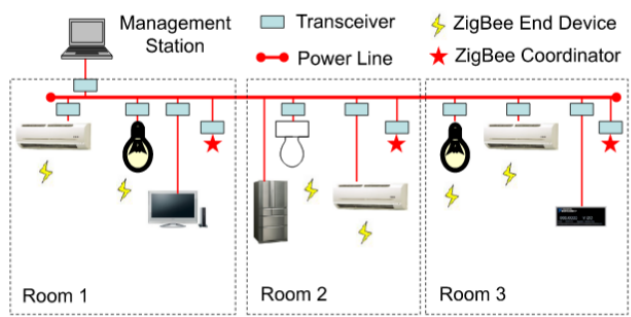

Fig. 6: Smart home's main components

This aspect of making decisions allows in many cases to considerably reduces cost and save energy [99].

In addition, smart homes devices generate an immense amount of data. This data is related to the home owner's sensitive information that must be well protected to prevent cyber threats. Indeed, these threats in a digital environment such as the smart home have been studied in many research studies, such as denial of service, ransomware, identity theft, information leakage, as pointed out and classified in [244], [126], [18], [28], [70], [179]. These threats lead us to consider the security of smart homes and how it should be ensured as pointed out in [232], [153]. Thus, many approaches have been proposed to ensure the security in smart homes, and among these approaches is the blockchain.

### B. Smart Homes and Blockchain

Since the creation of the Bitcoin virtual currency, the blockchain (BC) technology has gained wide attention. This technology has gained much interest and development because of the ambitious security features that it offers. Many surveys in the literature have studied the advantages of using BC technology in smart city applications in general, and in smart homes more specifically. The integration of BC in a smart environment such as smart homes requires certain adaptations [137], which leads to the necessity of highlighting some rules that will be used as detailed in the list below:

- **Transactions**: are the communication of the different interconnected devices in the smart home. This communication refers to various types such as storing, monitoring, and accessing.
- **Local blockchain**: the local BC is used mainly to track the transactions explained in the previous point. Generally, the local BC is private, which allows the owner of the home to set control policies. This latter through a policy header manages the incoming and outgoing transactions. The immutability of BC is preserved by hashing the previous block in the block header, which results in having two headers in each block: policy header and block header. Finally, linking the devices transactions' is made through the use of an immutable ledger in BC.
- **Home miner**: a miner in the smart home context is the device responsible for allowing and blocking incoming and outgoing transactions to the home and from the home. Thus, it authenticates, authorises and audits transactions.

It can be used as a stand-alone device or it can integrate with the home's internet gateway. The miner is the entity responsible for handling all communications related to the home.

- **Local storage**: the storage is necessary to complete the miner work or to store data locally using a backup device called local storage.

In the case of a smart home using BC technology, the scenario goes as follows: first, the initiation starts by adding devices. In this case, the owner should update the latest blocks policy header because the main objective of the policy header is authorising devices. This authorisation is possible through the miner which uses the policy header in the latest block in the BC. The transactions made between different devices might be realised by using direct communication between devices, or by using external entities. Direct communication is ensured using a shared key that needs to be allocated by the miner.

Indeed, major issues have to be considered when considering Blockchain technology to secure the smart home environment [45]: the IoT devices connected to the smart home are subject to a connection and disconnection, thus their numbers might change. The necessity to tolerate and mitigate cyberattacks on the smart home network. The IoT devices are limited in terms of computing power and energy consumption. Hence, there is a need for high performance while preserving the privacy and security.

### C. Blockchain for Smart Homes Existing Surveys

Many surveys dealing with blockchain-based smart home solutions can be found in the literature. In the table XXXIV, we summarised the most recent surveys, by mentioning their advantages and their disadvantages.

The Authors in [200] have worked on a systematic literature review. Their review dealt mainly with the blockchain as a technology for IoT integration. In the so doing, they have presented different applications areas. Then, they organised the available literature review as per the selected areas. After that, two main usage patterns were presented: device manipulation and data management.

In [186], the authors have surveyed the latest trends in using the blockchain technology in smart homes systems. By discussing the various smart homes based blockchain solutions. In this context, two case studies that show how blockchain may empower smart homes transactions related to are presented.

Authors in [17] have highlighted in their survey a comprehensive literature review of the recent security and privacy challenges related to IoT based on their layers. They also explored the impact of integrating blockchain technology with other new technologies such as machine learning. Also, they proposed a framework of IoT security and privacy requirements via blockchain.

Authors in [184] have presented an exhaustive literature review to present the huge impact of enabling smart home application blockchain based has transformed the smart environments besides the underlying issues about the IoT adaptations. Furthermore, a taxonomy highlighting the strengths, weaknesses, opportunities, and threats (SWOT) of a blockchain-based IoT environment. Also, they highlighted the necessity of considering security challenges when dealing with blockchain-based applications for smart homes. Besides, the essential implementation requirements of blockchain in the IoTs have been presented. Finally, a state-of-the-art framework of IoT while adopting security features and decentralised storage requirements of the blockchain.

Authors in [93] presented a comprehensive survey of Blockchain-based Industry 4.0 applications. This survey focused mainly on potential application design, security and privacy requirements, as well as corresponding attacks on Blockchain systems with potential countermeasures. Also, an in-depth analysis and a classification of security and privacy techniques used in these applications to enhance the advancement of security features have been presented. Furthermore, challenges of integrating BC into industrial applications that help design secure Blockchain-based applications as future directions are presented.

### D. Security Requirements of Smart Home Applications

Many requirements have been identified for ensuring a secure smart home-based application, as pointed out in [3], [141]. We identify such requirements as follows (RSSH for requirements for secure smart home applications). As discussed earlier in this part, a smart home is made up mainly from the communication of multiple connected devices that are controlled and monitored through a gateway. Such communication is always exposed to many cyber threats related to data leakage and privacy breaches.

Due to the heterogeneity of the connected devices in the smart home, many difficulties are faced, especially in the absence of security standards. For this reason, users of the smart home cannot benefit from different services. Thus, security is one of the main concerns to be considered. The following describes the security requirements for gateways in smart homes:

- RSSH1: **Authentication**: authentication is about preventing an attacker from acting maliciously within a normal network from outside the smart home network. Thus, BC is used to ensure constantly that the network is a valid member in order to enable necessary corrective actions.
- RSSH2: **Accessibility**. this item refers to the ability of granting access to only legible users of the smart home.
- RSSH3: **Privacy and confidentiality**. these two aspects deal with maintaining the confidentiality of the data generated within the smart home network. It is known that the data generated is sensitive, thus the access to it should be only by authorised entities. The use of symmetric key encryption between different nodes to verify transactions is one of the mechanisms used to ensure those aspects.
- RSSH4: **Integrity**: this item deal with preserving the transmitted data sent or received within the smart home network from any alteration or falsification. . The hash function is used to ensure this item, and it allows the tracking and checking of precisely what data is recorded.

TABLE XXXIV: **Blockchain for Smart homes** Existing Surveys

| Ref | Scope | Description | (+)Pros/(-)Cons |
|-----|-------|-------------|-----------------|
| [200] | Systematic review | ● Blockchain as a technology for IoT integration , analysis of the current research trends on the usage of Blockchain-related approaches and technologies in an IoT context, | (+) Coverage of the novelties in different application domains. <br> (+) Organising the available literature according to the categorization of novelties. <br> (+) The explanation of two main usage patterns: device manipulation and data management (open marketplace solution). <br> (+) Report on the development level of some of the presented solutions is included. |
| [186] | Systematic review | ● Prerequisites of a smart home to adopt blockchain technology. <br> ● Two case studies to show how blockchain empowers smart home transactions by eliminating the middleman: Smart home energy trading in smart grids and smart home data sharing. <br> ● Major challenges pertaining to interoperability protocols, security and privacy, data collection and sharing, data analytics, and latency.. | (-) Only few solutions are presented. <br><br> (+) Presented solutions are deeply explained. <br> (+) Potentially promising areas for future research are presented |
| [99] | Review of proposed solutions | ●solutions for the incorporation of IoT and blockchain in power systems, particularly in the distribution level, residential section, smart buildings, smart homes, energy hubs schemes | (+) Industrial companies are presented. |
| [17] | Comprehensive literature review | ● Recent security and privacy challenges related to IoT based on their layers. <br> ● A framework of IoT security and privacy requirements via blockchain has been proposed. | (+) The impact of integrating blockchain technology with other new technologies such as machine learning. |
| [184] | Comprehensive literature review | ● Studies related to how well blockchain has transformed the smart environments connected with IoT sensors and the underlying issues about IoT adaptation in smart environments. <br> ● Presentation of the facts of blockchain applications such as Bitcoin or Ethereum based works and pinpoint the necessities and security challenges. <br> ● Moreover, the essential implementation requirements of blockchain in the IoTs have been presented. | (+) A taxonomy highlighting the strengths, weaknesses, opportunities, and threats (SWOT) of a blockchain-based IoT environment. <br> (+) A state-of-the-art framework of IoT while adopting security features and decentralized storage requirements of the blockchain. |
| [93] | Comprehensive and state-of-the-art survey | ● Blockchain-based Industry 4.0 applications, focusing on potential application design, security and privacy requirements, as well as corresponding attacks on Blockchain systems with potential countermeasures have been discussed. <br> ● Open issues of integrating Blockchain technology into industrial applications that help design secure Blockchain-based applications as future directions are presented. | (+) Blockchain issues and challenges are presented. <br> (+) An in-depth analysis and a classification of security and privacy techniques used in these applications to enhance the advancement of security features. |

- RSSH5: **Availability**. this item is set to ensure that the services presented within a smart home network are always in time when they are requested by the user. This item is accomplished by limiting acceptable transactions by devices and mine.

*E. Requirements of Smart Home Applications to Adopt Blockchain Technology*

- **Solution building and testing**. This is related to the software and hardware requirements. First, a suitable BC platform has to be carefully chosen, (e.g., permissioned) and its consensus algorithm. The best choice is usually the private blockchain platform with its computationally less expensive consensus protocol. Secondly, hardware requirements have to be determined properly, in order to be able to implement the prototype in the real smart home environment. A performance matrix and Acceptance

Test Procedure (ATP) may assist to achieve the desired blockchain platform.

- **Rollout and integration**. This is due to the necessity of keeping both traditional systems and new BC systems. The integration of both systems might lead to a non-suitable adaptation.

- **Usability**. ensuring a good user interface and user experience (UX UI) and a web-based dashboard to facilitate the interaction between the system and the user.

*F. Summary of Blockchain for Secure smart home solutions*

In this section, we detail the most trending solutions related to the use of blockchain in smart home environments for security purposes. The categorisation of the solutions will be according to: their main contribution to the blockchain, the BC type used in the proposed solution, the consensus, the storage, the validation tools, and the limits if available. In

TABLE XXXV: Benefits of blockchain to Smart home Applications

| | |
|---|---|
| **Decentralization** | Using blockchain technology IoT can move and promote from centralization to decentralization which means a peer-to-peer distributed network structure that eliminates the exclusive of powerful company controls over IoT data processing and storing. Furthermore, decentralization enables IoT scalability to be improved [128], [188], [34], [154]. |
| **Identity** | In blockchain technology, each participant (peer) can identify any other devices or participants (peers) included in the distributed network of IoT, with immutable and unique identifiers for each peer. This means that blockchain technology can provide IoT with high authentication and authorization distributed peers for its applications [128]. |
| **Autonomy** | Blockchain technology opens opportunities for shifting to the next generation features of IoT applications, which will benefit the participants of IoT applications [154]. |
| **Reliability** | Which is the feature that the blockchain can bring to the IoT that enables each peer to authenticate data each time to detect if it is tampered with. In addition, it enables the traceability and accountability of sensor data. [128]. |
| **Security** | Blockchain increases the security of the IoT through the validation of all transactions (data exchange) by smart contracts, which will be optimised by the current IoT protocol [136]. |
| **Market of service** | Through the benefits, that blockchain brings to IoT, the IoT services varieties and coverage will be expanded rapidly in services and data economic fields, as well as decreasing the cost of centralization model installation and maintenance [133]. |
| **Secure Code deployment** | In which blockchain provides the immutable storage of the safety and security codes in the systems enabling manufacturers to trace states with confidence. |

[141], a blockchain-based smart home gateway architecture for preventing data forgery. The main contribution is the use of decentralized, using gateway and Private/ Consortium and Ethereum and Encryption method(Hash values and a verification process) and Cloud storage and Solidity and Smart contracts and additional computational complexity by the BC operations.

The work in [59] is about a suggested method that ensures data privacy in a smart home using BC technology, called smart home based on the IoT-Blockchain (SHIB). To validate this proposed method, an experimental scenario was used which includes Ganache, Remix, and web3.js has been built among the user, service provider, and smart home to evaluate the performance of the smart contract in the SHIB. One main limitation of this work is that the additional computational complexity of the blockchain operation is not studied.

Authors in [293] proposed an IoT architecture for the smart home based BC and smart contract to improve efficiency, storage and energy cost. The architecture is entailed over three tiers: smart contract triggered by the IoT device in the smart home if certain conditions are met, private BC in each smart home, and public BC that connects various houses. In [271] Private Blockchain-Based Secure Access Control for Smart Home Systems Private (PBAC) has been proposed. Security analysis and performance evaluation have been studied as well.

The authors in [183] focused their proposed work mainly on energy consumption for the smart home. They suggested an efficient lightweight integrated BC (ELIB) model for IoT security and privacy. The proposed BC system is a public one that was implemented using Hyperledger Fabric, PBFT consensus, Cloud storage, and Hyperledger composer. However, the storage of patients data in the blockchain network is heavy.

In the same context of energy consumption, authors in [5] proposed a BC enabled distributed demand side management in a community energy system with smart homes public. The proposed method has been developed using Hyperledger Fabric PBFT and cloud storage. One limitation of this approach is that the storage of patients' data in the blockchain network is heavy.

From the above table, we observe that most of the solutions proposed for smart home applications use private BC instead of public BC. The reason behind that is that the public BC presents many challenges related to high latency, full openness, low usability, low throughput, and low scalability. While private and consortium BCs have a trust-based model, low decentralization, and fewer security issues. This issue may hamper the adoption of BC in smart home. However, as the new consensus algorithms and privacy-preserving techniques are emerging, blockchain is expected to provide different levels of decentralization with divergent access policies in a scalable fashion to offer more reliable and secure smart environments. Moreover, smart contracts can provide intelligence to the BC for anomaly detection or execution of business logic for enabling safe and reliable smart environment services.

### G. Mapping of existing solutions to Security requirements and Future Directions

In Table XXXVII, we present a mapping between the solutions presented in Section VIII-F and the requirements presented in VIII-D. This mapping aims at highlighting the solutions presented as per the security requirements. One main finding of this section is that integrity is not always verified or checked in the presented works. This is a point to say that adopting BC for smart homes cannot guarantee full security.

Integrating BC technology into the smart home enabled solutions is motivated by many benefits, among which security is the top one. Yet, many challenges related to smart home

TABLE XXXVI: Summary of **Blockchain for Secure smart home** solutions

| Ref | Contribution/ Purpose | Blockchain Type | Framework | Consensus | Storage | Validation Tools | Limits |
|---|---|---|---|---|---|---|---|
| [141] | A blockchain-based smart home gateway architecture for preventing data forgery. Main contribution is the use of decentralized, using gateway | Private/ Consortium | Ethereum | Encryption method(Hash values and a verification process) | Cloud storage | Solidity + Smart contracts | additional computational complexity by the BC operations |
| [59] | smart home based the IoT-Blockchain (SHIB) | Private | Ethereum | PoW | storage devices | smart contracts that are Access Control Contract (ACC) Judge Contract (JC) and Register Contract (RC) | -additional computational complexity by blockchain operation is not studied. |
| [293] | Iot architecture in smart home environment based on blockchain and smart contract to improve efficiency, storage and energy cost. | Private | Hyperledger Fabric | Cluster Heads are the miners | Local storage | smart contract built within the IoT devices, | -How cluster are formed is not described, -No validation is done |
| [271] | Private Blockchain-Based Secure Access Control for Smart Home Systems (PBAC) | Private | NaN | policy header with an access control list | local storage/cloud storage | access control | security analysis is theoretical +detailed performace study and communication overhead are done |
| [183] | An efficient Lightweight integrated Blockchain (ELIB) model for IoT security and privacy | Public | Hyperledger Fabric | PBFT | Cloud storage | Hyperledger composer + Caliper | The storage of patients data in blockchain network is a heavy. |
| [5] | A Blockchain Enabled Distributed Demand Side Management in Community Energy System With Smart Homes | Public | Hyperledger Fabric | PBFT | Cloud storage | Hyperledger composer + Caliper | The storage of patients data in blockchain network is a heavy. |
| [87] | Blockchain security defence mechanism for IoT smart homes. Smart contracts for implementing registration of users and hardware elements | Private | Ethereum network | PBFT | Cloud storage | smart contracts for implementing registration of users | . |

BC enabled were captured and studied. The author in [251] has discussed the **design challenges** related to developing BC capabilities for a smart home were studied in an experimental setup. One of the main findings of this work pointed out that the design practices of the BC systems still have a low maturity level. Mainly because it is still lacking standards, governance models and means for quantitative analyses. One of the open issues related to the smart homes BC based is the **security challenges**. Networks configured in smart homes, collect and store multiple data, including sensitive information from residents. The miner in the smart home case is the entity responsible for handling all communications related to the home. The authors in [186] have pointed out that the impersonation attacks are one of the most critical attacks for smart homes that need a lot of attention. Since the efficiency and the speed of making sure that the nodes of the smart home network are not involved in potential impersonation attacks. This is usually checked by consensus among all mining nodes in the BC network. However, this requires enormous computational power and hence we need research on efficient consensus algorithms that not only reduce the computational power but also enhance the speed. Also, we need to consider the **overhead** on the resources-constrained IoT devices that are known to be lightweight devices that need time to verify the transmitted data. The use of BC requires transactions' validation and insertion block time, which results in responses delays. Thus, the traditional BC mechanism is not suitable for time-sensitive smart home IoT applications, where a minimum delay results in severe consequences. Besides, there is the issue of **interoperability** that is due to the different formats of data collected from various types of IoT devices. This heterogeneity makes implementing the system quite challenging. For example, the different consensus algorithms adopted by BC systems have a huge impact on implementing an interoperability protocol.

## IX. Conclusions

This survey explores the use of blockchain in smart cities. First, we start by reviewing existing surveys and we explain the distinction and therefore the motivation of this survey. Then we give an overview of blockchain. After that, we reviewed blockchain based solutions for the main smart city applications, namely smart healthcare, smart transportation, smart agriculture, supply chain management, smart grid, and smart homes. For each smart city application we first reviewed the existing surveys of that area, then we discussed the requirements and the benefits of blockchain to this area. Then, we discussed the blockchain based solutions and we conclude by a mapping of the reviewed solutions to the specified requirements and giving possible future directions. Although several solutions have been developed around blockchain for smart city applications, several challenges still need to be resolved before the real deployment of such solutions. Storage, throughput, scalability, and interoperability are some of the main challenges facing blockchain-enabled solutions. These challenges need to be carefully considered when proposing a blockchain-based solution.

## Acknowledgements

The authors wish to thank the anonymous reviewers for their valuable suggestions.

## References

[1] Khizar Abbas et al. "A Blockchain and Machine Learning-Based Drug Supply Chain Management and Recommendation System for Smart Pharmaceutical Industry". In: *Electronics* 9.5 (2020), p. 852.

[2] Juhar Abdella and Khaled Shuaib. "Peer to peer distributed energy trading in smart grids: A survey". In: *Energies* 11.6 (2018), p. 1560.

[3] Maha AbuNaser and Ahmad AA Alkhatib. "Advanced survey of blockchain for the internet of things smart home". In: *2019 IEEE Jordan international joint conference on electrical engineering and information technology (JEEIT)*. IEEE. 2019, pp. 58–62.

[4] *Action 9, the food Blockchain*. https://www.carrefour.com / en / group / food - transition / food - blockchain. (Accessed on 17/1/2021). 2018.

[5] Muhammad Afzal et al. "Blockchain enabled distributed demand side management in community energy system with smart homes". In: *IEEE Access* 8 (2020), pp. 37428–37439.

[6] Cornelius C Agbo, Qusay H Mahmoud, and J Mikael Eklund. "Blockchain technology in healthcare: a systematic review". In: *Healthcare*. Vol. 7. 2. Multidisciplinary Digital Publishing Institute. 2019, p. 56.

[7] Shubhani Aggarwal et al. "A new secure data dissemination model in internet of drones". In: *ICC 2019-2019 IEEE International Conference on Communications (ICC)*. IEEE. 2019, pp. 1–6.

[8] Shubhani Aggarwal et al. "Blockchain for smart communities: Applications, challenges and opportunities". In: *Journal of Network and Computer Applications* 144 (2019), pp. 13–48.

[9] Shubhani Aggarwal et al. "Energychain: Enabling energy trading for smart homes using blockchains in smart grid ecosystem". In: *Proceedings of the 1st ACM MobiHoc Workshop on Networking and Cybersecurity for Smart Cities*. 2018, pp. 1–6.

[10] *Agridigital. What is AgriDigital?* Accessed 26/2/2021. 2019. URL: https://knowledgebase.agridigital.io/hc/en-us/articles/226461448-What-is-AgriDigital-.

[11] Amanda Ahl et al. "Review of blockchain-based distributed energy: Implications for institutional development". In: *Renewable and Sustainable Energy Reviews* 107 (2019), pp. 200–211.

[12] Sabbir Ahmed, Mohammad Saidur Rahman, Mohammad Saiedur Rahaman, et al. "A blockchain-based architecture for integrated smart parking systems". In: *2019 IEEE International Conference on Pervasive Computing and Communications Workshops (PerCom Workshops)*. IEEE. 2019, pp. 177–182.

[13] Tareq Ahram et al. "Blockchain technology innovations". In: *2017 IEEE technology & engineering management conference (TEMSCON)*. IEEE. 2017, pp. 137–141.

[14] Abdullah Al Omar et al. "Privacy-friendly platform for healthcare data in cloud based on blockchain environment". In: *Future generation computer systems* 95 (2019), pp. 511–521.

[15] Kazi Masudul Alam et al. "A Blockchain-based Land Title Management System for Bangladesh". In: *Journal of King Saud University - Computer and*

TABLE XXXVII: Mapping of **Blockchain for Smart Homes** solutions to security requirements

| Ref | RSSH1 (Auth) | RSSH2 (AC) | RSSH3 (Priv) | RSSH4 (Intg) | RSSH5 (Avai) |
|-----|-----|-----|-----|-----|-----|
| [141] | ✓ | ✓ | ✓ | ✓ | ✓ |
| [59] | ✓ | ✓ | ✓ | ✗ | ✓ |
| [293] | ✓ | ✓ | ✓ | ✗ | ✓ |
| [271] | ✓ | ✓ | ✓ | ✗ | ✓ |
| [183] | ✓ | ✓ | ✓ | ✗ | ✓ |
| [5] | ✓ | ✓ | ✓ | ✗ | ✓ |

*Information Sciences* (2020). ISSN: 1319-1578. DOI: https://doi.org/10.1016/j.jksuci.2020.10.011. URL: https://www.sciencedirect.com/science/article/pii/S1319157820304912.

[16] Cohn Alan, West Travis, and Parker Chelsea. "Smart after all: blockchain, smart contracts, parametric insurance, and smart energy grids". In: *Georgetown Law Technology Review* 1 (2 2017). Innovative Solutions for Energy Transitions, 273–304.

[17] Omar Alfandi et al. "A survey on boosting IoT security and privacy through blockchain". In: *Cluster Computing* (2020), pp. 1–19.

[18] Bako Ali and Ali Ismail Awad. "Cyber and physical security vulnerability assessment for IoT-based smart homes". In: *sensors* 18.3 (2018), p. 817.

[19] Muhammad Salek Ali et al. "Applications of blockchains in the Internet of Things: A comprehensive survey". In: *IEEE Communications Surveys & Tutorials* 21.2 (2018), pp. 1676–1717.

[20] Johnston Alistair. *Spanish Company Sends Olive Oil to the Blockchain With Smart Contract Quality Control*. 2018.

[21] Moayad Aloqaily et al. "Design guidelines for blockchain-assisted 5G-UAV networks". In: *IEEE Network* 35.1 (2021), pp. 64–71.

[22] Marica Amadeo et al. "Named data networking for IoT: An architectural perspective". In: *2014 European Conference on Networks and Communications (EuCNC)*. IEEE. 2014, pp. 1–5.

[23] Gurtu Amulya and Johny Jestin. "Potential of blockchain technology in supply chain management: a literature review". In: *International Journal of Physical Distribution & Logistics Management* 49.9 (2019), pp. 881–900.

[24] Nitish Andola et al. "SpyChain: A Lightweight Blockchain for Authentication and Anonymous Authorization in IoD". In: *Wireless Personal Communications* (2021), pp. 1–20.

[25] Merlinda Andoni et al. "Blockchain technology in the energy sector: A systematic review of challenges and opportunities". In: *Renewable and Sustainable Energy Reviews* 100 (2019), pp. 143–174.

[26] Pinna Andrea and Ibba Simona. "A Blockchain-Based Decentralized System for Proper Handling of Temporary Employment Contracts". In: *Science and Information Conference*. 2019. DOI: https://doi.org/10.1007/978-3-030-01177-2_88.

[27] Elli Androulaki et al. "Hyperledger fabric: a distributed operating system for permissioned blockchains". In: *Proceedings of the thirteenth EuroSys conference*. 2018, pp. 1–15.

[28] Abdullahi Arabo. "Cyber security challenges within the connected home ecosystem futures". In: *Procedia Computer Science* 61 (2015), pp. 227–232.

[29] *ArbolMarket*. (Accessed on 15/3/2021). 2019. URL: https://www.arbolmarket.com.

[30] Oumaima Attia et al. "An Iot-blockchain architecture based on hyperledger framework for healthcare monitoring application". In: *2019 10th IFIP International Conference on New Technologies, Mobility and Security (NTMS)*. IEEE. 2019, pp. 1–5.

[31] Asaph Azaria et al. "Medrec: Using blockchain for medical data access and permission management". In: *2016 2nd International Conference on Open and Big Data (OBD)*. IEEE. 2016, pp. 25–30.

[32] Mahmoud M Badr et al. "Smart parking system with privacy preservation and reputation management using blockchain". In: *IEEE Access* 8 (2020), pp. 150823–150843.

[33] Saeed Asadi Bagloee et al. "Tradable mobility permit with Bitcoin and Ethereum–A Blockchain application in transportation". In: *Internet of Things* 8 (2019), p. 100103.

[34] Arshdeep Bahga and Vijay K Madisetti. "Blockchain platform for industrial internet of things". In: *Journal of Software Engineering and Applications* 9.10 (2016), p. 533.

[35] Jiabin Bao et al. "A survey of blockchain applications in the energy sector". In: *IEEE Systems Journal* (2020).

[36] Shihan Bao et al. "Pseudonym management through blockchain: Cost-efficient privacy preservation on intelligent transportation systems". In: *IEEE Access* 7 (2019), pp. 80390–80403.

[37] Ezedin Barka et al. "Towards a trusted unmanned aerial system using blockchain for the protection of critical infrastructure". In: *Transactions on Emerging Telecommunications Technologies* (2019), e3706.

[38] Mohamed Baza et al. "B-ride: Ride sharing with privacy-preservation, trust and fair payment atop public blockchain". In: *IEEE Transactions on Network Science and Engineering* (2019).

[39] Mohamed Baza et al. "Incentivized and secure blockchain-based firmware update and dissemination for autonomous vehicles". In: *Connected and Autonomous Vehicles in Smart Cities*. CRC press, 2020, pp. 475–493.

[40] Rafael Belchior et al. "A survey on blockchain interoperability: Past, present, and future trends". In: *ACM Computing Surveys (CSUR)* 54.8 (2021), pp. 1–41.

[41] Müßigmann Benjamin, von der Gracht Heiko, and Hartmann Evi. "Blockchain Technology in Logistics and Supply Chain Management—A Bibliometric Literature Review From 2016 to January 2020". In: *IEEE Transactions on Engineering Management* 67.4 (2020), pp. 988–1007. DOI: 10.1109/TEM.2020.2980733.

[42] Bharat Bhushan et al. "Blockchain for smart cities: A review of architectures, integration trends and future research directions". In: *Sustainable Cities and Society* (2020), p. 102360.

[43] *Blockchain technologies offer transparency that could improve human rights practices*. 2019. URL: https://www.openglobalrights.org/blockchain-technologies-offer-transparency-that-could-improve-human-rights-practices/.

[44] B. BORDEL et al. "A Blockchain-based Water Control System for the Automatic Management of Irrigation Communities". In: *2019 IEEE International Conference on Consumer Electronics (ICCE)*. 2019, pp. 1–2. DOI: 10.1109/ICCE.2019.8661940.

[45] Sotirios Brotsis et al. "On the suitability of blockchain platforms for IoT applications: Architectures, security, privacy, and performance". In: *Computer Networks* 191 (2021), p. 108005.

[46] Richard Gendal Brown. "The corda platform: An introduction". In: *Retrieved* 27 (2018), p. 2018.

[47] *Carrefour report activity 2018*. https://www.carrefour.com/sites/default/files/2020-01/carrefour-ra2018-en_0.pdf. (Accessed on 17/1/2021). 2018.

[48] Fran Casino, Thomas K Dasaklis, and Constantinos Patsakis. "A systematic literature review of blockchain-based applications: current status, classification and open issues". In: *Telematics and Informatics* 36 (2019), pp. 55–81.

[49] Shuchih Ernest Chang and Chi-Yin Chang. "Application of blockchain technology to smart city service: A case of ridesharing". In: *2018 IEEE International Conference on Internet of Things (iThings) and IEEE Green Computing and Communications (GreenCom) and IEEE Cyber, Physical and Social Computing (CPSCom) and IEEE Smart Data (SmartData)*. IEEE. 2018, pp. 664–671.

[50] Rajat Chaudhary et al. "BEST: Blockchain-based secure energy trading in SDN-enabled intelligent transportation system". In: *Computers & Security* 85 (2019), pp. 288–299.

[51] Gertrude. Chavez-Dreyfuss. *Sweden tests blockchain technology for land registry*. (Accessed on 25/2/2022). 2016. URL: https://www.reuters.com/article/us-sweden-blockchain-idUSKCN0Z22KV.

[52] Pan Chen et al. "Research on scalability of blockchain technology: Problems and methods". In: *Journal of Computer Research and Development* 55.10 (2018), p. 2099.

[53] Emeka Chukwu and Lalit Garg. "A systematic review of blockchain in healthcare: Frameworks, prototypes, and implementations". In: *IEEE Access* 8 (2020), pp. 21196–21214.

[54] Mandolla Claudio et al. "Building a digital twin for additive manufacturing through the exploitation of blockchain: A case analysis of the aircraft industry". In: *Computers in Industry* 109 (2019), pp. 134 –152. ISSN: 0166-3615. DOI: https://doi.org/10.1016/j.compind.2019.04.011. URL: http://www.sciencedirect.com/science/article/pii/S0166361518308741.

[55] *Composer*. https://hyperledger.github.io/composer/latest/introduction/introduction.html. Accessed: 2020-06-25.

[56] Gaby G Dagher et al. "Ancile: Privacy-preserving framework for access control and interoperability of electronic health records using blockchain technology". In: *Sustainable cities and society* 39 (2018), pp. 283–297.

[57] Xiaohai Dai et al. "Jidar: A jigsaw-like data reduction approach without trust assumptions for bitcoin system". In: *2019 IEEE 39th International Conference on Distributed Computing Systems (ICDCS)*. IEEE. 2019, pp. 1317–1326.

[58] Zakaria Dakhli, Zoubeir Lafhaj, and Alan Mossman. "The Potential of Blockchain in Building Construction". In: *Buildings* 9.4 (2019), p. 77. ISSN: 2075-5309. DOI: 10.3390/buildings9040077. URL: http://dx.doi.org/10.3390/buildings9040077.

[59] Thanh Long Nhat Dang and Minh Son Nguyen. "An approach to data privacy in smart home using blockchain technology". In: *2018 International Conference on Advanced Computing and Applications (ACOMP)*. IEEE. 2018, pp. 58–64.

[60] Pigini Danny and Conti Massimo. "NFC-Based Traceability in the Food Chain." In: *Sustainability* 9.10 (2017), p. 1910.

[61] Costa Davide et al. *Quadrans White Paper*. Tech. rep. Quadrans Foundation, 2018.

[62] Prashar Deepak et al. "Blockchain-Based Traceability and Visibility for Agricultural Products: A Decentralized Way of Ensuring Food Safety in India". In: *Sustainability* 12.8 (2020).

[63] Konstantinos Demestichas et al. "Blockchain in Agriculture Traceability Systems: A Review". In: *Applied Sciences* 10.12 (June 2020), p. 4113. ISSN: 2076-3417. DOI: 10.3390/app10124113. URL: http://dx.doi.org/10.3390/app10124113.

[64] Xinyang Deng and Tianhan Gao. "Electronic payment schemes based on blockchain in VANETs". In: *IEEE Access* 8 (2020), pp. 38296–38303.

[65] D Di Francesco Maesa and P Mori. "Blockchain 3.0 applications survey". In: *J. Parallel Distrib. Comput* 138 (2020), pp. 99–114.

[66] Mao Dianhui et al. "Innovative blockchain-based approach for sustainable and credible environment in food trade: A case study in shandong province, china". In: *Sustainability* 9 (2018), p. 3149. ISSN: 2071-1050. DOI: 10.3390/su10093149. URL: http://dx.doi.org/10.3390/su10093149.

[67] George Drosatos and Eleni Kaldoudi. "Blockchain applications in the biomedical domain: a scoping review". In: *Computational and structural biotechnology journal* 17 (2019), pp. 229–240.

[68] Ashutosh Dhar Dwivedi et al. "A decentralized privacy-preserving healthcare blockchain for IoT". In: *Sensors* 19.2 (2019), p. 326.

[69] Ala Ekramifard et al. "A systematic literature review of integration of blockchain and artificial intelligence". In: *Blockchain Cybersecurity, Trust and Privacy*. Springer, 2020, pp. 147–160.

[70] Adel S Elmaghraby and Michael M Losavio. "Cyber security challenges in Smart Cities: Safety, security and privacy". In: *Journal of advanced research* 5.4 (2014), pp. 491–497.

[71] Florentina Magda Enescu et al. "Implementing Blockchain Technology in Irrigation Systems That Integrate Photovoltaic Energy Generation Systems". In: *Sustainability* 12.4 (2020), p. 1540. ISSN: 2071-1050. DOI: 10.3390/su12041540. URL: http://dx.doi.org/10.3390/su12041540.

[72] Ayman Esmat et al. "A novel decentralized platform for peer-to-peer energy trading market with blockchain technology". In: *Applied Energy* 282 (2021), p. 116123.

[73] Christian Esposito et al. "Blockchain: A panacea for healthcare cloud-based data security and privacy?" In: *IEEE Cloud Computing* 5.1 (2018), pp. 31–37.

[74] Valentina Fabi, Giorgia Spigliantini, and Stefano Paolo Corgnati. "Insights on smart home concept and occupants' interaction with building controls". In: *Energy Procedia* 111 (2017), pp. 759–769.

[75] Kai Fan et al. "Medblock: Efficient and secure medical data sharing via blockchain". In: *Journal of medical systems* 42.8 (2018), p. 136.

[76] Ahmed Farouk et al. "Blockchain platform for industrial healthcare: Vision and future opportunities". In: *Computer Communications* (2020).

[77] *Fast Healthcare Interoperability Resources*. https://www.hl7.org/fhir/. Accessed: 2020-11-15.

[78] Qi Feng et al. "A survey on privacy protection in blockchain system". In: *Journal of Network and Computer Applications* 126 (2019), pp. 45–58.

[79] Tiago M Fernández-Caramés and Paula Fraga-Lamas. "A Review on the Use of Blockchain for the Internet of Things". In: *Ieee Access* 6 (2018), pp. 32979–33001.

[80] Mohamed Amine Ferrag et al. "Blockchain technologies for the internet of things: Research issues and challenges". In: *IEEE Internet of Things Journal* 6.2 (2018), pp. 2188–2204.

[81] João Carlos Ferreira et al. "A blockchain and gamification approach for smart parking". In: *First International Conference on Intelligent Transport Systems*. Springer. 2018, pp. 3–14.

[82] Simone Figorilli et al. "A Blockchain Implementation Prototype for the Electronic Open Source Traceability of Wood along the Whole Supply Chain". In: *Sensors* 18.9 (2018), p. 3133. ISSN: 1424-8220. DOI: 10.3390/s18093133. URL: http://dx.doi.org/10.3390/s18093133.

[83] *Foodcoin*. Accessed on 26/2/2021. 2018. URL: https://en.bitcoinwiki.org/wiki/Foodcoin.

[84] Antonucci Francesca et al. "A review on blockchain applications in the agri-food sector". In: *Journal of the science of food and agriculture* 99 (14 2019), pp. 6129–6138. DOI: 10.1002/jsfa.9912.

[85] *Gartner Report*. https://www.gartner.com/en/information - technology/insights/blockchain. Accessed: 2022-05-05.

[86] Chunpeng Ge, Xinshu Ma, and Zhe Liu. "A semi-autonomous distributed blockchain-based framework for UAVs system". In: *Journal of Systems Architecture* 107 (2020), p. 101728.

[87] Konstantinos M Giannoutakis et al. "A blockchain solution for enhancing cybersecurity defence of IoT". In: *2020 IEEE International Conference on Blockchain (Blockchain)*. IEEE. 2020, pp. 490–495.

[88] Greenspan Gideon. "Multichain private blockchain—white paper". In: *Accessed: Jun 28* (2015), p. 2021.

[89] Andrija Goranović et al. "Blockchain applications in microgrids an overview of current projects and concepts". In: *IECON 2017-43rd Annual Conference of the IEEE Industrial Electronics Society*. IEEE. 2017, pp. 6153–6158.

[90] Kristen N Griggs et al. "Healthcare blockchain system using smart contracts for secure automated remote patient monitoring". In: *Journal of medical systems* 42.7 (2018), p. 130.

[91] Rajesh Gupta, Aparna Kumari, and Sudeep Tanwar. "A taxonomy of blockchain envisioned edge-as-a-connected autonomous vehicles". In: *Transactions on Emerging Telecommunications Technologies* 32.6 (2021), e4009.

[92] Abdelatif Hafid, Abdelhakim Senhaji Hafid, and Mustapha Samih. "Scaling blockchains: A comprehensive survey". In: *IEEE Access* 8 (2020), pp. 125244–125262.

[93] Khizar Hameed et al. "A taxonomy study on securing Blockchain-based Industrial applications: An overview, application perspectives, requirements, attacks, countermeasures, and open issues". In: *Journal of Industrial Information Integration* (2022), p. 100312.

[94] Mohamed Haouari et al. "A novel proof of useful work for a blockchain storing transportation transactions". In: *Information Processing & Management* 59.1 (2022), p. 102749.

[95] Naveed Ul Hassan, Chau Yuen, and Dusit Niyato. "Blockchain technologies for smart energy systems: Fundamentals, challenges, and solutions". In: *IEEE Industrial Electronics Magazine* 13.4 (2019), pp. 106–118.

[96] Anton Hasselgren et al. "Blockchain in healthcare and health sciences—A scoping review". In: *International Journal of Medical Informatics* 134 (2020), p. 104040.

[97] Jigna J Hathaliya and Sudeep Tanwar. "An exhaustive survey on security and privacy issues in Healthcare 4.0". In: *Computer Communications* 153 (2020), pp. 311–335.

[98] Marko Hölbl et al. "A systematic review of the use of blockchain in healthcare". In: *Symmetry* 10.10 (2018), p. 470.

[99] Heliasadat Hosseinian et al. "Blockchain outlook for deployment of IoT in distribution networks and smart homes". In: *International Journal of Electrical and Computer Engineering* 10.3 (2020), p. 2787.

[100] W Hu et al. "A blockchain-based smart contract trading mechanism for energy power supply and demand network". In: *Advances in Production Engineering & Management* 14.3 (2019), pp. 284–296.

[101] Juma Husam, Shaalan Khaled, and Kamel Ibrahim. "A Survey on Using Blockchain in Trade Supply Chain Solutions". In: *IEEE Access* 7 (2019), pp. 184115–184132. DOI: 10.1109/ACCESS.2019.2960542.

[102] Hassan Mansur Hussien et al. "A systematic review for enabling of develop a blockchain technology in healthcare application: taxonomy, substantially analysis, motivations, challenges, recommendations and future direction". In: *Journal of medical systems* 43.10 (2019), p. 320.

[103] Allison Ian. *Maersk and IBM Want 10 Million Shipping Containers on the Global Supply Blockchain by Year-end*. http://www.ibtimes.co.uk/maersk-ibm-aim-get-10-million-shipping-containers-onto-globalsupply-blockchain-by-year-end-1609778. (Accessed on 17/1/2021). 2017.

[104] IATA. *Future Of The Airline Industry 2035*. https://www.iata.org/contentassets/690df4ddf39b47b5a075bb5dff30e1d8/iata-future-airline-industry-pdf.pdf. 2018.

[105] *IBM Blockchain. Now delivering value around the world*. Accessed 26/2/2021. 2019. URL: https://www.ibm.com/blockchain.

[106] Rateb Jabbar et al. "Blockchain technology for healthcare: Enhancing shared electronic health record interoperability and integrity". In: *2020 IEEE International Conference on Informatics, IoT, and Enabling Technologies (ICIoT)*. IEEE. 2020, pp. 310–317.

[107] Rateb Jabbar et al. "Blockchain Technology for Intelligent Transportation Systems: A Systematic Literature Review". In: *IEEE Access* (2022).

[108] Grecuccio Jacopo et al. "Combining Blockchain and IoT: Food-Chain Traceability and Beyond". In: *Energies* 13.15 (2020), p. 3820.

[109] Faisal Jamil et al. "A novel medical blockchain model for drug supply chain integrity management in a smart hospital". In: *Electronics* 8.5 (2019), p. 505.

[110] Faisal Jamil et al. "Towards a Remote Monitoring of Patient Vital Signs Based on IoT-Based Blockchain Integrity Management Platforms in Smart Hospitals". In: *Sensors* 20.8 (2020), p. 2195.

[111] Barnard Janette. *The Missing Link in the Food Chain: Blockchain, San Francisco: White Paper*. https://resources.decisionnext.com/hubfs/PDFs/missing-link-food-chain-blockchain.pdf. (Accessed on 17/1/2021). 2017.

[112] Muhammad Umar Javed et al. "Blockchain-based secure data storage for distributed vehicular networks". In: *Applied Sciences* 10.6 (2020), p. 2011.

[113] HS Jennath et al. "Parkchain: a blockchain powered parking solution for smart cities". In: *Frontiers in Blockchain* 2 (2019), p. 6.

[114] Emanuel Ferreira Jesus et al. "A survey of how to use blockchain to secure internet of things and the stalker attack". In: *Security and Communication Networks* 2018 (2018).

[115] Duan Jiang et al. "A Content-Analysis Based Literature Review in Blockchain Adoption within Food Supply Chain". In: *International journal of environmental research and public health* 17.5 (2020), p. 1784.

[116] Hai Jin, Xiaohai Dai, and Jiang Xiao. "Towards a novel architecture for enabling interoperability amongst multiple blockchains". In: *2018 IEEE 38th International Conference on Distributed Computing Systems (ICDCS)*. IEEE. 2018, pp. 1203–1211.

[117] Olamide Jogunola et al. "Demonstrating blockchain-enabled peer-to-peer energy trading and sharing". In: *2019 IEEE Canadian Conference of Electrical and Computer Engineering (CCECE)*. IEEE. 2019, pp. 1–4.

[118] Richard Joseph et al. "BlockWheels-A Peer to Peer Ridesharing Network". In: *2021 5th International Conference on Intelligent Computing and Control Systems (ICICCS)*. IEEE. 2021, pp. 166–171.

[119] Wang Jun et al. "The outlook of blockchain technology for construction engineering management". In: *Frontiers of Engineering Management* (1 2017), pp. 67–75.

[120] Andreas Kamilaris, Agusti Fonts, and Francesc X. Prenafeta-Bold. "The rise of blockchain technology in agriculture and food supply chains". In: *Trends in Food Science Technology* 91 (2019), pp. 640–652. ISSN: 0924-2244. DOI: https://doi.org/10.1016/j.tifs.2019.07.034. URL: https://www.sciencedirect.com/science/article/pii/S0924224418303686.

[121] Eung Seon Kang et al. "A blockchain-based energy trading platform for smart homes in a microgrid". In: *2018 3rd international conference on computer and communication systems (ICCCS)*. IEEE. 2018, pp. 472–476.

[122] Jiawen Kang et al. "Enabling localized peer-to-peer electricity trading among plug-in hybrid electric vehicles using consortium blockchains". In: *IEEE Transactions on Industrial Informatics* 13.6 (2017), pp. 3154–3164.

[123] D Kapoor, RB Vyas, and D Dadarwal. "An Overview on Pharmaceutical Supply Chain: A Next Step towards Good Manufacturing Practice. Drug Des Int Prop Int J 1 (2)-2018". In: *DDIPIJ. MS. ID* 107 ().

[124] David Katz. "Plastic bank: launching social plastic revolution". In: *Field Actions Science Reports. The journal of field actions* (19 2019), 96–99.

[125] Gagandeep Kaur and Charu Gandhi. "Scalability in blockchain: Challenges and solutions". In: *Handbook of Research on Blockchain Technology*. Elsevier, 2020, pp. 373–406.

[126] Houssain Kettani and Robert M Cannistra. "On cyber threats to smart digital environments". In: *Proceedings of the 2nd International Conference on Smart Digital Environment*. 2018, pp. 183–188.

[127] Salah Khaled et al. "Blockchain-Based Soybean Traceability in Agricultural Supply Chain". In: *IEEE Access* 7 (2019), pp. 73295–73305. DOI: 10.1109/ACCESS.2019.2918000.

[128] Minhaj Ahmad Khan and Khaled Salah. "IoT security: Review, blockchain solutions, and open challenges". In: *Future Generation Computer Systems* 82 (2018), pp. 395–411.

[129] Salam Khanji and Sameer Assaf. "Boosting ridesharing efficiency through blockchain: Greenride application case study". In: *2019 10th International Conference on Information and Communication Systems (ICICS)*. IEEE. 2019, pp. 224–229.

[130] Asma Khatoon. "A blockchain-based smart contract system for healthcare management". In: *Electronics* 9.1 (2020), p. 94.

[131] Seyednima Khezr et al. "Blockchain technology in healthcare: A comprehensive review and directions for future research". In: *Applied sciences* 9.9 (2019), p. 1736.

[132] Navid Khoshavi, Gabrielle Tristani, and Arman Sargolzaei. "Blockchain Applications to Improve Operation and Security of Transportation Systems: A Survey". In: *Electronics* 10.5 (2021), p. 629.

[133] Sesaria Kikitamara, MCJD van Eekelen, and Dipl Ing Jan-Peter Doomernik. "Digital Identity Management on Blockchain for Open Model Energy System". PhD thesis. Master's Thesis, 23 Apr, 2017.

[134] Mihui Kim and Youngmin Kim. "Multi-Blockchain Structure for a Crowdsensing-Based Smart Parking System". In: *Future Internet* 12.5 (2020), p. 90.

[135] Uri Klarman et al. "bloxroute: A scalable trustless blockchain distribution network whitepaper". In: *IEEE Internet Things J.* (2018).

[136] Charalampos S Kouzinopoulos et al. "Implementing a Forms of Consent Smart Contract on an IoT-based Blockchain to promote user trust". In: *2018 Innovations in Intelligent Systems and Applications (INISTA)*. IEEE. 2018, pp. 1–6.

[137] B Krishna, P Rajkumar, and Venkateshwarlu Velde. "Integration of blockchain technology for security and privacy in internet of things". In: *Materials Today: Proceedings* (2021).

[138] Sowmya Kudva et al. "Pebers: Practical ethereum blockchain based efficient ride hailing service". In: *2020 IEEE International Conference on Informatics, IoT, and Enabling Technologies (ICIoT)*. IEEE. 2020, pp. 422–428.

[139] Vijay Kumar, Carina Hallqvist, and Daniel Ekwall. "Developing a Framework for Traceability Implementation in the Textile Supply Chain". In: *Systems* 5.2 (2017), p. 33. ISSN: 2079-8954. DOI: 10.3390/systems5020033. URL: http://dx.doi.org/10.3390/systems5020033.

[140] Aron Laszka et al. "TRANSAX: A blockchain-based decentralized forward-trading energy exchanged for transactive microgrids". In: *2018 IEEE 24th International Conference on Parallel and Distributed Systems (ICPADS)*. IEEE. 2018, pp. 918–927.

[141] Younghun Lee et al. "A blockchain-based smart home gateway architecture for preventing data forgery". In: *Human-centric Computing and Information Sciences* 10.1 (2020), pp. 1–14.

[142] Ao Lei et al. "Blockchain-based dynamic key management for heterogeneous intelligent transportation systems". In: *IEEE Internet of Things Journal* 4.6 (2017), pp. 1832–1843.

[143] Changle Li et al. "Vehicle position correction: A vehicular blockchain networks-based GPS error sharing framework". In: *IEEE Transactions on Intelligent Transportation Systems* 22.2 (2020), pp. 898–912.

[144] Hongbiao Li, Fan Xiao, and Lixin Yin. "Application of Blockchain Technology in Energy Trading: a Review". In: *Frontiers in Energy Research* 9 (2021), p. 130.

[145] Hui Li et al. "FADB: A fine-grained access control scheme for VANET data based on blockchain". In: *IEEE Access* 8 (2020), pp. 85190–85203.

[146] Lun Li et al. "Creditcoin: A privacy-preserving blockchain-based incentive announcement network for communications of smart vehicles". In: *IEEE Transactions on Intelligent Transportation Systems* 19.7 (2018), pp. 2204–2220.

[147] Wanxin Li et al. "Blockchain-enabled Identity Verification for Safe Ridesharing Leveraging Zero-Knowledge Proof". In: *2020 3rd International Conference on Hot Information-Centric Networking (HotICN)*. IEEE. 2020, pp. 18–24.

[148] Wanxin Li et al. "Privacy-preserving traffic management: A blockchain and zero-knowledge proof inspired approach". In: *IEEE Access* 8 (2020), pp. 181733–181743.

[149] Yinan Li et al. "Design and management of a distributed hybrid energy system through smart contract and blockchain". In: *Applied Energy* 248 (2019), pp. 390–405.

[150] Yuhong Li et al. "A blockchain-assisted intelligent transportation system promoting data services with privacy protection". In: *Sensors* 20.9 (2020), p. 2483.

[151] Z. Li et al. "Consortium Blockchain for Secure Energy Trading in Industrial Internet of Things". In: *IEEE Transactions on Industrial Informatics* 14.8 (2018), pp. 3690–3700. DOI: 10.1109/TII.2017.2786307.

[152] Siyi Liao et al. "Securing Collaborative Environment Monitoring in Smart Cities Using Blockchain Enabled Software-Defined Internet of Drones". In: *IEEE Internet of Things Magazine* 4.1 (2021), pp. 12–18.

[153] Huichen Lin and Neil W Bergmann. "IoT privacy and security challenges for smart home environments". In: *Information* 7.3 (2016), p. 44.

[154] Iuon-Chang Lin and Tzu-Chun Liao. "A survey of blockchain security issues and challenges." In: *IJ Network Security* 19.5 (2017), pp. 653–659.

[155] Iuon-Chang Lin et al. "Traceability of Ready-to-Wear Clothing through Blockchain Technology." In: *sustainability* 12.18 (2020), p. 7491.

[156] W. Lin et al. "Blockchain Technology in Current Agricultural Systems: From Techniques to Applications". In: *IEEE Access* 8 (2020), pp. 143920–143937. DOI: 10.1109/ACCESS.2020.3014522.

[157] Yu-Pin Lin et al. "Blockchain: The Evolutionary Next Step for ICT E-Agriculture". In: *Environments* 4.3 (2017), p. 50. ISSN: 2076-3298. DOI: 10.3390/environments4030050. URL: http://dx.doi.org/10.3390/environments4030050.

[158] Chao Liu et al. "Blockchain based energy trading model for electric vehicle charging schemes". In: *International Conference on Smart Grid Inspired Future Technologies*. Springer. 2018, pp. 64–72.

[159] Rui Liu and Yongqi Ge. "Smart home system design based on Internet of Things". In: *2017 12th International Conference on Computer Science and Education (ICCSE)*. IEEE. 2017, pp. 444–448.

[160] Xingchen Liu et al. "A blockchain-based trust management with conditional privacy-preserving announcement scheme for VANETs". In: *IEEE Internet of Things Journal* 7.5 (2019), pp. 4101–4112.

[161] Federico Lombardi et al. "A blockchain-based infrastructure for reliable and cost-effective IoT-aided smart grids". In: (2018).

[162] Yang Lu. "The blockchain: State-of-the-art and research challenges". In: *Journal of Industrial Information Integration* 15 (2019), pp. 80–90.

[163] Shuyun Luo et al. "Blockchain-Based Task Offloading in Drone-Aided Mobile Edge Computing". In: *IEEE Network* 35.1 (2021), pp. 124–129.

[164] Thelma B. Machado, Leonardo Ricciardi, and M. Beatriz P P Oliveira. "Blockchain technology for the management of food sciences researches". In: *Trends in Food Science & Technology* 102 (2020). ISSN: 0924-2244. DOI: https://doi.org/10.1016/j.tifs.2020.03.043. URL: https://www.sciencedirect.com/science/article/pii/S0924224419310507.

[165] Yash Madhwal and Panfilov Peter. "Blockchain And Supply Chain Management: Aircrafts' Parts' Business Case". In: *The 28th DAAAM International Symposium*. DAAAM International. 2017, pp. 1051–1056.

[166] ElMessiry Magdi and ElMessiry Adel. "Blockchain Framework for Textile Supply Chain Management." In: *International Conference on Blockchain*. Lecture Notes in Computer Science, Springer. 2018, pp. 213–227.

[167] Lukas Malina et al. "Lightweight Ring Signatures for Decentralized Privacy-preserving Transactions." In: *ICETE (2)*. 2018, pp. 692–697.

[168] Tieman Marco and Darun Mohd Ridzuan. "Leveraging Blockchain Technology for Halal Supply Chains". In: *Islam and Civilisational Renewal* 8 (2017), pp. 547–550.

[169] De Clercq Matthieu, Vats Anshu, and Alvaro Biel. *Agriculture 4.0: the Future of Farming Technology*. In collabration with Oliver Wyman. 2018. URL: https://www.worldgovernmentsummit.org/api/publications/document?id=95df8ac4-e97c-6578-b2f8-ff0000a7ddb6.

[170] Marco Mazzoni, Antonio Corradi, and Vincenzo Di Nicola. "Performance evaluation of permissioned blockchains for financial applications: The ConsenSys Quorum case study". In: *Blockchain: Research and applications* 3.1 (2022), p. 100026.

[171] Thomas McGhin et al. "Blockchain in healthcare applications: Research challenges and opportunities". In: *Journal of Network and Computer Applications* 135 (2019), pp. 62–75.

[172] Carlo R.W. de Meijer. *What is holding back blockchain adoption and what should be done?* (Accessed on 10/2/2022). 2021. URL: https://www.finextra.com/blogposting/21337/what-is-holding-back-blockchain-adoption-and-what-should-be-done.

[173] Esther Mengelkamp et al. "A blockchain-based smart grid: towards sustainable local energy markets". In: *Computer Science-Research and Development* 33.1-2 (2018), pp. 207–214.

[174] Khaleel Mershad. "SURFER: A secure SDN-based routing protocol for internet of vehicles". In: *IEEE Internet of Things Journal* 8.9 (2020), pp. 7407–7422.

[175] Khaleel Mershad, Omar Cheikhrouhou, and Leila Ismail. "Proof of accumulated trust: A new consensus protocol for the security of the IoV". In: *Vehicular Communications* 32 (2021), p. 100392.

[176] Caro Miguel Pincheira et al. "Blockchain-based traceability in agri-food supply chain management: a practical implementation". In: *IoT vertical and tropical summit on agriculture.* Institute of Electrical and Electronics Engineers. 2018, 1–4.

[177] Branka Mikavica and Aleksandra Kostić-Ljubisavljević. "Blockchain-based solutions for security, privacy, and trust management in vehicular networks: a survey". In: *The Journal of Supercomputing* (2021), pp. 1–56.

[178] Tomas Mikula and Rune Hylsberg Jacobsen. "Identity and access management with blockchain in electronic healthcare records". In: *2018 21st Euromicro conference on digital system design (DSD)*. IEEE. 2018, pp. 699–706.

[179] Zlatogor Minchev and Luben Boyanov. "Smart Homes Cyberthreats Identification Based on Interactive Training". In: *Proceedings of International Conference on Application of Information and Communication Technology and Statistics in Economy and Education (ICAICTSEE)*. International Conference on Application of Information and Communication … 2013, p. 72.

[180] Tripoli Mischa and Schmidhuber Josef. *Emerging Opportunities for the Application of Blockchain in the Agri-Food Industry*. FAO and ICTSD: Rome and Geneva, http://www.fao.org/3/CA1335EN/ca1335en.pdf. (Accessed on 17/1/2021). 2018.

[181] Ferrag Mohamed Amine et al. "Security and Privacy for Green IoT-Based Agriculture: Review, Blockchain Solutions, and Challenges". In: *IEEE Access* 8 (2020), pp. 32031–32053. DOI: 10.1109/ACCESS.2020.2973178.

[182] Bhabendu Kumar Mohanta et al. "Blockchain technology: A survey on applications and security privacy challenges". In: *Internet of Things* 8 (2019), p. 100107.

[183] Sachi Nandan Mohanty et al. "An efficient Lightweight integrated Blockchain (ELIB) model for IoT security and privacy". In: *Future Generation Computer Systems* 102 (2020), pp. 1027–1037.

[184] Sana Moin et al. "Securing IoTs in distributed blockchain: Analysis, requirements and open issues". In: *Future Generation Computer Systems* 100 (2019), pp. 325–343.

[185] Muhammad Baqer Mollah et al. "Blockchain for future smart grid: A comprehensive survey". In: *IEEE Internet of Things Journal* (2020).

[186] Md Moniruzzaman et al. "Blockchain for smart homes: Review of current trends and research challenges". In: *Computers & Electrical Engineering* 83 (2020), p. 106585.

[187] M. Safdar Munir, Imran Sarwar Bajwa, and Sehrish Munawar Cheema. "An intelligent and secure smart watering system using fuzzy logic and blockchain". In: *Computers & Electrical Engineering* 77 (2019), pp. 109–119. ISSN: 0045-7906. DOI: https://doi.org/10.1016/j.compeleceng.2019.05.006.

[188] URL: https://www.sciencedirect.com/science/article/pii/S0045790618328581.

[188] Fedor Muratov et al. "YAC: BFT Consensus Algorithm for Blockchain". In: *arXiv preprint arXiv:1809.00554* (2018).

[189] Ahmed S Musleh, Gang Yao, and SM Muyeen. "Blockchain applications in smart grid–review and frameworks". In: *IEEE Access* 7 (2019), pp. 86746–86757.

[190] Aung Myo Min and Chang Yoon Seok. "Traceability in a food supply chain: Safety and quality perspectives". In: *Food Control* 39 (2014), pp. 172–184. ISSN: 0956-7135. DOI: https://doi.org/10.1016/j.foodcont.2013.11.007. URL: http://www.sciencedirect.com/science/article/pii/S0956713513005811.

[191] N. Kumari et al. "Design of Electronic Voting Portal Using Blockchain Technology". In: *Journal of Electronic Design Technology* 11 (2020).

[192] Tri Nguyen, Risto Katila, and Tuan Nguyen Gia. "A Novel Internet-of-Drones and Blockchain-based System Architecture for Search and Rescue". In: *arXiv preprint arXiv:2108.00694* (2021).

[193] Tri Hong Nguyen, Juha Partala, and Susanna Pirttikangas. "Blockchain-based mobility-as-a-service". In: *2019 28th International Conference on Computer Communication and Networks (ICCCN)*. IEEE. 2019, pp. 1–6.

[194] Morris Nicky. *Maersk/IBM complete supply chain blockchain pilot*. https://www.ledgerinsights.com/enterprises-invest-blockchain-compliance/. (Accessed on 17/1/2021). 2019.

[195] *Office of the National Coordinator for Health Information Technology*. https://www.healthit.gov/. Accessed: 2020-11-15.

[196] Daiki Ogawa, Koichi Kobayashi, and Yuh Yamashita. "Blockchain-Based Distributed Optimization for Energy Management Systems". In: *2019 IEEE International Conference on Industrial Cyber Physical Systems (ICPS)*. IEEE. 2019, pp. 706–711.

[197] Se-Chang Oh et al. "Implementation of blockchain-based energy trading system". In: *Asia Pacific Journal of Innovation and Entrepreneurship* (2017).

[198] Chuka Oham et al. "B-FERL: Blockchain based framework for securing smart vehicles". In: *Information Processing & Management* 58.1 (2021), p. 102426.

[199] Friha Othmane et al. "Internet of Things for the Future of Smart Agriculture: A Comprehensive Survey of Emerging Technologies". In: *IEEE/CAA Journal of Automatica Sinica* 8.JAS-2020-1035 (2021), p. 718. ISSN: 2329-9266. DOI: 10.1109/JAS.2021.1003925. URL: http://www.ieee-jas.net/article/id/6510bc39-48c2-4754-88de-8156b50f7ed8.

[200] Alfonso Panarello et al. "Blockchain and iot integration: A systematic survey". In: *Sensors* 18.8 (2018), p. 2575.

[201] Jema Sharin PankiRaj, Abdulsalam Yassine, and Salimur Choudhury. "An auction mechanism for profit maximization of peer-to-peer energy trading in smart grids". In: *Procedia Computer Science* 151 (2019), pp. 361–368.

[202] Nidhi Pathak, Anandarup Mukherjee, and Sudip Misra. "AerialBlocks: Blockchain-Enabled UAV Virtualization for Industrial IoT". In: *IEEE Internet of Things Magazine* 4.1 (2021), pp. 72–77.

[203] Alejandro Ranchal Pedrosa and Giovanni Pau. "ChargeItUp: On blockchain-based technologies for autonomous vehicles". In: *Proceedings of the 1st Workshop on Cryptocurrencies and Blockchains for Distributed Systems*. 2018, pp. 87–92.

[204] Seung Jae Pee et al. "Blockchain based smart energy trading platform using smart contract". In: *2019 International Conference on Artificial Intelligence in Information and Communication (ICAIIC)*. IEEE. 2019, pp. 322–325.

[205] Chunrong Peng et al. "Blockchain for vehicular internet of things: Recent advances and open issues". In: *Sensors* 20.18 (2020), p. 5079.

[206] Doriane Perard et al. "Erasure code-based low storage blockchain node". In: *2018 IEEE International Conference on Internet of Things (iThings) and IEEE Green Computing and Communications (GreenCom) and IEEE Cyber, Physical and Social Computing (CPSCom) and IEEE Smart Data (SmartData)*. IEEE. 2018, pp. 1622–1627.

[207] GONCZOL PETER et al. "Blockchain Implementations and Use Cases for Supply Chains-A Survey". In: *IEEE Access* 8 (2020), pp. 11856–11871. DOI: 10.1109/ACCESS.2020.2964880.

[208] Claudia Pop et al. "Blockchain based decentralized management of demand response programs in smart energy grids". In: *Sensors* 18.1 (2018), p. 162.

[209] Deepak Prashar et al. "Integrating IOT and blockchain for ensuring road safety: An unconventional approach". In: *Sensors* 20.11 (2020), p. 3296.

[210] Ameyaw Prince Donkor and de Vries Walter Timo. "Transparency of Land Administration and the Role of Blockchain Technology, a Four-Dimensional Framework Analysis from the Ghanaian Land Perspective". In: *Land* (12 2020), p. 491.

[211] *Provenance. A Platform for Business.* Accessed 26/2/2021. 2019. URL: https://www.provenance.org/business/platform.

[212] Yazdan Ahmad Qadri et al. "The Future of Healthcare Internet of Things: A Survey of Emerging Technologies". In: *IEEE Communications Surveys & Tutorials* 22.2 (2020), pp. 1121–1167.

[213] Maciel M Queiroz, Renato Telles, and Silvia H Bonilla. "Blockchain and supply chain management integration: a systematic review of the literature". In: *Supply Chain Management: An International Journal* 25.2 (2019), pp. 241–254.

[214] *Quorumchain consensus.* https://github.com/ConsenSys/quorum/wiki. Accessed: 2020-11-15.

[215] Mohammad Saidur Rahman, Ibrahim Khalil, and Mohammed Atiquzzaman. "Blockchain-Powered Policy Enforcement for Ensuring Flight Compliance in Drone-Based Service Systems". In: *IEEE Network* 35.1 (2021), pp. 116–123.

[216] TP Abdul Rahoof and VR Deepthi. "HealthChain: A Secure Scalable Health Care Data Management System Using Blockchain". In: *International Conference on Distributed Computing and Internet Technology*. Springer. 2020, pp. 380–391.

[217] Wasim Ahmad Raja et al. "The Role of Blockchain Technology in Aviation Industry". In: *IEEE Aerospace and Electronic Systems Magazine* (2020).

[218] Geetanjali Rathee et al. "A blockchain framework for securing connected and autonomous vehicles". In: *Sensors* 19.14 (2019), p. 3165.

[219] Danda B Rawat et al. "Blockchain enabled named data networking for secure vehicle-to-everything communications". In: *IEEE Network* 34.5 (2020), pp. 185–189.

[220] *Recereum.* (Accessed on 15/3/2021). 2017. URL: http://recereum.com.

[221] Dakshita Reebadiya et al. "Blockchain-based Secure and Intelligent Sensing Scheme for Autonomous Vehicles Activity Tracking Beyond 5G Networks". In: *Peer-to-Peer Networking and Applications* (2021), pp. 1–18.

[222] George Reno Varghese et al. "Food quality traceability prototype for restaurants using blockchain and food quality data index." In: *Journal of Cleaner Production* 240.8 (2019), p. 118021.

[223] Ana Reyna et al. "On blockchain and its integration with IoT. Challenges and opportunities". In: *Future generation computer systems* 88 (2018), pp. 173–190.

[224] Derrick Rountree and Ileana Castrillo. *The basics of cloud computing: Understanding the fundamentals of cloud computing in theory and practice*. Newnes, 2013.

[225] Tara Salman et al. "Security services using blockchains: A state of the art survey". In: *IEEE Communications Surveys & Tutorials* 21.1 (2018), pp. 858–880.

[226] Yevhenii Semenko and Damien Saucez. "Distributed privacy preserving platform for ridesharing services". In: *International Conference on Security, Privacy and Anonymity in Computation, Communication and Storage*. Springer. 2019, pp. 1–14.

[227] Ghazanfar Shahgholian. "A brief review on microgrids: Operation, applications, modeling, and control". In: *International Transactions on Electrical Energy Systems* (2021), e12885.

[228] Pradip Kumar Sharma, Neeraj Kumar, and Jong Hyuk Park. "Blockchain-based distributed framework for automotive industry in a smart city". In: *IEEE Transactions on Industrial Informatics* 15.7 (2018), pp. 4197–4205.

[229] Surbhi Sharma and Baijnath Kaushik. "A survey on internet of vehicles: Applications, security issues &

solutions". In: *Vehicular Communications* 20 (2019), p. 100182.

[230] Bingqing Shen, Jingzhi Guo, and Yilong Yang. "MedChain: efficient healthcare data sharing via blockchain". In: *Applied sciences* 9.6 (2019), p. 1207.

[231] Shuyun Shi et al. "Applications of Blockchain in Ensuring the Security and Privacy of Electronic Health Record Systems: A Survey". In: *Computers & Security* (2020), p. 101966.

[232] Zaied Shouran, Ahmad Ashari, and Tri Priyambodo. "Internet of things (IoT) of smart home: privacy and security". In: *International Journal of Computer Applications* 182.39 (2019), pp. 3–8.

[233] Chang SHUCHIH E. and Chen YICHIAN. "When Blockchain Meets Supply Chain: A Systematic Literature Review on Current Development and Potential Applications". In: *IEEE Access* 8 (2020), pp. 62478–62494. DOI: 10.1109/ACCESS.2020.2983601.

[234] Maninderpal Singh, Gagangeet Singh Aujla, and Rasmeet Singh Bali. "Odob: One drone one block-based lightweight blockchain architecture for internet of drones". In: *IEEE INFOCOM 2020-IEEE Conference on Computer Communications Workshops (INFOCOM WKSHPS)*. IEEE. 2020, pp. 249–254.

[235] Rajani Singh, Ashutosh Dhar Dwivedi, and Gautam Srivastava. "Internet of Things Based Blockchain for Temperature Monitoring and Counterfeit Pharmaceutical Prevention". In: *Sensors* 20.14 (2020), p. 3951.

[236] Sushil Kumar Singh, Yi Pan, and Jong Hyuk Park. "Blockchain-enabled Secure Framework for Energy-Efficient Smart Parking in Sustainable City Environment". In: *Sustainable Cities and Society* (2021), p. 103364.

[237] Joao Sousa and Alysson Bessani. "From Byzantine consensus to BFT state machine replication: A latency-optimal transformation". In: *2012 Ninth European Dependable Computing Conference*. IEEE. 2012, pp. 37–48.

[238] You Sun et al. "A decentralizing attribute-based signature for healthcare blockchain". In: *2018 27th International conference on computer communication and networks (ICCCN)*. IEEE. 2018, pp. 1–9.

[239] Council of Supply Chain Management Professionals (CSCMP). *Supply chain Management, Terms and Glossary*. (Accessed on 17/1/2021). 2013. URL: ww.cscmp.org/CSCMP/Educate/SCM_Definitions_and_Glossary_of_Terms/CSCMP/Educate/SCM_Definitions_and_Glossary_of_Terms.aspx?hkey=60879588-f65f-4ab5-8c4b-6878815ef92.

[240] *Swachhcoin*. (Accessed on 15/3/2021). 2018. URL: http://swachhcoin.com/.

[241] Toqeer Ali Syed et al. "A comparative analysis of blockchain architecture and its applications: Problems and recommendations". In: *IEEE Access* 7 (2019), pp. 176838–176869.

[242] Yawen Tan, Jiajia Liu, and Nei Kato. "Blockchain-Based Key Management for Heterogeneous Flying Ad-Hoc Network". In: *IEEE Transactions on Industrial Informatics* (2020).

[243] Sudeep Tanwar, Karan Parekh, and Richard Evans. "Blockchain-based electronic healthcare record system for healthcare 4.0 applications". In: *Journal of Information Security and Applications* 50 (2020), p. 102407.

[244] Noshina Tariq, Farrukh Aslam Khan, and Muhammad Asim. "Security Challenges and Requirements for Smart Internet of Things Applications: A Comprehensive Analysis". In: *Procedia Computer Science* 191 (2021), pp. 425–430.

[245] Agrawal Tarun Kumar, Sharma Ajay, and Kumar Vijay. "Blockchain-Based Secured Traceability System for Textile and Clothing Supply Chain". In: *In Artificial Intelligence for Fashion Industry in the Big Data Era*. Springer, 2018, 197–208.

[246] Agrawal Tarun Kumar, Koehl Ludovic, and Campagne Christine. "A secured tag for implementation of traceability in textile and clothing supply chain." In: *The International Journal of Advanced Manufacturing Technology* 99 (2018), 2563–2577.

[247] Mohamed Torky and Aboul Ella Hassanein. "Integrating blockchain and the internet of things in precision agriculture: Analysis, opportunities, and challenges". In: *Computers and Electronics in Agriculture* 178 (2020), p. 105476. ISSN: 0168-1699. DOI: https://doi.org/10.1016/j.compag.2020.105476. URL: https://www.sciencedirect.com/science/article/pii/S0168169919324329.

[248] Choi Tsan-Ming et al. "The mean-variance approach for global supply chain risk analysis with air logistics in the blockchain technology era". In: *Transportation Research Part E: Logistics and Transportation Review* 127 (2019), pp. 178 –191. ISSN: 1366-5545. DOI: https://doi.org/10.1016/j.tre.2019.05.007. URL: http://www.sciencedirect.com/science/article/pii/S1366554519302601.

[249] Žiga Turk and Robert Klinc. "Potentials of Blockchain Technology for Construction Management". In: *Procedia Engineering* 196 (2017). Creative Construction Conference 2017, CCC 2017, 19-22 June 2017, Primosten, Croatia, pp. 638 –645. ISSN: 1877-7058. DOI: https://doi.org/10.1016/j.proeng.2017.08.052. URL: http://www.sciencedirect.com/science/article/pii/S187770581733179X.

[250] Md Ashraf Uddin et al. "Continuous patient monitoring with a patient centric agent: A block architecture". In: *IEEE Access* 6 (2018), pp. 32700–32726.

[251] Kristiina Valtanen. "Design challenges of developing a blockchain-enabled smart home". In: *2021 Conference on Information Communications Technology and Society (ICTAS)*. IEEE. 2021, pp. 34–39.

[252] Hargaden Vincent et al. "The Role of Blockchain Technologies in Construction Engineering Project

Management". In: *2019 IEEE International Conference on Engineering, Technology and Innovation (ICE/ITMC)*. 2019, pp. 1–6. DOI: 10.1109/ICE.2019.8792582.

[253] Jayneel Vora et al. "BHEEM: A blockchain-based framework for securing electronic health records". In: *2018 IEEE Globecom Workshops (GC Wkshps)*. IEEE. 2018, pp. 1–6.

[254] Chao Wang et al. "A survey: applications of blockchain in the Internet of Vehicles". In: *EURASIP Journal on Wireless Communications and Networking* 2021.1 (2021), pp. 1–16.

[255] Di Wang and Xiaohong Zhang. "Secure data sharing and customized services for intelligent transportation based on a consortium blockchain". In: *IEEE Access* 8 (2020), pp. 56045–56059.

[256] Hao Wang and Yujiao Song. "Secure cloud-based EHR system using attribute-based cryptosystem and blockchain". In: *Journal of medical systems* 42.8 (2018), p. 152.

[257] Naiyu Wang et al. "When energy trading meets blockchain in electrical power system: The state of the art". In: *Applied Sciences* 9.8 (2019), p. 1561.

[258] Qianlong Wang et al. "TrafficChain: A blockchain-based secure and privacy-preserving traffic map". In: *IEEE Access* 8 (2020), pp. 60598–60612.

[259] Qin Wang et al. "Blockchain for the IoT and industrial IoT: A review". In: *Internet of Things* (2019), p. 100081.

[260] Yingli Wang et al. "Making sense of blockchain technology: How will it transform supply chains?" In: *International Journal of Production Economics* 211 (2019), pp. 221–236. ISSN: 0925-5273. DOI: https://doi.org/10.1016/j.ijpe.2019.02.002. URL: http://www.sciencedirect.com/science/article/pii/S0925527319300507.

[261] Yuntao Wang et al. "Challenges and solutions in autonomous driving: A blockchain approach". In: *IEEE Network* 34.4 (2020), pp. 218–226.

[262] Pornpit Wongthongtham et al. "Blockchain-enabled Peer-to-Peer energy trading". In: *Computers & Electrical Engineering* 94 (2021), p. 107299.

[263] Qi Xia et al. "BBDS: Blockchain-based data sharing for electronic medical records in cloud environments". In: *Information* 8.2 (2017), p. 44.

[264] QI Xia et al. "MeDShare: Trust-less medical data sharing among cloud service providers via blockchain". In: *IEEE Access* 5 (2017), pp. 14757–14767.

[265] Junfeng Xie et al. "A survey of blockchain technology applied to smart cities: Research issues and challenges". In: *IEEE Communications Surveys & Tutorials* 21.3 (2019), pp. 2794–2830.

[266] Lixia Xie et al. "Blockchain-based secure and trustworthy Internet of Things in SDN-enabled 5G-VANETs". In: *IEEE Access* 7 (2019), pp. 56656–56666.

[267] Hang Xiong et al. "Blockchain Technology for Agriculture: Applications and Rationale". In: *Frontiers in Blockchain* 3 (2020), p. 7. ISSN: 2624-7852. DOI: 10.3389/fbloc.2020.00007. URL: https://www.frontiersin.org/article/10.3389/fbloc.2020.00007.

[268] Xu Xiwei, Weber Ingo, and Mark Staples. "Case Study: AgriDigital". In: *Architecture for Blockchain Applications*. 2019. DOI: https://doi.org/10.1007/978-3-030-03035-3_12.

[269] Haitao Xu et al. "Edge Computing Resource Allocation for Unmanned Aerial Vehicle Assisted Mobile Network With Blockchain Applications". In: *IEEE Transactions on Wireless Communications* 20.5 (2021), pp. 3107–3121.

[270] Yibin Xu and Yangyu Huang. "Segment blockchain: A size reduced storage mechanism for blockchain". In: *IEEE Access* 8 (2020), pp. 17434–17441.

[271] Jingting Xue, Chunxiang Xu, and Yuan Zhang. "Private Blockchain-Based Secure Access Control for Smart Home Systems". In: *KSII Transactions on Internet & Information Systems* 12.12 (2018).

[272] Xiaodong Yang et al. "Medical Data Sharing Scheme Based on Attribute Cryptosystem and Blockchain Technology". In: *IEEE Access* 8 (2020), pp. 45468–45476.

[273] Yao-Tsung Yang et al. "Blockchain-based traffic event validation and trust verification for VANETs". In: *IEEE Access* 7 (2019), pp. 30868–30877.

[274] Zhe Yang et al. "Blockchain-based decentralized trust management in vehicular networks". In: *IEEE Internet of Things Journal* 6.2 (2018), pp. 1495–1505.

[275] Abbas Yazdinejad et al. "Enabling drones in the internet of things with decentralized blockchain-based security". In: *IEEE Internet of Things Journal* 8.8 (2020), pp. 6406–6415.

[276] Kayikci Yaşanur et al. "Food supply chain in the era of Industry 4.0: blockchain technology implementation opportunities and impediments from the perspective of people, process, performance, and technology". In: *Production Planning & Control* 0.0 (2020), pp. 1–21.

[277] Abdullah Yildizbasi. "Blockchain and renewable energy: Integration challenges in circular economy era". In: *Renewable Energy* 176 (2021), pp. 183–197.

[278] Wenchi Ying, Suling Jia, and Wenyu Du. "Digital enablement of blockchain: Evidence from HNA group". In: *International Journal of Information Management* 39 (2018), pp. 1–4. ISSN: 0268-4012. DOI: https://doi.org/10.1016/j.ijinfomgt.2017.10.004. URL: http://www.sciencedirect.com/science/article/pii/S0268401217308423.

[279] Haixia You, Haochen Hua, and Junwei Cao. "A Smart Contract-based Energy Trading Strategy in Energy Internet". In: *2019 IEEE International Conference on Energy Internet (ICEI)*. IEEE. 2019, pp. 478–483.

[280] Yong Yuan and Fei-Yue Wang. "Towards blockchain-based intelligent transportation systems". In: *2016 IEEE 19th international conference on*

*intelligent transportation systems (ITSC)*. IEEE. 2016, pp. 2663–2668.

[281] Tsang Yung Po et al. "Blockchain-Driven IoT for Food Traceability with an Integrated Consensus Mechanism." In: *IEEE Access*. IEEE. 2019, 129000–129017.

[282] Bessem Zaabar et al. "HealthBlock: A secure blockchain-based healthcare data management system". In: *Computer Networks* 200 (2021), p. 108500.

[283] Bessem Zaabar et al. "Secure and Privacy-aware Blockchain-based Remote Patient Monitoring System for Internet of Healthcare Things". In: *2021 17th International Conference on Wireless and Mobile Computing, Networking and Communications (WiMob)*. IEEE. 2021, pp. 200–205.

[284] Pengjie Zeng et al. "A Scheme of Intelligent Traffic Light System Based on Distributed Security Architecture of Blockchain Technology". In: *IEEE Access* 8 (2020), pp. 33644–33657.

[285] Can Zhang et al. "BSFP: blockchain-enabled smart parking with fairness, reliability and privacy protection". In: *IEEE Transactions on Vehicular Technology* 69.6 (2020), pp. 6578–6591.

[286] David Zhang. "Application of Blockchain Technology in Incentivizing Efficient Use of Rural Wastes: A case study on Yitong System". In: *Energy Procedia* 158 (2019). Innovative Solutions for Energy Transitions, pp. 6707–6714. ISSN: 1876-6102. DOI: https://doi.org/10.1016/j.egypro.2019.01.018. URL: https://www.sciencedirect.com/science/article/pii/S187661021930027X.

[287] Jun Zhang et al. "Blockchain based intelligent distributed electrical energy systems: needs, concepts, approaches and vision". In: *Zidonghua Xuebao/acta Automatica Sinica* 43.9 (2017), pp. 1544–1554.

[288] Peng Zhang et al. "FHIRChain: applying blockchain to securely and scalably share clinical data". In: *Computational and structural biotechnology journal* 16 (2018), pp. 267–278.

[289] Xiaohong Zhang and Di Wang. "Adaptive traffic signal control mechanism for intelligent transportation based on a consortium blockchain". In: *IEEE Access* 7 (2019), pp. 97281–97295.

[290] Xuefei Zhang et al. "Blockchain based secure package delivery via ridesharing". In: *2019 11th International Conference on Wireless Communications and Signal Processing (WCSP)*. IEEE. 2019, pp. 1–6.

[291] Fan Zhi-Ping, Wu Xue-Yan, and Cao Bing-Bing. "Considering the traceability awareness of consumers: should the supply chain adopt the blockchain technology?." In: *Annals of Operations Research* (2020), pp. 1–24.

[292] Qiheng Zhou et al. "Solutions to scalability of blockchain: A survey". In: *Ieee Access* 8 (2020), pp. 16440–16455.

[293] Yiyun Zhou et al. "Improving iot services in smart-home using blockchain smart contract". In: *2018 IEEE International Conference on Internet of Things (iThings) and IEEE Green Computing and Communications (GreenCom) and IEEE Cyber, Physical and Social Computing (CPSCom) and IEEE Smart Data (SmartData)*. IEEE. 2018, pp. 81–87.

[294] Peng Zhuang, Talha Zamir, and Hao Liang. "Blockchain for cybersecurity in smart grid: A comprehensive survey". In: *IEEE Transactions on Industrial Informatics* 17.1 (2020), pp. 3–19.

[295] Haider Dhia Zubaydi et al. "A review on the role of blockchain technology in the healthcare domain". In: *Electronics* 8.6 (2019), p. 679.